%% file: lopezMarti_cham.tex
\documentclass[structabstract]{aa}  
\usepackage{graphicx}
\usepackage{natbib}
\bibpunct{(}{)}{;}{a}{}{,} 
\usepackage{txfonts}
\usepackage{lscape}
\usepackage{longtable,lscape}
\usepackage{color}

\begin{document}
   \title{Proper motions of young stars in Chamaeleon}

   \subtitle{I. A Virtual Observatory study of spectroscopically confirmed members}

 \author{Bel\'en L\'opez Mart\'{\i}
          \inst{1}
          \and
          Francisco Jim\'enez-Esteban \inst{1,2,3}
          \and 
          Amelia Bayo  \inst{4}
          \and
          David Barrado \inst{1,5}
	  \and 
	  Enrique Solano \inst{1,2}
	  \and
	  Carlos Rodrigo \inst{1,2}
          }

   \institute{
        Centro de Astrobiolog\'{\i}a (INTA-CSIC), Departamento de Astrof\'{\i}sica, P.O. Box 78, E-28261 Villanueva de la Ca\~nada, Madrid, Spain\\
	\email{belen@cab.inta-csic.es}
	\and
	Spanish Virtual Observatory, Spain
	\and
	Saint Louis University, Madrid Campus, Division of Science and Engineering, Avenida~del~Valle 34, E-28003 Madrid, Spain
	 \and
          European Southern Observatory, Alonso de C\'ordova 3107, Vitacura, Santiago, Chile         
	\and
	Calar Alto Observatory, Centro Astron\'omico Hispano-Alem\'an, C/ Jes\'us Durb\'an Rem\'on 2-2, E-04004 Almer\'{\i}a, Spain
	}

   \date{Received; accepted }

 
  \abstract
   {The study of the motion of the members of a given open cluster or stellar association provides key information about their formation and early evolution. The Chamaeleon cloud complex constitutes one of the closest and best studied low-mass star-forming regions in the Galaxy. 
   }
   {We want to provide further evidence of the origin of the proposed stellar members of Chamaeleon and to identify interlopers from the foreground $\epsilon$~Cha and $\eta$~Cha associations. 
  }
   {We compile lists of spectroscopically confirmed members of Chamaeleon~I and II, $\epsilon$~Cha and $\eta$~Cha, and of background objects in the same line of sight.  Using Virtual Observatory tools,
we cross-match these lists with the UCAC3 catalogue to get the proper motions of the objects. In the vector point diagram, we identify the different moving groups, and use this information to study the membership of proposed candidate members of the associations from the literature. For those objects with available radial velocities, we compute their Galactic space velocities. We look for correlations between the known properties of the objects and their proper motions.
   }
   {The members of the dark clouds exhibit clearly different proper motions from those of the foreground associations and of the background stars. The data suggest that Chamaeleon~II could have different dynamical properties from Chamaeleon~I. Although the two foreground clusters $\epsilon$ and $\eta$~Chamaeleontis constitute two different proper motion groups, they have similar spatial motions, which are different from the spatial motion of Chamaeleon~I. On the other hand, the space motions of the Chamaeleon~II stars look more similar to those of the foreground clusters than to the Chamaeleon~I stars, but the numbers are low. We find no correlations between the proper motions and the properties of the objects in either of the clouds.
   }
 {On the basis of proper motion, Chamaeleon~I and II constitute two physical entities unrelated to the foreground $\epsilon$ and $\eta$~Chamaeleontis clusters, but with the available data it is unclear to what extent the stellar populations in both clouds are physically connected to each other. 
}


   \maketitle

%
\section{Introduction}\label{sec:intro}

The kinematic properties of the members of a given stellar association hold important clues to its history. Some formation models predict that the early dynamical evolution of the parent proto-stellar cluster should lead to mass-dependent kinematic distributions, and, eventually, to an efficient mass segregation \citep{kroupa2003}, while other numerical simulations predict similar kinematic properties over the whole mass spectrum \citep{e.g. bate2003,bate2012}. Various authors have used radial velocity measurements to study the kinematic properties of young low-mass objects \citep[e.g.][]{jeffries2006,maxted2008}, and a number of surveys have used proper motions to  identify and confirm new low-mass members in young associations and clusters \citep[e.g.][]{moraux2001,kraus2007,bouy2009,caballero2010}. In particular, our recent proper motion study of the \object{Lupus} star-forming region shows that it is possible to distinguish between probable low-mass members of the complex and likely contaminants using kinematic information from available astrometric catalogues \citep[][hereafter LJS11]{volupus}.

The \object{Chamaeleon} complex is one of the closest and best studied low-mass star-forming regions. At an estimated distance in the range 115-215~pc, it is composed of three dark clouds, named Chamaeleon~I, II and III  \citep{schwartz1977,schwartz1991}. The oldest one is \object{Chamaeleon~I}, with a mean age of about 2~Myr; it contains more than 300 known young stars \citep[see][for the latest census]{luhman2008a}, most  of them clustered in two cloud cores containing two intermediate-mass stars, \object{HD~97048} and \object{HD~97300}. \object{Chamaeleon~II} seems to be at an earlier stage of evolution than Chamaeleon~I, because it contains more embedded than visual objects \citep{gauvin1992}. Surveys by several authors have identified more than 60 stars in this cloud \citep[][and references therein]{spezzi2007,spezzi2008,alcala2008}. No active star formation seems to be taking place in \object{Chamaeleon~III}.

The Chamaeleon clouds have also been the target of kinematic studies, generally based on radial velocities and focused on the brightest stars in the complex \citep[e.g.][]{dubath1996}. The exceptions are a few very low-mass stars and brown dwarf candidates in Chamaeleon~I \citep[][and references therein]{joergens2006}. Proper motion and parallax information from Hipparcos has been used to assess the origin of the star formation in the clouds and their relation to other neighbouring associations \citep{sartori2003}. X-ray observations during the past two decades, combined with Hipparcos astrometry, have shown that the  young stars observed towards the Chamaeleon sky area can be grouped into several distinct  associations: the Chamaeleon dark cloud complex itself, and the foreground (at around 100~pc) \object{$\epsilon$ Chamaeleontis} (\object{$\epsilon$~Cha}) and \object{$\eta$ Chamaeleontis} (\object{$\eta$~Cha}) young associations \citep{feigelson2003,mamajek1999}. The observations have shown that these two associations are older than the Chamaeleon star-forming complex ($\sim$15-20~Myr), but it is unclear whether there is a relation between the two, or whether they are somehow related to the dark clouds.

In this paper, we make use of proper motions from the Third US Naval Observatory CCD Astrograph Catalog \citep[UCAC3;][]{zacharias2010} and of Virtual Observatory\footnote{The VO is a project designed to provide the astronomical community with the data access and the research tools 
needed to enable the exploration of the digital, multiwavelength universe resident in the astronomical data archives. http://www.ivoa.net} (VO) tools to investigate the kinematic properties of members and candidate members of the Chamaeleon dark clouds. Our goals are to provide further evidence of a common origin for all these objects and to test their association with the dark clouds and between the clouds themselves.

%
\section{Available data}\label{sec:data}

\subsection{Object compilation}\label{sec:cats}

We compiled a list of Chamaeleon confirmed members (through spectroscopy) from the literature. For Chamaeleon~I, we retrieved the list from the most recent census by \citet{luhman2007}, and completed it with the eight new members reported by \citet{luhman2008}. The final list contained 304 objects down to $R=24$~mag, with extinctions in the range $A_V\simeq0$-20~mag). For Chamaeleon~II, we used the list of confirmed members from \citet{spezzi2008}, which includes all objects studied in previous works. In total, we had 41 Chamaeleon~II spectroscopically confirmed members, with $R$-band magnitudes down to about 23~mag and extinctions in the range $A_V\simeq0$-11~mag. 

To assess how clearly we can separate the different populations based on the available proper motions, and to check the possible presence of interlopers from the foreground clusters in our Chamaeleon~I and II lists, we compiled two lists of members for the $\epsilon$~Cha and $\eta$ Cha associations, merging the catalogues provided by \citet{luhman2004b} and \citet{fernandez2008} for both groups, and adding the objects from the lists by \citet{luhman2004c} in $\eta$~Cha and by \citet{silva2009} in $\epsilon$~Cha. Membership of these objects in the corresponding cluster has been confirmed by several different means, including spectroscopy and proper motions. The final compiled lists contained 36 stars in $\epsilon$~Cha and 20 in $\eta$~Cha down to $R\sim18$~mag. No estimations of extinction are available for these stars, but their location outside areas of dense dust suggests that it is very low in most cases. 

For control purposes, we also considered the list of 88 background sources from Table~4 of \citet{luhman2007}.  These objects were initially proposed in the literature as Chamaeleon~I members, but they were later discarded through spectroscopy.

We  note that the magnitude limit of most of these compilations is much fainter than the UCAC3 magnitude limit (see Sect.~\ref{sec:xm} below); therefore,  a significant fraction of objects are not expected to have counterparts in that catalogue. Since the faintest objects are also the 
lowest mass ones, and given that the substellar mass limit is estimated to be around $R\sim21$~mag at the age and distance of the Chamaeleon clouds, this implies that only stellar members of the associations will have measured proper motions. In particular, the objects at the substellar boundary studied by \citet{wfichai} in Chamaeleon~I and \citet{barrado2004} in Chamaeleon~II are not included in UCAC3.

In a second step in our study, we also investigated those candidate members of the Chamaeleon associations from the literature whose true nature is still unclear. This list included 63 X-ray emitting stars unveiled by the ROSAT mission \citep{alcala1995} that remained unclassified, as well as some objects from other previous works still lacking spectroscopic confirmation of youth: the list of uncertain objects provided by \citet[][his Table~8; 21 objects]{luhman2004} for Chamaeleon~I, and the candidate members of Chamaeleon~II without spectroscopic confirmation of membership from \citet[][11 objects]{spezzi2008} and the DENIS survey \citep[][41 objects]{vuong2001}. 

In total, our list of unconfirmed candidates amounted to 136 objects (63 ROSAT stars and 73 objects from other works), but the vast majority of them are too faint to be included in the proper motion study, and only a handful of them have proper motion measurements available in UCAC3, as explained in Sect.~\ref{sec:cand}.

\subsection{Proper motion data}\label{sec:xm}

\addtocounter{table}{1} 
\addtocounter{table}{1} 
\addtocounter{table}{1} 

To get proper motion measurements, we cross-matched our compiled catalogues with the UCAC3 catalogue, available within the Virtual Observatory. This is an all-sky survey containing about 100 million objects, 95\% of them with proper motions, covering a dynamical range of about 8-16~mag in a single bandpass between V and R. Its positional accuracy is about 15 to 100 mas per coordinate, depending on magnitude. The proper motion errors range from 1 to 10 mas/yr, depending on magnitude and observing history. 

As in our previous work on Lupus (LJS11), we followed a VO-based methodology to cross-match and analyse the data. We made use of the Multiple Cone Search utility of TOPCAT\footnote{TOPCAT is an interactive graphical viewer and editor for tabular data that allows the user to examine, analyse, combine, and edit astronomical tables. http://www.star.bris.ac.uk/$\sim$mbt/topcat/}. A matching radius of 2$\arcsec$ was used. We retained all sources whose proper motion errors were not set to zero, which had been computed using more than two epoch positions, and which had an object classification flag (ot) between 0 and 3. After purging the data in this way, the number of counterparts with proper motion data in Chamaeleon~I and II were 81 and 25 (corresponding to the 27\% and 61\% of the total number of stars included in our compiled lists), respectively. In addition, UCAC3 provided proper motions for 19 objects in $\epsilon$~Cha and 12 in $\eta$~Cha, respectively (53\% and 60\% of the compiled lists, respectively). As for the background sources, the number of objects with available proper motions was 52 (59\%). These samples were further cleaned by removing bad proper motion measurements and interlopers identified during the analysis (see Sect.~\ref{sec:mgs} for details).  The proper motion data for the members of the dark clouds, the foreground clusters, and the contaminants are presented in Tables~\ref{tab:mem}, \ref{tab:clus}, and \ref{tab:cont}, respectively. 

In LJS11, a comparison of the cross-matching results of the Lupus catalogues with different astrometric catalogues (USNO-B1, SuperCOSMOS, PPMX) led to the conclusion that UCAC3 provided the best compromise between the number of counterparts (dependent among others on the magnitude limit of the catalogue) and the typical proper motion errors. For the present work, we further compared the analysis presented in Sect.~\ref{sec:groups} with the results obtained using the recently released PPMXL catalogue \citep{roeser2010}. Although this catalogue is deeper in magnitude than UCAC3, and thus included much more sources from our initial member lists, it was not possible to distinguish the different populations due to the larger errors in the proper motions for most of the objects. Thus, we concluded that UCAC3 is still the best option available for our study.

\subsection{Complementary data}\label{sec:data}

\subsubsection{Radial velocities}\label{sec:rvdata}

Aiming at better insight into the kinematics of the members of the Chamaeleon associations, we combined the UCAC3 proper motions with radial velocity measurements from the literature \citep[][]{covino1997,joergens2001,biazzo2012a} and from the SIMBAD database. We only considered measurements with errors better than about 30\%. If several measurements were available for the same star, the average was computed. Known and suspected spectroscopic binaries were discarded.

Our final sample contains radial velocity information for 18 stars in Chamaeleon~I and nine in Chamaeleon~II. These data are summarized in Table~\ref{tab:mem}. We also retrieved radial velocity data for 15 stars in the $\epsilon$~Cha association and four stars in the $\eta$~Cha association, which are listed in Table~\ref{tab:clus}. We also considered radial velocities from \citet{covino1997} for the candidate members of the associations discussed in Sect.~\ref{sec:rosat}.

\addtocounter{table}{1} 

\subsubsection{Photometry}\label{sec:phot}

We compiled multiwavelength photometry for the confirmed members and candidate members of the Chamaeleon associations discussed in this work. Most of the data were retrieved using  the latest version of the VO SED Analyzer tool\footnote{http://svo2.cab.inta-csic.es/svo/theory/vosa/} 
\citep[VOSA;][]{bayo2008,bayo2012}. They include ultraviolet photometry from GALEX \citep{bianchi2000}, optical photometry ($B$, $V$ and $I$) from the Tycho-2 and DENIS catalogues \citep{hog2000,denis2005}, near-infrared photometry from 2MASS \citep{skrutskie2006}, and mid-infrared photometry from AKARI and WISE  \citep{ishihara2010,cutri2012}.

More photometry data from the literature were retrieved through  the VizieR catalogue service \citep{ochsenbein2000}. For Chamaeleon~II members and candidate members, we retrieved the compilation of optical and infrared photometry by \citet{alcala2008}, which includes Johnson $U$-band data from \citet{hughes1992}, Cousins $R_CI_C$ and Sloan $z$ photometry from \citet{spezzi2007}, and mid- and far-infrared photometry from IRAS and from the {\em ``cores to disks''} ({\em c2d}) {\em Spitzer} legacy programme \citep{evans2003}.\footnote{The {\em c2d} data are now retrievable through VOSA \citep{bayo2012}.} 
For Chamaeleon~I members and candidate members, we retrieved the  {\em Spitzer}/IRAC and MIPS 24~$\mu$m photometry provided by \citet{luhman2008}. Optical photometry from the survey by \citet{wfichai} was also included in the data compilation for this cloud, as well as IRAS fluxes for three objects. The photometric information is summarized in Table~\ref{tab:phot}.\footnote{Available online.}

\section{Kinematic groups towards Chamaeleon}\label{sec:mgs}

   \begin{figure*}[t]
   \centering
  \includegraphics[width=18cm]{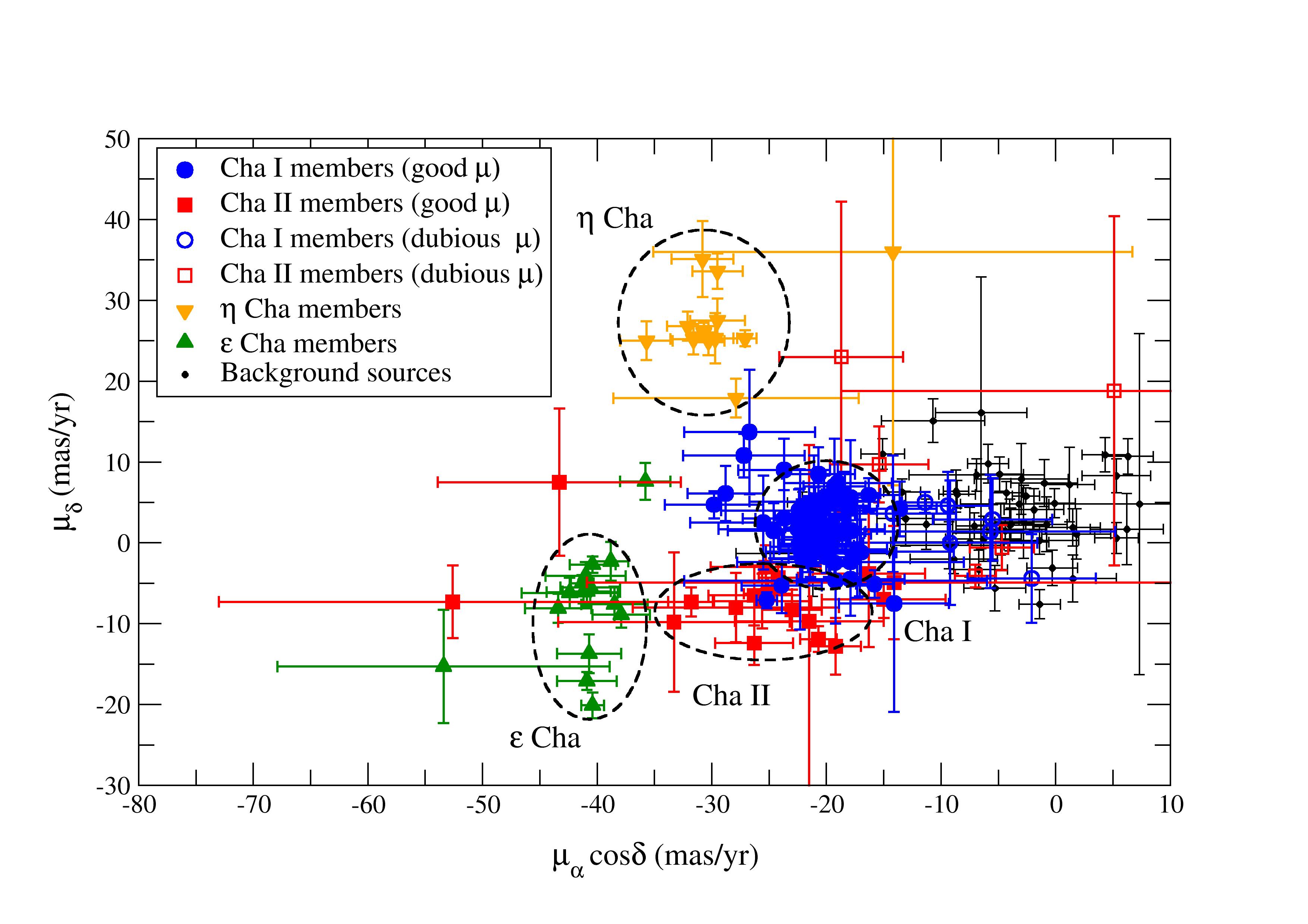}\hfill 
      \caption{\footnotesize
	       UCAC3 vector point diagram for the objects seen towards the Chamaeleon sky area. Only spectroscopically confirmed members are considered. Sources belonging to different kinematic associations and to the background have been plotted with different symbols and colours. Objects with reliable and dubious proper motions have been indicated with solid and open symbols, respectively (see Sect.~\ref{sec:groups} for details). To help the eye, the moving groups have been labelled and marked with ellipses. Some interlopers (objects with discrepant proper motions that are more likely members of other associations) are identified (see discussion in Sect.~\ref{sec:inter}).
	        }
         \label{fig:pm}
   \end{figure*}

\subsection{Identification of proper motion groups}\label{sec:groups}

\begin{figure}[t]
   \centering
  \includegraphics[width=10.5cm]{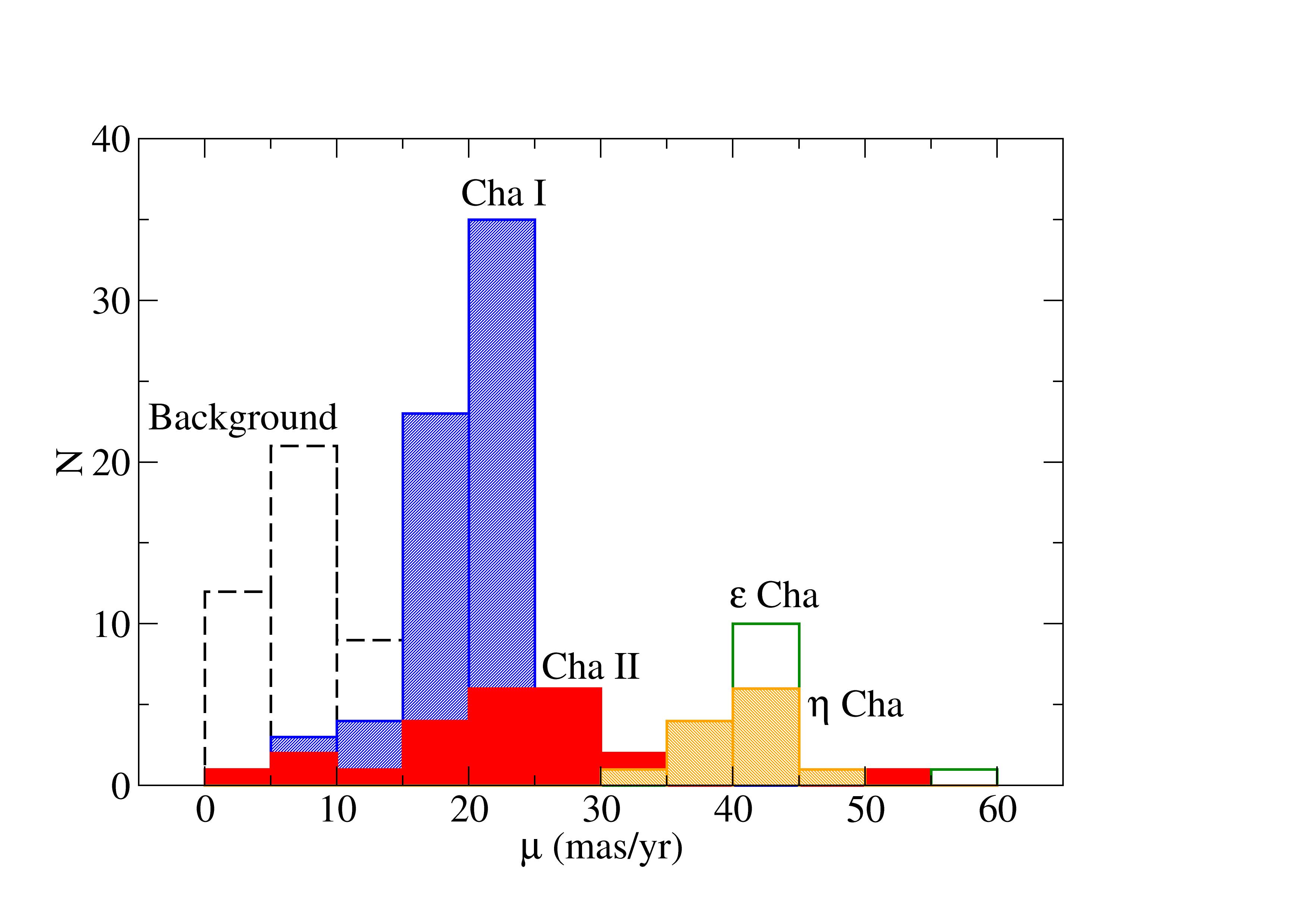}\hfill
      \caption{\footnotesize
	       Total proper motion histograms for the different moving groups: Cha~I (blue hashed histogram), Cha~II (filled red histogram), $\epsilon$~Cha (green  empty histogram), $\eta$~Cha (orange hashed histogram) and background sources (black dashed-line histogram).  Only objects with reliable proper motions are considered.
 }
         \label{fig:hist}
   \end{figure}

\begin{figure*}[ht]
   \centering
     \includegraphics[width=18cm]{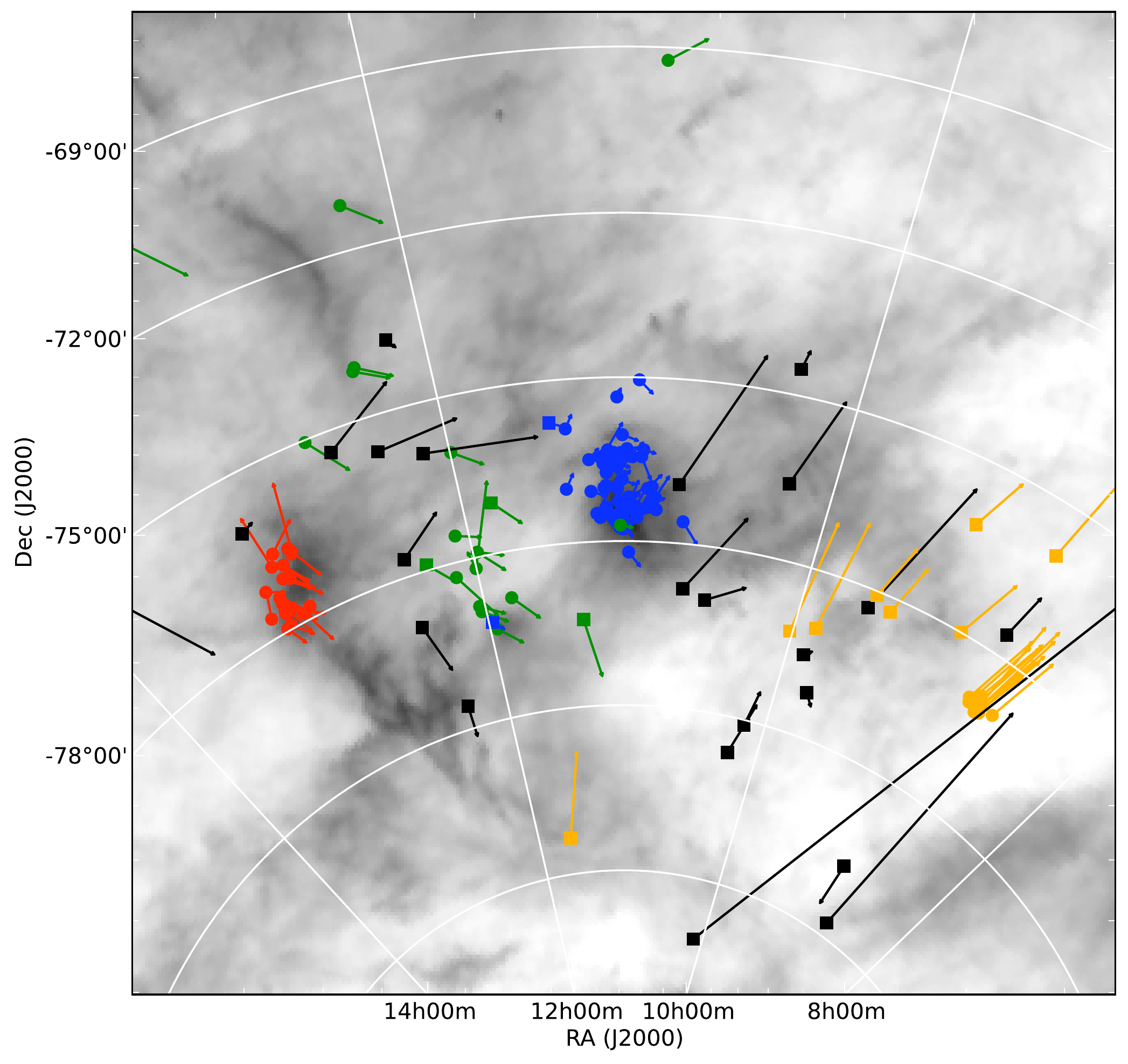}\hfill\\
      \caption{\footnotesize
	       Current spatial location of the members of the Chamaeleon associations (circles) and of the ROSAT sources discussed in Sect.~\ref{sec:rosat} (squares). Colours as in Fig.~\ref{fig:pm}. Also shown are the expected displacement of these objects within 10$^5$~Myr. The background image is a dust map by Schlegel et al. \citep{schlegel1998}. Only objects with reliable proper motions are considered.
	       }
         \label{fig:map}
   \end{figure*}

Figure~\ref{fig:pm} shows the vector point diagram for the young sources seen towards the Chamaeleon sky area that are included in the UCAC3 astrometric catalogue. In Fig.~\ref{fig:hist} we can see the histograms of the total proper motion modulus for the Chamaeleon~I and II members, for the members of the foreground associations, and for the known background contaminants. As expected, the young objects are clearly separated from the older background objects in these plots, and several distinct kinematic groups can be seen, which correspond to different spatial locations in the sky (see Fig.~\ref{fig:map}).

The spectroscopically confirmed members of Chamaeleon~I (hereafter ``the \object{Cha\,I moving group}'') are clustered in the same area of the vector-point diagram, roughly around the position ($-19, +2$)~mas/yr. The Chamaeleon~II objects, on the other hand, are seen mostly clustered around the position ($-25, -7$)~mas/yr, slightly shifted from the Chamaeleon~I sources (Fig.~\ref{fig:pm}). This seems to suggest that these objects (hereafter ``the \object{Cha\,II moving group}'') are moving, on average, faster than the Chamaeleon~I stars in the tangential direction. This is also seen in the shift of the peak of the total proper motion distribution in Chamaeleon~II with respect to Chamaeleon~I (Fig.~\ref{fig:hist}). If real, this result is interesting, because all estimations in the literature place Chamaeleon~II at the same distance or further away than Chamaeleon~I; hence, the difference in proper motion between both populations cannot be attributed to distance. However, the difference in mean proper motions is of little significance owing to the low number of Chamaeleon~II sources with measured proper motions, and to the large dispersion. 

The two foreground associations form in turn two distinct kinematic groups in the vector point diagram, which are clearly different from those in the dark clouds. The mean UCAC3 proper motions for these two groups are in good agreement with previously published values \citep{kharchenko2005,wu2009}. Interestingly, the total proper motion is very similar in both associations, as shown by the histograms in Fig.~\ref{fig:hist}. As shown later in Sect.~\ref{sec:rv}, the radial velocities of the stars in both clusters are also very similar.

Finally, with a few exceptions, the objects classified as background contaminants have lower proper motion values than any of the young moving groups and are clustered in a different location of the vector point diagram. This confirms our ability to distinguish members and non-members of the Chamaeleon associations with the UCAC3 data. Even so, the separation between the groups is not complete, and there is some overlapping between young stars and contaminants and between young stars from different groups. 

A few spectroscopically confirmed members from both Chamaeleon clouds present UCAC3 proper motions that are clearly discrepant from those of their attributed associations. To check the reliability of the reported measurements, we followed the same procedure as in LJS11: Using {\em Aladin}, we visually inspected these objects, blinking and comparing two sets of images separated several decades in time, from the optical POSS and the near-infrared 2MASS surveys \citep{skrutskie2006}. The 2MASS sources and the UCAC3 counterparts were superimposed on the images to assess the reliability of the cross-match. Besides, we used other available astro-photometric databases, namely the Astrographic Catalogue AC2000.2 \citep{urban1998} and the SuperCOSMOS Sky Survey \citep{hambly2001} to verify the peculiar proper motions, and the USNO-B Catalog \citep{monet2003} and the PPMX and PPMXL catalogues \citep[][2010]{roeser2008} for comparison. This procedure showed that the clearly discrepant proper motions were mostly caused by errors in the measurements, mainly from centroiding errors or source confusion. A rough estimation of the proper motions of the objects, based on the comparison of the positions measured by the different catalogues, and with the motion of other nearby cloud members, suggests that most of them have proper motions in relatively good agreement with the corresponding moving groups.

The reliability of the proper motions is difficult to assess in some cases (e.g. \object{HD~104237}D and E) because these sources are part of visual binaries or multiples not always resolved in the images. And in a few cases (e.g. \object{CW~Cha}), we could not test the proper motions in this way because SuperCOSMOS provides no data for the stars, or else the data were wrong. The sources with clearly wrong proper motions (7 from Chamaeleon~I and 2 from Chamaeleon~II) were removed from the vector point diagram of Fig.~\ref{fig:pm}. They were also excluded from Tables~\ref{tab:mem} and \ref{tab:clus}. Objects with dubious proper motions (because they could not be tested or the checks were unconclusive) were not removed from the tables, but a flag was added. 

We also identified four spectroscopically confirmed members (\object{2MASS~J11183572-7935548}, \object{CM~Cha}, \object{Sz~60W}, and \object{CP-68 1388}) whose UCAC3 proper motions seem to be right but do not agree with the means of their attributed associations; these interlopers are discussed in Sect.~\ref{sec:inter} below.

\subsection{Stars with Hipparcos and Tycho measurements}\label{sec:hip}

\begin{table*}[t]
\label{tab:hip}
\caption{Hipparcos and Tycho proper motions and Hipparcos parallaxes for Chamaeleon stars}
\centering
\begin{tabular}{l c r c r c c c c}
\hline
\hline\noalign{\smallskip}
  \multicolumn{1}{l}{} &
  \multicolumn{2}{c}{UCAC3} &
  \multicolumn{2}{c}{Hipparcos/Tycho} &
  \multicolumn{1}{c}{} &
  \multicolumn{1}{c}{} &  
  \multicolumn{1}{c}{} &
  \multicolumn{1}{c}{} \\
  \multicolumn{1}{l}{Name} &
  \multicolumn{1}{c}{$\mu_{\alpha}\cos\delta$} &
  \multicolumn{1}{c}{$\mu_{\delta}$} &
  \multicolumn{1}{c}{$\mu_{\alpha}\cos\delta$} &
  \multicolumn{1}{c}{$\mu_{\delta}$} &
  \multicolumn{1}{c}{$\pi$} &
  \multicolumn{1}{c}{$d$} &  
  \multicolumn{1}{c}{Reference$^{\mathrm{a}}$} &
  \multicolumn{1}{c}{Remarks$^{\mathrm{b}}$} \\
  \multicolumn{1}{l}{} &
  \multicolumn{1}{c}{(mas/yr)} &
  \multicolumn{1}{c}{(mas/yr)} &
  \multicolumn{1}{c}{(mas/yr)} &
  \multicolumn{1}{c}{(mas/yr)} &
  \multicolumn{1}{c}{(mas)} &
  \multicolumn{1}{c}{(pc)} &
  \multicolumn{1}{c}{} &
  \multicolumn{1}{c}{} \\
\noalign{\smallskip}\hline\noalign{\smallskip}

  \multicolumn{7}{c}{Chamaeleon~I cloud}\\
  \noalign{\smallskip}
\object{HD~93828}  &  --25.2\,$\pm$\,0.9  & --7.1\,$\pm$\,1.0   & $-26.72\pm0.99$ & $-8.91\pm0.84$  & $5.70\pm0.96$ & $175\pm30$ & H &  \\
\object{HD~97048}  &  --17.8\,$\pm$\,0.8  & 1.5\,$\pm$\,0.8     & $-18.64\pm0.72$ &  $+2.18\pm0.62$  & $6.31\pm0.62$ & $158\pm16$ &  H &  \\
\object{HD~97300}  &  --17.7\,$\pm$\,1.0  & --2.7\,$\pm$\,1.1   & $-21.63\pm0.94$ & $-0.72\pm0.78$  & $5.60\pm0.88$ & $178\pm28$ &  H & d \\
\object{CV~Cha}    &  --21.3\,$\pm$\,1.1  & 1.9\,$\pm$\,1.0     & $-20.82\pm2.95$ & $+1.32\pm2.78$   & $6.97\pm2.86$ & $143\pm59$ & H &  \\
\object{CW~Cha}    &  --13.6\,$\pm$\,3.2  & 4.2\,$\pm$\,3.4     & $-26.05\pm3.82$ & $-6.46\pm3.55$  & $3.92\pm3.56$ & $255\pm232$ &  H & d \\
\noalign{\smallskip}\hline\noalign{\smallskip}

  \multicolumn{7}{c}{$\epsilon$~Chamaeleontis cluster}\\
  \noalign{\smallskip}
\object{CP-68~1388}      & --35.8\,$\pm$\,2.2  & 7.6\,$\pm$\,2.3  &  $-36.3\pm1.9$   &   $+4.7\pm1.6$	 &				 &	      & T &  \\
\object{T~Cha}           & --41.2\,$\pm$\,3.2  & --6.\,$\pm$\,1.4 &  $-33.45\pm3.87$ &   $-9.36\pm2.97$  &   $6.11\pm2.96^{\mathrm{c}}$  & $164\pm79$ & H & d \\ 
\object{DW~Cha}          & --40.4\,$\pm$\,2.0  & --2.\,$\pm$\,1.0 &  $-41.7\pm1.62$  &   $+0.13\pm1.4$   &   $11.06\pm1.52$		 &  $90\pm12$ & H & \\ 
\object{RX~J1159.7-7601} & --40.9\,$\pm$\,2.1  & --6.\,$\pm$\,1.8 &  $-40.53\pm1.66$ &   $-5.83\pm1.49$  &   $9.89\pm1.72$		 & $101\pm18$ & H & \\ 
\object{HD~104467}	 & --41.3\,$\pm$\,1.0  & --5.\,$\pm$\,1.0 &  $-39.8\pm1.7$   &   $-6.3\pm1.6$	 &				 &	      & T & \\
\object{HD~105923}	 & --37.9\,$\pm$\,2.5  & --8.\,$\pm$\,1.6 &  $-37.6\pm1.7$   &   $-7.7\pm1.6$	 &				 &	      & T & \\
\object{RX~J1239.4-7502} & --40.7\,$\pm$\,2.8  & --13\,$\pm$\,2.4 &  $-45.9\pm2.8$   &   $-13.6\pm2.6$   &				 &	      & T & \\
\object{CD-69~1055}	 & --40.9\,$\pm$\,2.6  & --17\,$\pm$\,1.1 &  $-40.6\pm2.1$   &   $-19.1\pm2.0$   &				 &	      & T &  \\
\object{MP~Mus} 	 & --40.4\,$\pm$\,1.0  & --20\,$\pm$\,1.6 &  $-40.8\pm1.6$   &   $-23.3\pm1.6$   &				 &	      & T &  \\
\noalign{\smallskip}\hline\noalign{\smallskip}

  \multicolumn{7}{c}{$\eta$~Chamaeleontis cluster}\\
  \noalign{\smallskip}
\object{EG~ Cha}   & $-$27.1\,$\pm$\,1.0  & 25.3\,$\pm$\,1.0 & $-29.5\pm2.1$    &  $27.6\pm1.6$     &  & &	   T  & \\
\object{HD~75505}  & $-$30.3\,$\pm$\,1.4  & 25.0\,$\pm$\,1.8 & $-29.9\pm2.0$    &  $27.9\pm1.7$     &  & &	   T & \\
  \noalign{\smallskip}\hline\noalign{\smallskip}

\end{tabular}
\begin{flushleft}
{\bf Notes.} \\
     $^{\mathrm{a}}$ H$=$Hipparcos \citep{vanleeuwen2007}; T$=$Tycho \citep{hog1998} \\      
     $^{\mathrm{b}}$ d$=$discrepant proper motion respect to UCAC3 \\
     $^{\mathrm{c}}$ The 1997 Main Hipparcos Catalogue provides a parallax $\pi=15.06$~mas for this object (d$\sim66$~pc).
\end{flushleft}
\end{table*}

Five Chamaeleon~I and three $\epsilon$~Cha bona fide members from our UCAC3 samples have proper motion and parallax measurements provided by the latest Hipparcos reduction \citep{vanleeuwen2007}. In addition, six $\epsilon$~Cha and two $\eta$~Cha members have proper motions from the Tycho catalogue \citep{hog1998}. These data are summarized in  Table~5, 
together with the parallax, if available, and the distance derived from it. We used this information as a further check of the accuracy of the UCAC3 measurements. 

We note that the proper motions of bright stars in UCAC3 are based on about 140 catalogues, including Hipparcos and Tycho. Therefore, we generally expect good agreement between the UCAC3 measurements and the proper motions listed in Table~5, 
as is indeed found. However, problems may have affected the measurements in some of the images used by UCAC3, either due to saturation, nebulosity, or the presence of nearby fainter stars. This seems to be the case for two Chamaeleon~I stars (\object{HD~97300} and \object{CW~Cha}). For these two objects, the Hipparcos proper motions are in much better agreement with membership 
in the Cha~I moving group than the UCAC3 values.  

In particular, the membership in the dark cloud of \object{HD~97300} (the intermediate-mass star seen towards the northern core of Chamaeleon~I) and its evolutionary status have sometimes been questioned. However, most authors agree that this is most likely a Herbig Ae/Be star similar to \object{HD~97048} in the southern Chamaeleon~I core. The proper motion seems to confirm that this star  is a member of the Chamaeleon~I cloud. 

It is also interesting to note that \object{HD~97048} has been included in the catalogue of potential young  runaway stars by \citet{tetzlaff2011}, albeit with not very high probability (61\%). However, we  
notice that the selection by these authors is only based on Hipparcos stars, which are relatively bright. Therefore, the sample used by \citet{tetzlaff2011} included only a handful of Chamaeleon~I members. Indeed, Table~5 
shows that the proper motion of this object is lower, in modulus, than the rest of Chamaeleon~I stars with Hipparcos or Tycho measurements, what could be interpreted as a hint for a different motion pattern.
However, when the larger sample of cloud members provided by the UCAC3 catalogue are considered, HD~97048 does not stand up as having remarkably different kinematical properties from the rest (see also the discussion in Sect.~\ref{sec:dist}). In view of our analysis, we think it unlikely that HD~97048 is actually running away from its birth place.

There is also overall good agreement between the individual parallaxes of the stars in Table~5 
and the usually quoted distances to the associations.The only possible exception is T~Cha, whose Hipparcos parallax suggests a greater distance. However, given the large error, this estimation is still consistent with membership to the $\epsilon$~Cha cluster (but see also discussion in Sect.~\ref{sec:vel}).

\subsection{Radial velocities of the members of the proper motion groups}\label{sec:rv}

   \begin{figure*}[t]
   \centering
    \includegraphics[width=\textwidth]{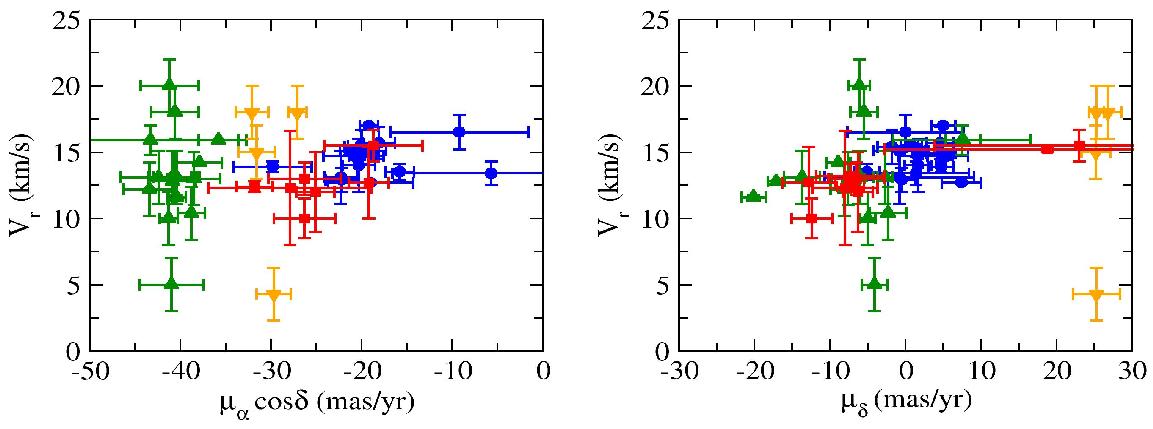}\hfill
       \caption{\footnotesize
	      Plots of radial velocity vs. proper motion components for stars in the Chamaeleon associations. Only objects with reliable proper motions are considered. Symbols and colours as in Fig.~\ref{fig:pm}.
	       }
         \label{fig:rv}
   \end{figure*}

Figure~\ref{fig:rv} shows two plots of the radial velocities versus the proper motion components of the stars. With a couple of exceptions, all the objects from the same proper motion group display similar radial velocities, and are therefore clustered in these diagrams. The radial velocity data therefore confirm that there is a physical relationship between these stars. However, given that all the associations have similar mean radial velocities, in the range 10-20~km/s, it is not possible to distinguish between members of the dark clouds and of the foreground associations on the basis of radial velocity alone. 

The largest individual deviations from the mean group values are observed for \object{EM~Cha} in the $\eta$~Cha cluster ($V_r=4.3\pm2$~km/s) and \object{GSC~9420-0948} in the $\epsilon$~Cha cluster ($V_r=5\pm2$~km/s). The radial velocities of these stars differ by more than 3$\sigma$ from the mean cluster values, which challenges their membership to the quoted associations. We therefore flag these objects in Table~\ref{tab:clus}. Also, \object{T~Cha} has a reported radial velocity discrepant in more than 2$\sigma$ from the mean value for the $\epsilon$~Cha association ($V_r=20.0\pm2.0$~km/s). The discrepancy may be related to the presence of an unresolved companion, as there is indeed one reported in the literature \citep{huelamo2011}.  We discuss this object further in Sect.~\ref{sec:vel}.

\begin{table*}[t]
\label{tab:groups}
\caption{Mean and weighted mean proper motion components and radial velocities for the moving groups identified towards the Chamaeleon sky area$^{\mathrm{a}}$}
\centering
\begin{tabular}{l l l l l l l l l}
\hline
\hline\noalign{\smallskip}
  \multicolumn{1}{l}{Group} &
  \multicolumn{4}{c}{Arithmetic means} &
  \multicolumn{4}{c}{Weighted means} \\
  \multicolumn{1}{l}{} &
  \multicolumn{1}{c}{$<\mu_{\alpha}\cos\delta>$} &
  \multicolumn{1}{c}{$<\mu_{\delta}>$} &
  \multicolumn{1}{c}{$<\mu>$} &  
  \multicolumn{1}{c}{$<V_r>$}  &
  \multicolumn{1}{c}{$<\mu_{\alpha}\cos\delta>$} &
  \multicolumn{1}{c}{$<\mu_{\delta}>$} &
  \multicolumn{1}{c}{$<\mu>$} &  
  \multicolumn{1}{c}{$<V_r>$}  \\
  \multicolumn{1}{l}{} &
 \multicolumn{1}{c}{(mas/yr)} &
 \multicolumn{1}{c}{(mas/yr)} &
  \multicolumn{1}{c}{(mas/yr)} &
  \multicolumn{1}{c}{(km/s)} &
 \multicolumn{1}{c}{(mas/yr)} &
 \multicolumn{1}{c}{(mas/yr)} &
  \multicolumn{1}{c}{(mas/yr)} &
  \multicolumn{1}{c}{(km/s)} \\
\noalign{\smallskip}\hline\noalign{\smallskip}
  \noalign{\smallskip}
 \object{Cha~I}           & $-$19.5\,$\pm$\,4.8  &   \hspace{0.15cm}+1.7\,$\pm$\,3.9  & 20.0\,$\pm$\,4.8 & 14.6\,$\pm$\,1.2 & $-$20.7\,$\pm$\,0.3 & \hspace{0.15cm}  +0.9\,$\pm$\,0.3 & 19.80\,$\pm$\,0.12 & 14.02\,$\pm$\,0.06 \\ 
 \object{Cha~II}          & $-$23.5\,$\pm$\,5.3  & \hspace{0.15cm}$-$7.7\,$\pm$\,2.9  & 24.9\,$\pm$\,5.3 &  12.9\,$\pm$\,1.6  & $-$23.9\,$\pm$\,0.7 & \hspace{0.15cm}$-$7.8\,$\pm$\,0.7 &  25.9\,$\pm$\,0.9  &	14.5\,$\pm$\,0.2	     \\ 
 \object{$\epsilon$~Cha}  & $-$36.8\,$\pm$\,11.7 & \hspace{0.15cm}$-$1.7\,$\pm$\,13.2 & 39.8\,$\pm$\,9.3 & 13.5\,$\pm$\,2.6 & $-$37.8\,$\pm$\,0.5 & \hspace{0.15cm}$-$1.5\,$\pm$\,0.7 &  38.5\,$\pm$\,0.5  & 11.7\,$\pm$\,0.2 \\ 
 \object{$\eta$~Cha}      & $-$29.1\,$\pm$\,4.9  &                 +27.4\,$\pm$\,4.9  & 40.4\,$\pm$\,3.4 & 17.0\,$\pm$\,1.4 & $-$29.9\,$\pm$\,0.5 &		    +25.9\,$\pm$\,0.5 & 38.69\,$\pm$\,0.03 &   14\,$\pm$\,1 \\  
\noalign{\smallskip}\hline\noalign{\smallskip}
\end{tabular}
\begin{flushleft}
{\bf Notes.} \\
    $^{\mathrm{a}}$  Interlopers and objects with uncertain proper motion measurements have been excluded from the calculations.  \\
\end{flushleft}
\end{table*}


\subsection{Interlopers}\label{sec:inter}

Once the proper motion lists are purged from inaccurate measurements, three true interlopers remain, namely one Chamaeleon~I object (\object{2MASS~J11183572-7935548}) and two Chamaeleon~II objects (\object{2MASS~J11183572-7935548} and \object{CM~Cha}), whose proper motions seem to be in better agreement with membership to the $\epsilon$~Cha group. 

The source \object{2MASS~J11183572-7935548} was included as a Chamaeleon~I member in the list of \citet{luhman2007}. However, in a later work, \citet{luhman2008b} reclassified it as an $\epsilon$~Cha member on the basis of its UCAC2 proper motion. Our analysis of the UCAC3 data confirms their result. This object is therefore included in Table~\ref{tab:clus} as a probable member of the $\epsilon$~Chamaeleontis cluster.

\object{CM~Cha} had been listed as a Chamaeleon~II member by \citet{spezzi2008}. However, it has UCAC3 proper motion components $\mu_{\alpha}\cos\delta=-43.3\pm10.6$ and $\mu_{\delta}=7.5\pm9.1$~mas/yr. These values are only marginally compatible with the Cha~II moving group, and seem to be in better agreement with membership in the foreground $\epsilon$~Cha association. Unfortunately, AC2000.2 provides no data for this star, and the data provided by SuperCOSMOS are wrong, probably because this object appears blended with other star in the POSS image, so we cannot test its proper motion directly in the way explained above. In the literature we find a reported proper motion of ($-60, 21$)~mas/yr for this object, with a quoted accuracy of 5-10 mas/yr  \citep{teixeira2000}. The PPMXL catalogue reports a similar value of ($-62, 9$)~mas/yr. These measurements are quite discrepant from the UCAC3 measurement, even when the large errors are considered, but indicate in any case that CM~Cha is probably located closer to the Sun than the Chamaeleon~II members. On the other hand, there is spectroscopic evidence that the object is young: \citet{spezzi2008} report an equivalent width for the Li\,I absorption line of $0.38~\AA$, which is similar to many other cloud members, but close to the lowest values measured by these authors. (Their range encompasses 0.35 to 0.61~$\AA$, and typical values in star forming regions are in the range 0.3-0.8~$\AA$.) They also report 
an H$\alpha$ equivalent width of $-29~\AA$, which is consistent with low accretion. This suggests that, though young, CM~Cha is among the most evolved objects in the Spezzi et al. sample. On the other hand, the reported radial velocity for this star ($15.9\pm1.1$~km/s; \citet{torres2006} quite well agrees
with the values measured for other stars in the Chamaeleon clouds, but also with the radial velocities reported for $\epsilon$~Cha members (see Sect.~\ref{sec:rv}). The space motion of this star (as computed using the UCAC3 proper motions) also seems to be in better agreement with this cluster (see Sect.~\ref{sec:vel} below). Taking all this into account, we tentatively reclassify CM~Cha as a member of the $\epsilon$~Cha young association, and list it accordingly in Table~\ref{tab:clus}. We note, however, that if the proper motion reported by \citet{teixeira2000} were correct, this star would probably be located even closer to the Sun than the $\epsilon$~Cha members, and thus could be an unrelated young star from the solar neighbourhood.

The situation is also ambiguous for \object{Sz~60W}, because the large error in $\mu_{\delta}$ makes its UCAC3 proper motion compatible with both the $\epsilon$~Cha and the Cha~II moving groups. This star is part of a binary object with a separation of 3.6$^{\prime\prime}$, and the secondary has no UCAC3 counterpart. This may affect the proper motion measurement of the primary, if the system is not completely resolved in the images used by UCAC3 to compute the proper motion, or if source confusion has occurred at some stage. Indeed, secondaries at separations lower than 10$^{\prime\prime}$ are often not resolved in UCAC3. 
More precise measurements are required to clarify the nature of Sz~60W, but we note that the spatial location of this source, within the dark cloud (in contrast with CM~Cha that is located in the outskirts), suggests a physical connection with Chamaeleon~II. For the time being, Sz~60W is listed in Table~\ref{tab:mem} as a member of this cloud.

In addition, we identify one object, namely \object{CP-68 1388}, whose reported UCAC3 proper motion differs from the rest of  $\epsilon$~Chamaeleontis members, and is seen detached from moving group in the vector-point diagram of Fig.~\ref{fig:pm}. This is due to the value of the proper motion in the $\delta$ direction ($\mu_{\delta}=7.6\pm2.3$~mas/yr), because its $\mu_{\alpha}\cos\delta$ value ($-35.8\pm2.2$~mas/yr) and its radial velocity (15.9~km/s) are in good agreement with those of other cluster members. This star was observed by Tycho, and the proper motions reported in Table~5 
agree closely with the UCAC3 values (although the Tycho $\mu_{\delta}$ value is somewhat lower, $\mu_{\delta}=4.70$~mas/yr). The different proper motion and the spatial location of CP-68 1388, to the North of Chamaeleon~I and more than $6^{\circ}$ away from the closest $\epsilon$~Chamaeleontis member, makes its membership to the cluster dubious. It is interesting, however, that this object is placed halfway between the $\epsilon$ and $\eta$~Chamaeleontis cluster in the vector point diagram. As shown later in Sect.~\ref{sec:vel}, the space velocities of CP-68 1388 agree well with other stars in both clusters. We therefore keep this object in the list of $\epsilon$~Cha members for the time being.

\object{VW~Cha} also deserves some comments. This star has been listed in the literature both as a Chamaeleon~I member \citep{luhman2007} and as a member of the $\epsilon$~Cha group \citep{fernandez2008}. We find that both its spatial location within the Chamaeleon~I southern cloud core and its UCAC3 proper motion ($\mu_{\alpha}\cos\delta=-19.7\pm1.3$~mas/yr, $\mu_{\delta}=-0.8\pm3.4$~mas/yr) are in better agreement with VW~Cha belonging to the dark cloud. The proper motion values reported by SIMBAD ($\mu_{\alpha}\cos\delta=-23$~mas/yr and $\mu_{\delta}=-7$~mas/yr; \citet{teixeira2000} agree well with our classification. This star also has a reported radial velocity \citep[$V_r=17.6\pm3.3$~km/s;][]{torres2006} that is slightly higher than the mean value for Chamaeleon~I, but similar to the values reported for other member stars (see Sect.~\ref{sec:rv}).

\subsection{Final member lists}\label{sec:final}

Once the proper motion lists are purged of incorrect measurements, and once the interlopers assigned to the associations they are likely to belong to, the definitive member lists for the Cha~I, Cha~II, $\epsilon$~Cha and $\eta$~Cha contain 73, 22, 21, and 12 stars, respectively. These are the objects presented in Tables~1 and 2. These lists still contain objects with dubious measurements that could not be proved wrong, as discussed in Sect.~\ref{sec:groups}.

After this analysis, we computed the  means of the proper motion and values for the different kinematic groups, which are listed in Table~6. 
Interlopers and objects with wrong or dubious proper motion measurements were not considered in the calculations. Because of the large proper motion errors for some of the objects, we also computed the weighted means, which are also given in Table~6. 
The values of both the arithmetic and weighted means for each parameter are fully consistent with each other, within the estimated errors. 

Table~6 
also lists the arithmetic and weighted mean radial velocities that we derive for three of the four groups. 
For the $\epsilon$ and $\eta$~Cha clusters, the mean radial velocities are in good agreement with previously published values \citep{kharchenko2005,wu2009}.

   \begin{figure*}[t]
   \centering
    \includegraphics[width=0.99\textwidth]{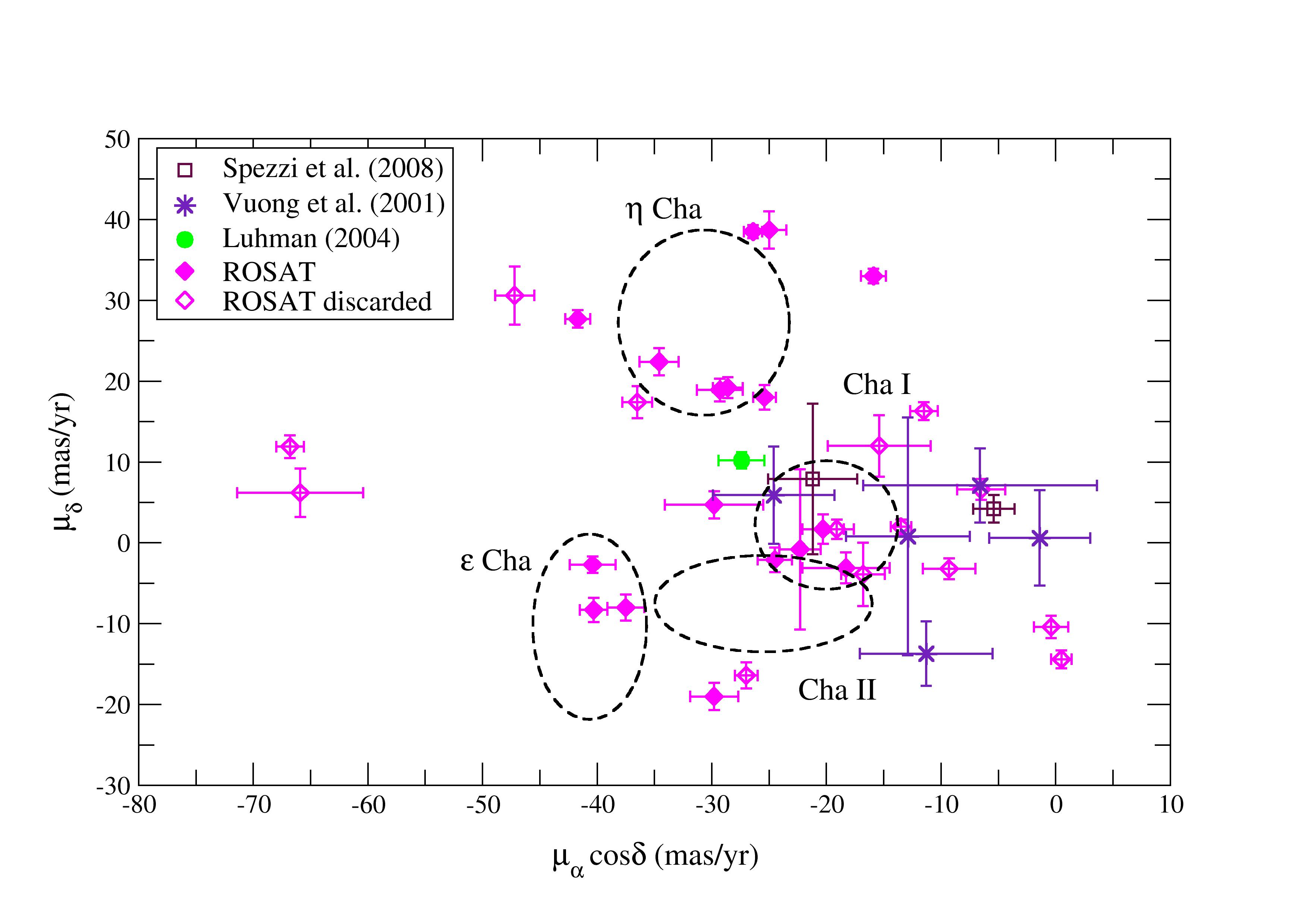}\hfill
\caption{\footnotesize
	       Vector point diagram for Chamaeleon candidate members from the literature. The ellipses indicate the approximate location of the kinematical groups identified in Fig.~\ref{fig:pm}. 	
	       }
         \label{fig:cand}
   \end{figure*}

%
\section{Proper motions of candidate Chamaeleon members}\label{sec:cand}

\addtocounter{table}{1}  

In the next step, we made an attempt to confirm candidate members of Chamaeleon~I and II on the basis of their proper motion. As in our Lupus study, we compared the proper motions of candidate members proposed in the literature with those of the confirmed members. 

\subsection{ROSAT stars}\label{sec:rosat}
				        									      
In the 1990s, the ROSAT mission unveiled a dispersed population of X-ray emitting stars towards the Chamaeleon area, but not confined to the dark clouds \citep{alcala1995}. \citet{alcala1997} provided the first spectroscopic evidence that these stars were young. \citet{covino1997} performed radial velocity measurements for most of these stars, showing that the sample included objects belonging to different kinematical populations. This result was confirmed for a subsample of these objects by \citet{frink1998} combining information from several proper motion catalogues. In addition, these authors used the Hipparcos parallaxes available for some of the stars to show that they were located at different distances(from 60 to about 200~pc). The different stellar populations identified by these authors were thus assigned to the Chamaeleon~I cloud, the foreground $\epsilon$~Cha cluster, or to an undetermined field population. However, many stars remained unclassified owing to the lack of proper motions. Only 18 out of the 81 stars studied by \citet{covino1997} are included in our compilations of members of the associations seen towards the Chamaeleon area. These objects  are flagged in Tables~\ref{tab:mem} and \ref{tab:clus}.

To clarify the nature of the remaining ROSAT sources, we performed the same proper motion analysis on them described in Sect.~\ref{sec:groups}. UCAC3 provides proper motion measurements for 38 of these objects, listed in Table~7. 
The radial velocities and evolutionary status of these stars according to \citet{covino1997} are also indicated. We show the vector point diagram for these objects in Fig.~\ref{fig:cand}, compared to the loci of the known young populations in the Chamaeleon area. We also show the spatial location of the ROSAT sources in Fig.~\ref{fig:map}.

The UCAC3 vector point diagram confirms that the ROSAT stars are a mixture of different kinematical populations. From their position on this diagram, some sources present completely discrepant proper motions from  any of the moving groups identified in Sect.~\ref{sec:groups}. The radial velocity measurements from \citet{covino1997} confirm that these stars follow a completely different motion pattern with respect to the Chamaeleon associations. Other sources can be nicely assigned to one of the moving groups in the Chamaeleon sky area; however, a few of them are also discarded because their radial velocities, as provided by \citet{covino1997}, are in complete disagreement with membership to any of the associations, or because their location in the sky does not coincide with that of the group with similar proper motion (especially in the case of the dark clouds). All the ROSAT objects with discrepant proper motions and/or radial velocities are classified as zero-age main sequence (ZAMS) stars or stars of unclear nature by these authors.

This left us with 13 stars (11 PMS stars and two stars of unclear nature) whose spatial location, proper motions and radial velocities (except for RXJ1150.4-7704, which has no available radial velocity measurement) all overlap with one of the Chamaeleon moving groups. We assigned two of these objects to the Cha~I moving group. The eleven other sources are classified as candidate members of the $\epsilon$~Cha association (three objects) or the $\eta$~Cha association (eight objects). We do not find any good Chamaeleon~II candidate members.  The membership of all these sources is indicated in Table~7. 

   \begin{figure*}[ht]
   \centering
  \includegraphics[width=9cm]{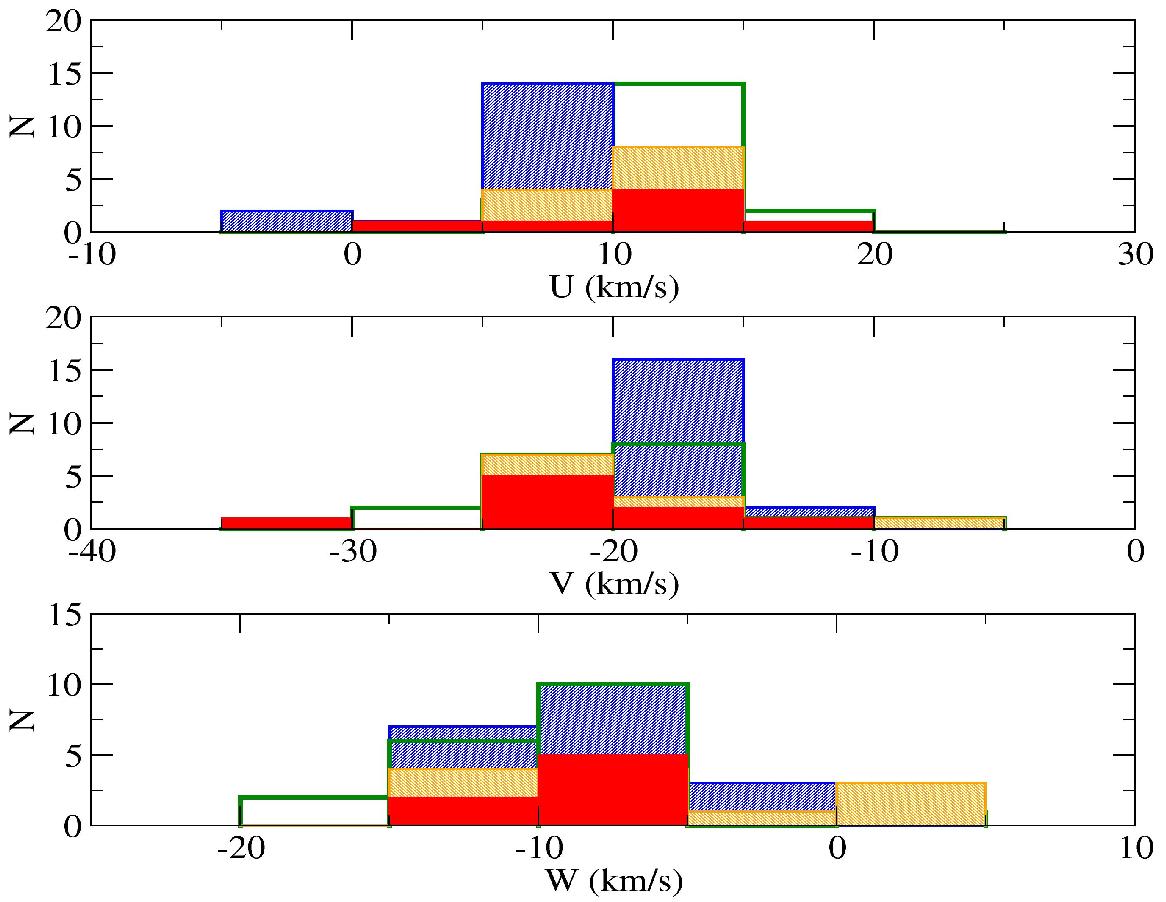}\hfill
  \includegraphics[width=9cm]{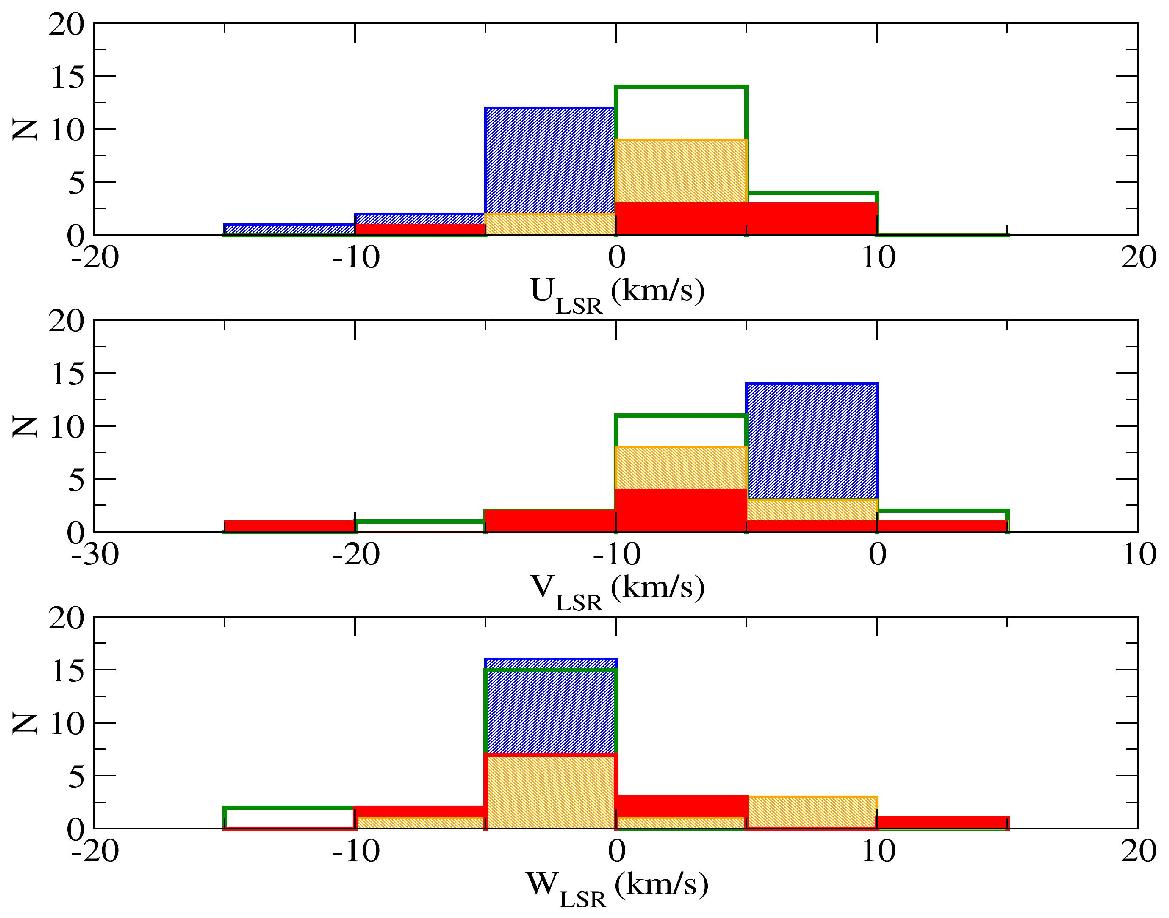}
      \caption{\footnotesize
	      Histograms of space velocities, uncorrected and corrected for the solar reflex motion with respect to the LSR (left and right panels, respectively) for stars in the different Chamaeleon associations. Symbols  as in Fig.~\ref{fig:hist}.
}
         \label{fig:vel}
   \end{figure*}

In two cases, namely \object{RXJ1123.2-7924} and \object{RX~J1158.5-7913}, membership in a single association cannot be clearly attributed. The spatial location of these two objects (two PMS stars according to the study by \citet{covino1997} seems more consistent with membership in the $\epsilon$~Cha moving group than in either of the dark clouds. However, this is not so straightforward from their UCAC3 proper motions.   \object{RXJ1123.2-7924}  is located in the vector point diagram closer to the Cha~II moving group than to the $\epsilon$~Cha members, but its proper motion values are still marginally consistent with membership in this group. Given its location in the sky, in the area between the Chamaeleon~I and the Chamaeleon~III clouds, and about 5$^{\circ}$ from the known Chamaeleon~II members, we tentatively assign this object to the $\epsilon$~Cha moving group, and list it in Table~7. 

\object{RX~J1158.5-7913} is seen towards the small cloud \object{[DB2002b]\,G300.23-16.89}, located halfway between the Chamaeleon~I and II clouds and about 2.8$^{\circ}$ south-east of the former. The star \object{T~Cha} and other members and candidate members of the $\epsilon$~Cha association are also seen in this area. However, the proper motion components of RX~J1158.5-7913 ($\mu_{\alpha}\cos\delta=-18.3\pm3.8$~mas/yr, $\mu_{\delta}=-3.1\pm1.9$~mas/yr) place this star in the overlapping area between Chamaeleon~I and II members in the vector point diagram, but  closer to the Chamaeleon~I mean value given in Table~6. 
This looks completely incompatible with membership in the $\epsilon$~Cha cluster, even though the radial velocity reported by Covino et al.  ($V=13.1\pm2$~km/s) is very close to the mean value we obtain for the $\epsilon$~Cha members (see Table~6).  
Because the proper motion and radial velocity look consistent with the mean values derived for the Cha~I moving group, we tentatively assign RX~J1158.5-7913 to this cloud for the time being.  More accurate kinematic data will help clarify the nature of this object.

\subsection{Other candidate members from the literature}\label{sec:prev}

\addtocounter{table}{1} 

Several works have provided lists of candidate members in the Chamaeleon clouds that still lack spectroscopic confirmation of youth. Unfortunately, most of the postulated objects are too faint to be included in the UCAC3 catalogue and, therefore, we cannot test their membership on the basis of their proper motion. There are, however, a few exceptions, which we discuss in this section. The UCAC3 proper motions for these candidates are listed in Table~\ref{tab:cand}, and they are plotted in the vector point diagram shown in Fig.~\ref{fig:cand}.

\citet{luhman2004} provides a list of objects whose membership in Chamaeleon~I is uncertain (cf. his Table~8), 21 of them lacking confirmation in later works. Only one of these objects has a valid UCAC3 counterpart, namely \object{C1-14}, an infrared source first quoted by \citet{prusti1991}, and also identified with the X-ray source \object{CHX~15a}.\footnote{In addition to \object{C1-14}, SIMBAD lists three other sources within 2$^{\prime\prime}$ (the value of our cross-matching radius) from the UCAC3 position: \object{CD-75~522}, \object{Glass~R}, and \object{WKK~F~32}. The former has a reported proper motion suggesting ia foreground star. From the intercomparison of their properties as listed in SIMBAD, we are led to think that  Glass~R,  WKK~F~32, and C1-14 must actually be the same object.}
This object is located within the northern cloud core of Chamaeleon I  and, as shown in Fig.~\ref{fig:cand}, its proper motion is compatible with the Cha~I moving group. The spectral type of this source is badly constrained, but quoted in the range F0-A7, which would make it one of the most massive objects in the dark cloud, together with HD~97300 and HD~97048.

In addition to the confirmed Chamaeleon~II members, \citet{spezzi2008} list a number of objects still lacking spectroscopic confirmation of membership to the cloud, but considered likely members on the basis of the multiwavelength photometric analysis reported in a previous work \citep{alcala2008}. Only two of these sources have UCAC3 counterparts, namely \object{IRAS~12589-7646} (\object{ISO-Cha~II~89}) and \object{IRAS~12448-7650}. The proper motion for the latter object clearly disagrees with membership in the Cha~II moving group, and strongly suggests that \object{IRAS~12448-7650} belongs to the background. As for  \object{IRAS~12589-7646}, its proper motion  ($\mu_{\alpha}\cos\delta=-21.2\pm3.9$~mas/yr, $\mu_{\delta}=7.9\pm3.9$~mas/yr) places this source in the area of the Cha~I moving group rather than Cha~II. From our own check, however, the proper motion seems to be around (5, 1.1)~mas/yr, which would place it in the same area as the background objects in the vector-point diagram. In view of these results, the classification of IRAS~12589-7646 as a Chamaeleon~II member is dubious, and we conservatively flag it as ``rejected'' in Table~\ref{tab:cand}. 

\citet{vuong2001} performed a search for candidate members of Chamaeleon~II based on DENIS $IJKs$ photometry. Only five objects from their list are included in UCAC3, namely their sources \object{[VCE2001]\,C18}, \object{C29}, \object{C64}, \object{X3}, and \object{X4}. As seen in Fig.~\ref{fig:cand}, their proper motions do not agree with membership in the Cha~II moving group. Therefore, it seems unlikely that these objects belong to the Chamaeleon~II dark cloud. The only possible exception is the X-ray source \object{[VCE2001]\,X4}, whose location in the vector point diagram is marginally compatible with those of the Chamaeleon~II members. While deceiving, this result is not totally surprising, since spectroscopic studies of about 30 sources from the \citet{vuong2001} selection also failed to confirm most of them as Chamaeleon~II members  \citep{barrado2004,spezzi2008}, thus showing the high contamination of the original sample (which amounted to 76 sources). None of the five DENIS sources discussed above with available proper motions was observed in these works. To date, there have only been five confirmed young stars from the \citet{vuong2001}) list \citep[see compiled list in][]{spezzi2008}, but only one, \object{[VCE2001]\,C61}, is included in UCAC3 and listed in Table~\ref{tab:mem}. The proper motion of this object is compatible with Chamaeleon~II membership.

   \begin{figure}[t]
   \centering
  \includegraphics[width=0.47\textwidth]{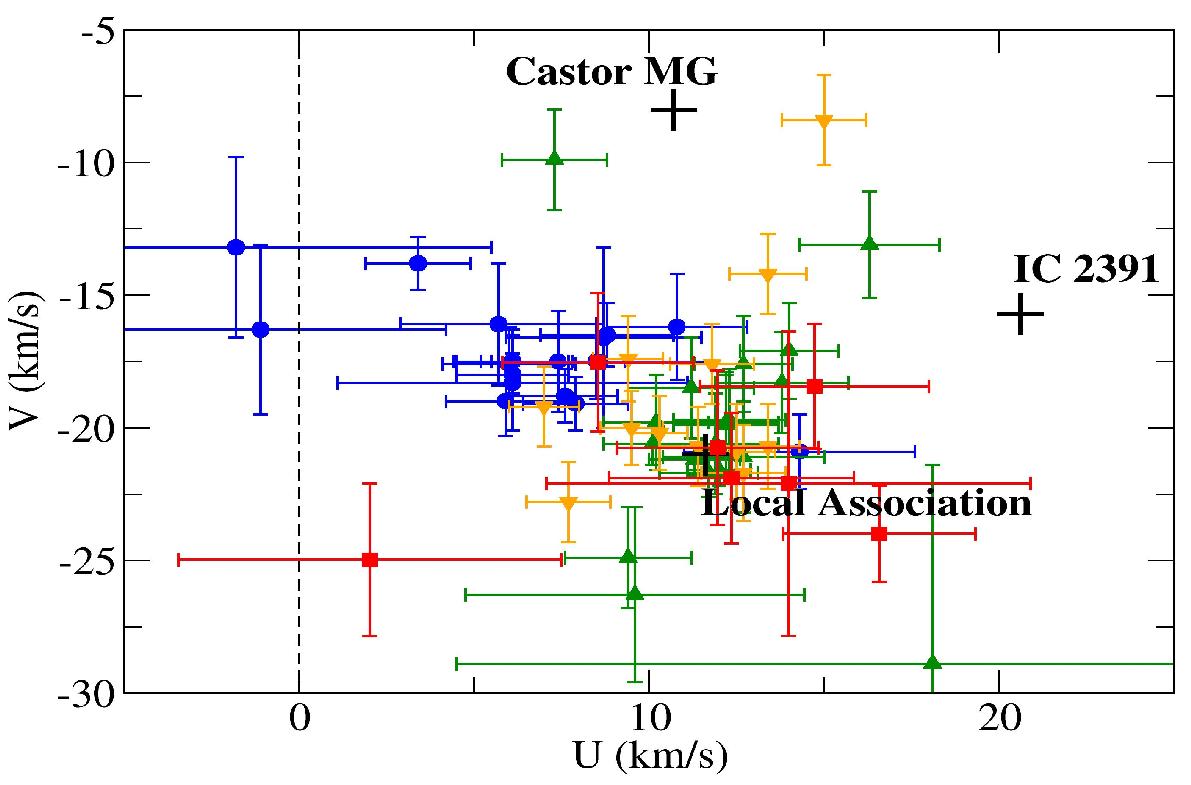}\\
  \includegraphics[width=0.47\textwidth]{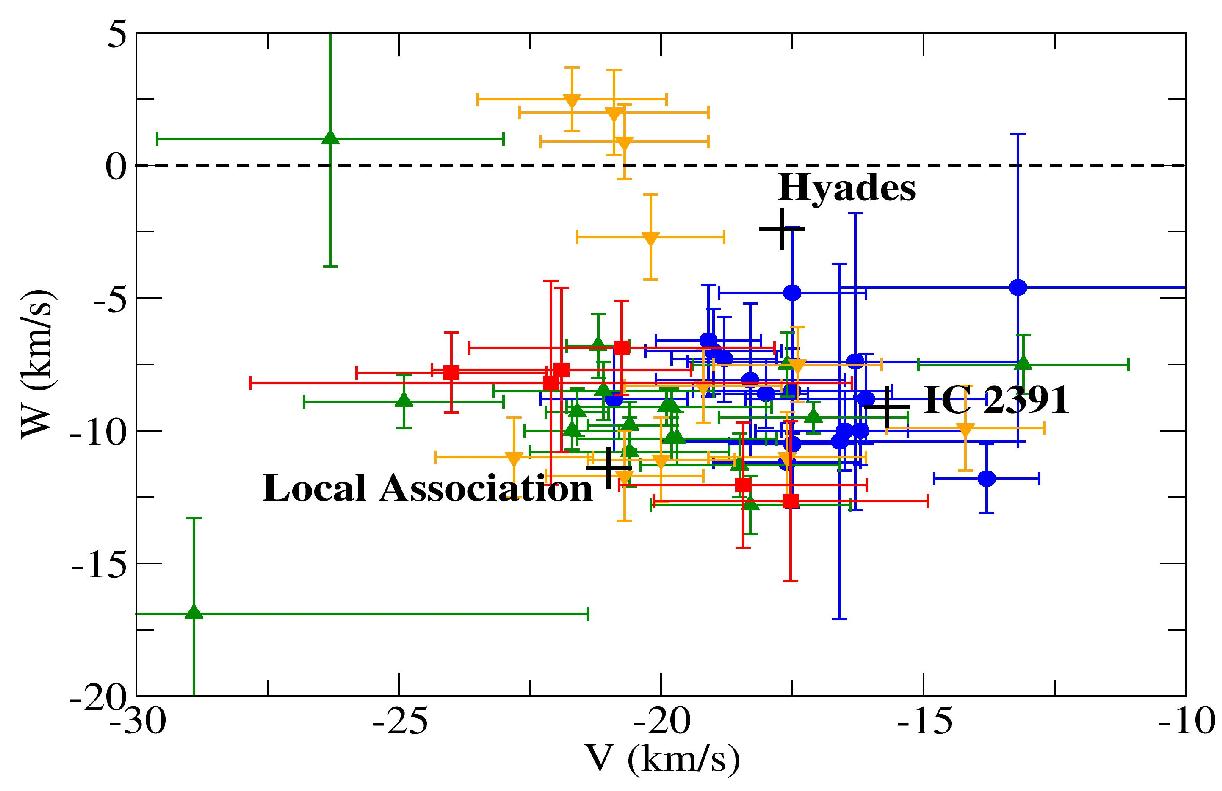}
      \caption{\footnotesize
	      ($U, V$) and ($V, W$) planes for stars in the different Chamaeleon association. The big crosses indicate the location of known nearby moving groups. Rest of symbols  and colours as in Fig.~\ref{fig:pm}.
}
         \label{fig:planes}
   \end{figure}

\section{Space velocities}\label{sec:vel}

\addtocounter{table}{1} 

We computed the galactic space velocities for the stars (members and candidate members) with available radial velocity measurements. For this purpose we made use of an IDL routine retrieved from the IDL Astronomy Library,\footnote{http://idlastro.gsfc.nasa.gov/ftp/pro/astro/gal\_uvw.pro} originally written by W. Landsman and later modified by S. Koposov. The procedure follows the formulation in \citet{johnson1987}, except for two details: the $U$ component is positive toward the Galactic \emph{anticentre}, and the Hipparcos transformation matrix is used. We modified the routine to include the computation of the velocity errors in the way prescribed by \citet{johnson1987}.

Since most of the stars lack parallax measurements, we assumed the canonical distances to the associations for this calculation: $160\pm15$~pc and $178\pm18$~pc to Chamaeleon~I and II, respectively \citep{whittet1997}, and $111\pm5$~pc and $97\pm5$~pc to the $\epsilon$~Cha and $\eta$~Cha clusters, respectively \citep{feigelson2003}. For \object{T~Cha} and \object{CV~Cha}, we used the distances inferred from their Hipparcos parallaxes as given in Table~5. 

The resulting space velocities are listed in Table~9, 
as well as the velocities corrected for the velocity of the Sun with respect to the local standard of rest (LSR), assumed to be ($U_{\odot}, V_{\odot}, W_{\odot}$)$=$($-8.5$, 13.38, 6.49)~km/s. Figure~\ref{fig:vel} shows the histograms for the space velocities of the stars in the different associations. 

Although this exercise is certainly affected by lots of uncertainties, especially in the stellar distances, the resulting space velocities confirm the membership of the stars in the different Chamaeleon associations to the young disk population, according to the prescriptions listed in \citet{leggett1992}. The histograms displayed in Fig.~\ref{fig:vel} further confirm the presence of different kinematical populations towards Chamaeleon. In particular, the stars in the Chamaeleon~I cloud seem to display, as a sample, a different motion pattern from the rest. This is most evident in the Galactic centre-anticentre direction ($U$ component), especially when the velocities are corrected with respect to the LSR (upper right panel of Fig.~\ref{fig:vel}): While the mean $U_{LSR}$-component of the Cha~I moving group is negative (i.e. pointing to the Galactic centre), for the rest of associations this component is positive (i.e. pointing to the Galactic anticentre). 

The stars in the two foreground clusters seem to have very similar space velocities, so the differences in proper motion observed in the vector point diagram of Fig.~\ref{fig:pm} seem to be mostly related to the different spatial location of both groups in the projected sky (separated by about 8.5$^{\circ}$). This fact, together with the similar ages and distances of both associations, hints at a connection between them, as already suggested in the literature \citep[e.g.][]{feigelson2003}. Indeed, the space velocities we obtain are in good agreement with published values \citep[e.g.][]{zuckerman2004}, and with the membership of both clusters to the so-called ``Local Association'' (or ``Pleiades moving group''), a coherent kinematic stream of young stars (ages $\sim$20-150~Myr) with embedded clusters and associations (such as the Pleiades, $\alpha$~Per or Scorpius-Centaurus) first postulated by \citet{eggen1975}. This is shown in Fig.~\ref{fig:planes}, where we show the location of the stars from Table~9 
in the ($U, V$) and ($V, W$) planes, together with the location of the mean space velocities for several nearby young moving groups \citep[ages 20-600~Myr;][]{montes2001}.

Interestingly, the stars in Chamaeleon~II seem to have space velocities more similar to those of the foreground clusters than to the Chamaeleon~I stars. However, the numbers are low, and the errors for most stars are large, as is the dispersion in the data. Better precision and statistics with larger samples are required to investigate the eventual relation of Chamaeleon~II with the $\epsilon$ and $\eta$~Cha associations, and to other young moving groups.

We close this section with some remarks on particular objects. First of all, we note that the usually quoted parallax for \object{T~Cha} is $15\pm3$~mas, corresponding to a distance of only $66\pm15$~pc. This is the value included in the Hipparcos Main Catalogue (ESA 1997). However, the new reduction performed by \citep{vanleeuwen2007} provides a parallax of $6\pm3$~mas, corresponding to a distance of $160\pm80$~pc. The first value would place this star closer to the Sun than most of the members of the $\epsilon$~Cha association, and with quite discrepant space velocities, even when the errors are considered. In contrast, the most recent distance value is compatible, within the error, with the canonical distance to the cluster, and provides space velocities in relatively good agreement with T~Cha being an $\epsilon$~Cha member. The same trend is observed if the space velocities are computed using the Hipparcos proper motions from Table~5 
instead of the UCAC3 ones. Nevertheless, T~Cha is seen detached from the group of $\epsilon$~Cha members in both planes represented in Fig.~\ref{fig:planes}, and, given the large uncertainties in the distance (i.e. in the space velocities), the membership of this star to the $\epsilon$~Cha  moving group cannot be assigned with certainty.

Second, we note that the space velocities of \object{CM~Cha} show relatively good agreement with those of other $\epsilon$~Cha members, when the mean distance to this cluster is used in the calculation. If the distance to Chamaeleon~II is used instead, the resulting space velocities are quite discrepant from those of the only confirmed Cha~II member with available space velocities. This further supports the hypothesis that CM~Cha is a member of the $\epsilon$~Cha association, as suggested in Sect.~\ref{sec:inter}. Even so, it must be noted that the velocity component in the direction of the Galactic poles ($W$) of CM~Cha is quite different from the rest of $\epsilon$~Cha members, as it is pointing towards the northern instead of the southern Galactic pole. We note, though, that a handful of $\eta$~Cha cluster members are also located in the same area of the ($V, W$) plane.
 
 We made the same check with the two ROSAT sources discussed at the end of Sect.~\ref{sec:rosat}.  \object{RX~J1158.5-7913}, a potential interloper from the Chamaeleon~I cloud in the sky area of the $\epsilon$~Cha members, has space velocities that are fully consistent with membership in the dark cloud, if the mean distance to it is used in the calculation. In contrast, the resulting space velocities when the distance to $\epsilon$~Cha is used are not consistent with membership in  this cluster. On the other hand, \object{RXJ1123.2-7924}, whose proper motion looks compatible with both  the $\epsilon$~Cha cluster and the Chamaeleon~II cloud, has space velocities fully compatible with the foreground association.

%
\section{Properties of the kinematic populations of the Chamaeleon dark clouds}\label{sec:prop}

   \begin{figure}[ht]
   \centering
      \includegraphics[width=9.1cm]{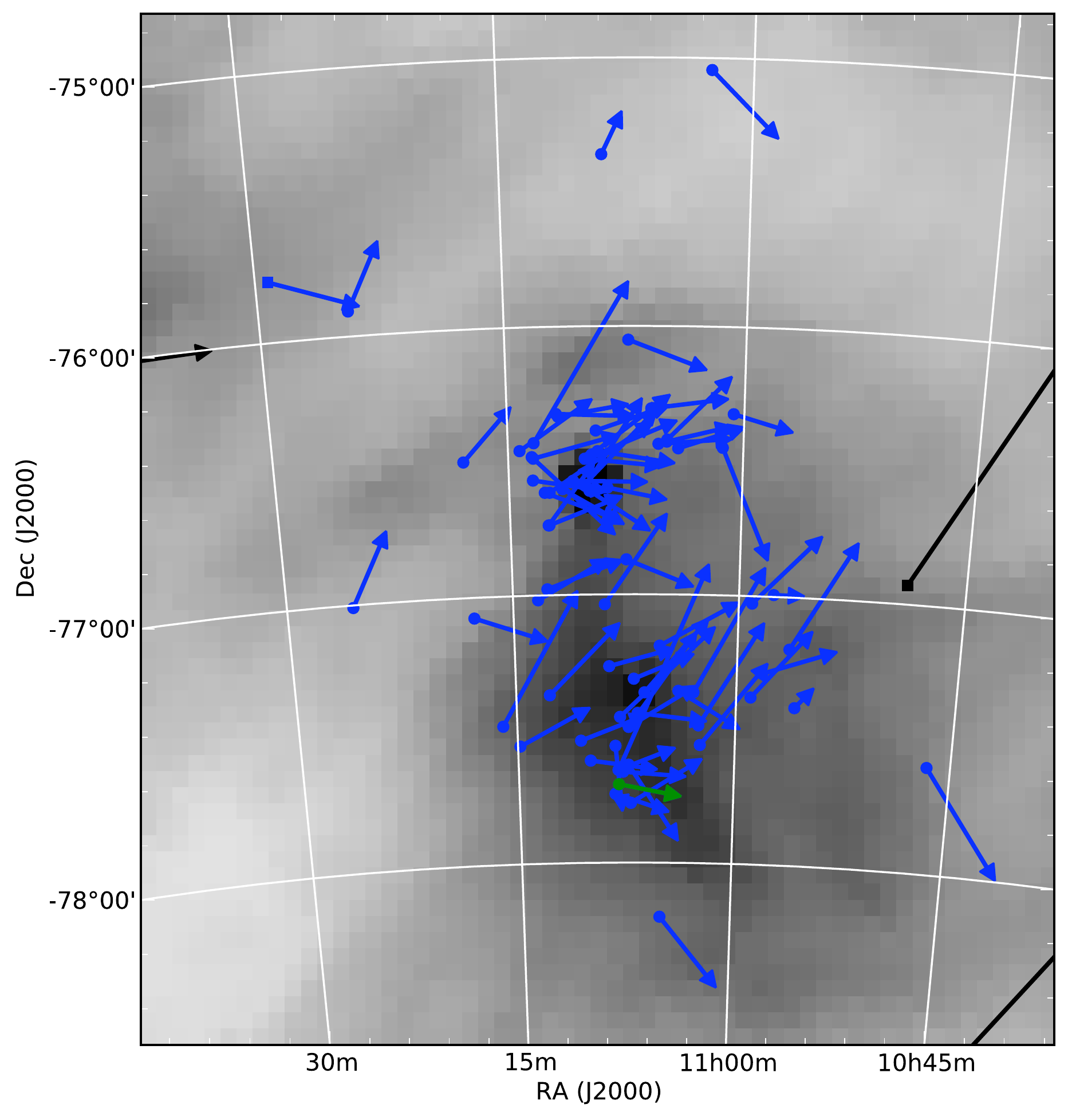}\hfill\\
     \includegraphics[width=9.1cm]{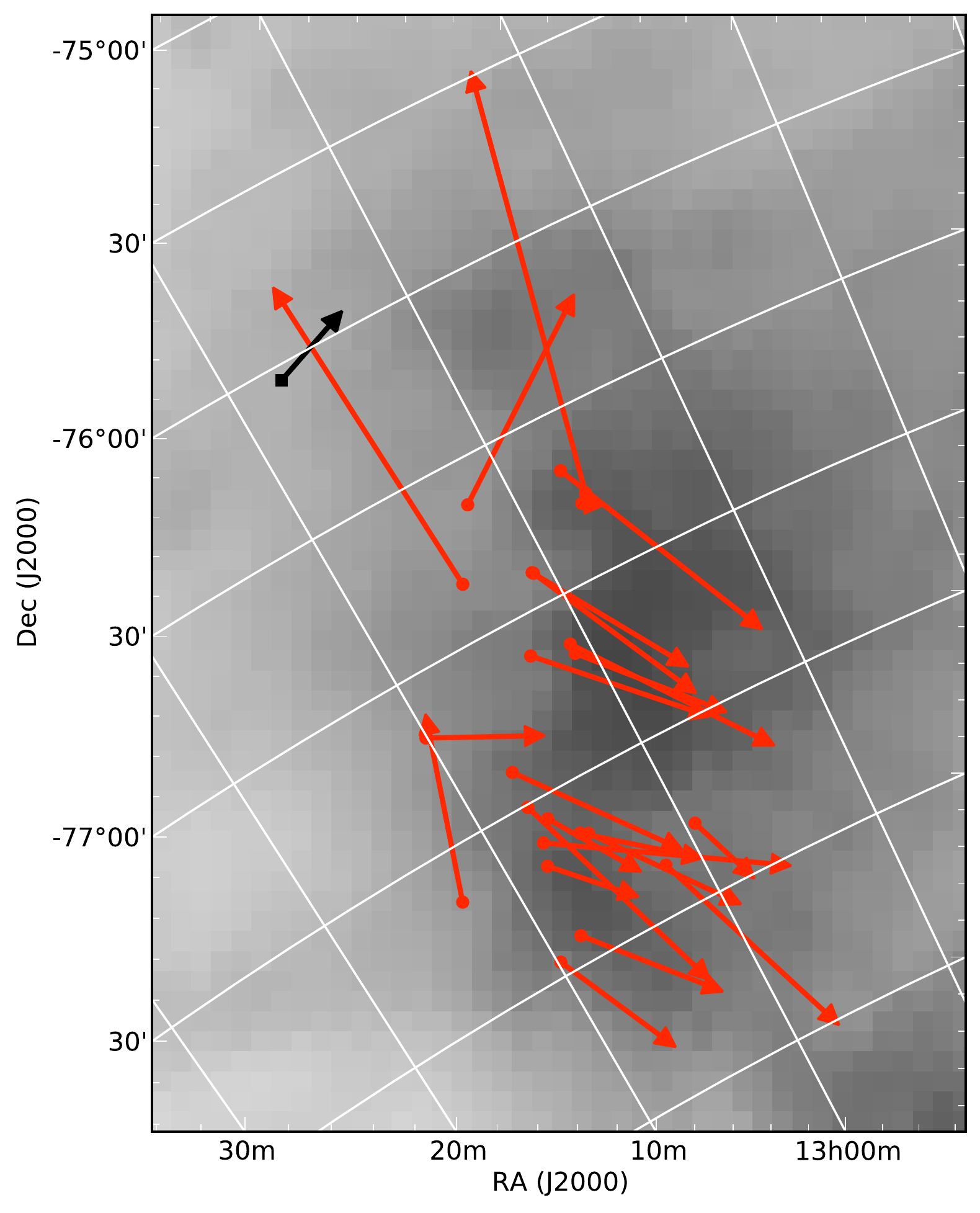}\hfill
     \caption{\footnotesize
	       Enlargements of the map from Fig.~\ref{fig:map} showing the Chamaeleon~I and II areas in more detail (upper and lower panels, respectively).
	       }
         \label{fig:dist}
   \end{figure}

   \begin{figure*}[t]
   \centering
  \includegraphics[width=15cm]{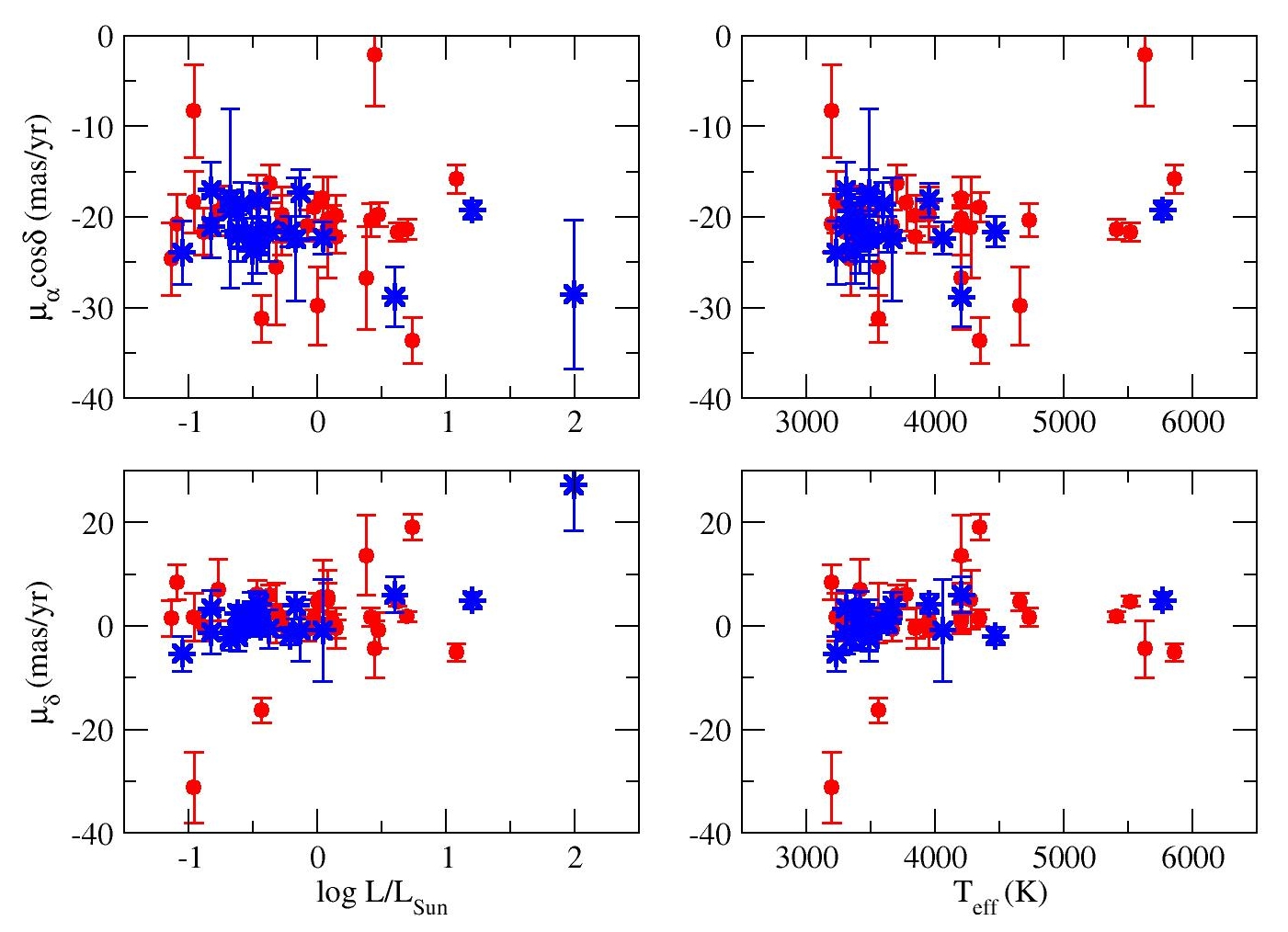}\hfill
      \caption{\footnotesize
	       Proper motion components versus bolometric luminosity (left) and effective temperature (right) for Chamaeleon~I members with and without disks (red circles and blue asterisks, respectively).
	       }
         \label{fig:pars_chai}
   \end{figure*}

   \begin{figure*}[t]
   \centering
  \includegraphics[width=15cm]{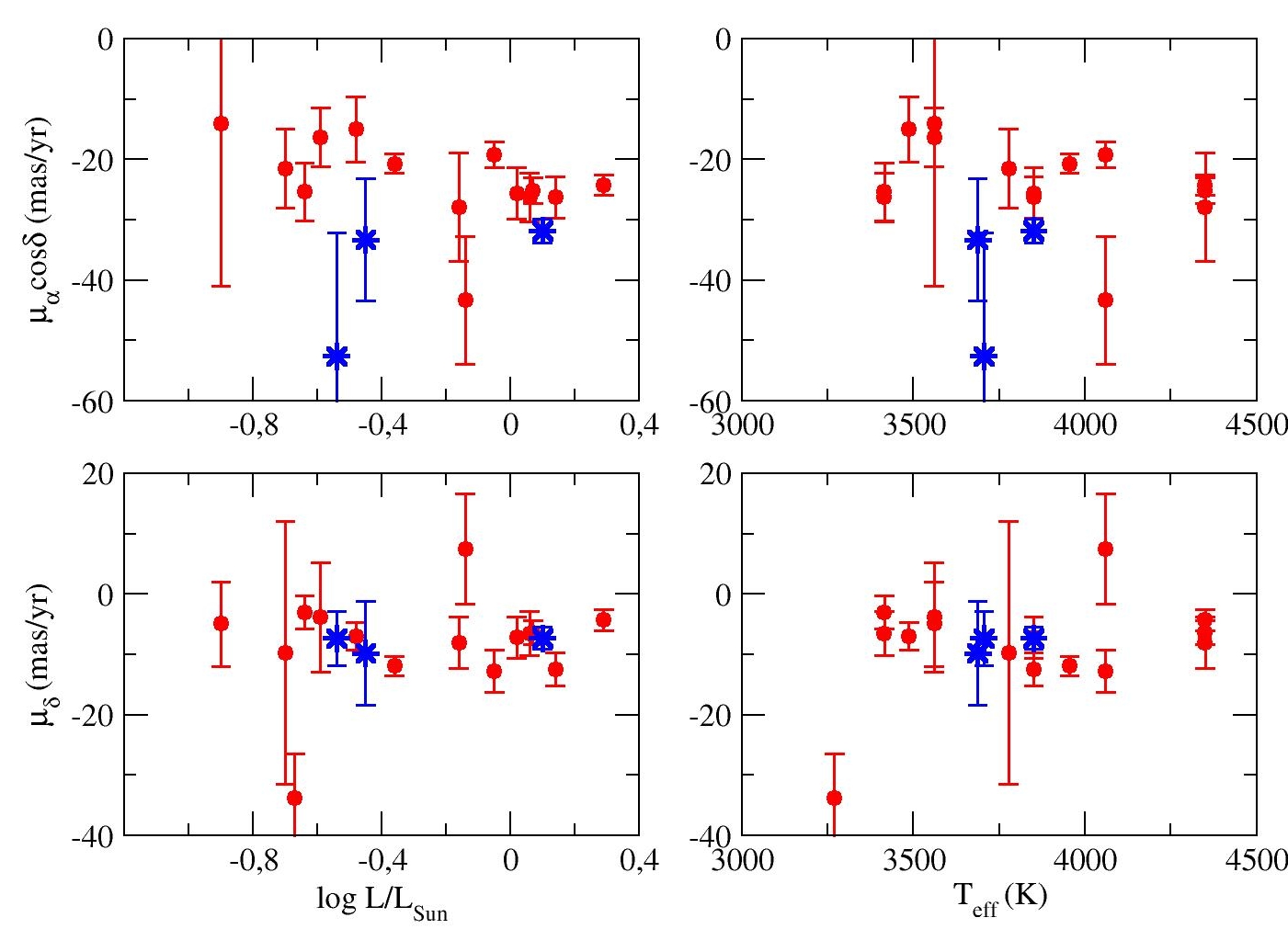}\hfill\\
      \caption{\footnotesize
	       Same as Fig.~\ref{fig:pars_chai} for Chamaeleon~II.}
         \label{fig:pars_chaii}
   \end{figure*}

\subsection{Spatial location}\label{sec:dist}

Figure~\ref{fig:dist} shows two enlargements of Fig.~\ref{fig:map} in  the location of the Chamaeleon~I and II clouds. As noted in previous works, the sources belonging to the Cha~I moving group tend to cluster around the two main cores containing the intermediate-mass stars HD~97300 (the northern core) and HD~97048 (the southern core). \citet{luhman2007} reports on a difference in isochronal ages of a few Myr between both cores, which he interprets as the result of two different bursts of star formation. We therefore checked whether any proper motion difference could be seen between the two cores. To this purpose, all the stars within a radius of $50^{\prime}$ from one of the intermediate-mass star were defined as members of the corresponding core. This distance was chosen after inspecting the distance distributions from both intermediate-mass stars. At the UCAC3 precision, we found no obvious difference between the proper motions of both groups. 

No evident difference is seen either between the in- and the off-cloud populations, neither in proper motion moduli nor in proper motion directions. This seems to confirm that both samples belong to the same structure. It also suggests that the location of the outer sources is not the consequence of these objects being ejected from their parental birth sites.

We performed a similar analysis in Chamaeleon~II, dividing the population in a northern group, located north-west of a dense core in the northern part of the cloud, and a southern group, located in and around a smallest core to the south. Again, we found no obvious difference between the proper motions of both groups. A comparison between the in- and off-cloud population is not possible in this case due to the low number of objects and the dubious proper motions of most of the sources located off-cloud (leftmost sources in the lower panel of Fig.~\ref{fig:dist}).

\subsection{Physical parameters}\label{sec:pars}

\addtocounter{table}{1} 

As a next step, we investigated the possible correlations between the proper motions and the physical parameters of the objects. \citet{luhman2008b} performed an analysis of the spectral energy distributions (SEDs) of confirmed members of Chamaeleon~I. A similar study of Chamaeleon~II was presented by \citet{alcala2008} and \citet{spezzi2008}. These works provide a classification of the confirmed members of the dark clouds into class III (diskless) sources, class II (disk) sources and so-called ``flat-spectrum sources'', which are thought to be halfway between class I (envelope) sources and class II sources. They also provide estimations of the 
effective temperature and bolometric luminosity of the objects. In total, we compiled values of physical parameters for 47 stars in Cha~I and 14 stars in Cha~II, whose properties are summarized in  Table~\ref{tab:pars}. Only objects with reliable UCAC3 proper motions are included. 

We used this information to look for correlations between the proper motions and the physical parameters of the objects. Figures~\ref{fig:pars_chai} and \ref{fig:pars_chaii} show the plots of the proper motion components versus $T_{eff}$
and $\log\,L/L_{\odot}$ 
for the objects in both clouds. No trends are seen in the proper motion with either of the considered parameters. Since these properties are directly correlated with mass in young low-mass stars, this result suggests that the motion of the objects does not depend on their mass to the precision that we can prove with the UCAC3 data. 

In Chamaeleon~I, where this can be investigated, we do not find any correlation either between proper motion and the presence of a (primordial) circumstellar disk, or for objects with and without disks (class II and class III sources) separately.

We note that similar studies in other star-forming regions also failed to find any correlation between the proper motion and the properties of the objects \citep[e.g.][LSJ11]{bouy2009}. More accurate proper motion measurements and better statistics are required to reveal any eventual trend with the physical properties of young low-mass stars in these regions.

%
\section{Conclusions}\label{sec:concl}

We have performed a kinematic study of the Chamaeleon clouds based on UCAC3 proper motions and radial velocities from the literature. Our analysis has led to the following conclusions:

\begin{itemize}

\item
With the UCAC3 data, we were able to separate the several distinct kinematic groups that coexist in the Chamaeleon sky area:  the Chamaeleon I and II dark clouds, the young $\epsilon$~Cha and $\eta$~Cha associations, and the field contaminants. 

\item
Compiled radial velocity data from the literature showed that all the young stars have similar radial velocities, regardless of the association the objects belong to. It is not possible to discern membership to the different groups seen towards Chamaeleon on the basis of radial velocity information alone.

\item
The stars in Chamaeleon~I and II may have different kinematical properties, but this result must be confirmed with more precise proper motions. If true, the Chamaeleon~II members would be moving faster in the tangential direction than the Chamaeleon~I members. Because they are located at the same or a further distance, this would question the physical connection between both clouds.

\item 
The kinematical analysis confirmed the membership of almost all the studied objects to the corresponding associations. We only identified two possible interlopers from the $\epsilon$~Cha association in our list of Chamaeleon~II members: CM~Cha and Sz~60W. With the available data, we reclassified CM~Cha as an $\epsilon$~Cha member. The nature of Sz~60W is more uncertain due to the big proper motion error, but given its spatial location and its isochronal age (2~Myr) we think it more likely that this object belongs to Chamaeleon~II. 

\item
We reanalysed a sample of X-ray detected stars towards the Chamaeleon sky area and confirmed that they constitute a mixture of different populations. After purging the sample of foreground and background contaminants and excluding the objects previously classified in the literature, we identified  two stars as likely Chamaeleon~I members, three as $\epsilon$~Cha members, and eight as $\eta$~Cha members. We did not identify any Chamaeleon~II members in this sample.

\item
We checked the proper motions of candidate members of the Chamaeleon dark clouds from the literature, and confirmed one object, C1-14, as a probable member of Chamaeleon~I. This would be one of the earliest type (i.e. most massive) stars in this dark cloud. In Chamaeleon~II, we checked the proper motions of two candidate members listed by \citet{spezzi2008} and of five candidate members from the list by \citet{vuong2001},  and found it unlikely that any of them belongs to the dark cloud. 

\item
For those objects with available radial velocities, we computed the corresponding Galactic space motions. Our results confirm the different spatial motion of Chamaeleon~I from the foreground associations. The data also hint at a difference in spatial motion between Chamaeleon~I and II, further suggesting  different origins for the populations in both clouds, but this has to be confirmed with better datasets. 

\item
The space velocities for the $\epsilon$ and $\eta$~Cha members and candidate members are in very good agreement with previous estimations in the literature, which identify these two clusters as part of the Local Association. Interestingly, the space motions of most of the Cha~II members with available radial velocity measurements also look quite similar to those of the Local Association stars. It would be highly desirable to have a larger sample of stars with estimated space velocities in this dark cloud, in order to investigate the eventual relation between Chamaeleon~II and the Local Association and to confirm or discard the different motion pattern from Chamaeleon~I.

\item
We investigated the relation between the proper motions and the published properties of the members and candidate members of Chamaeleon~I and II, such as spatial location, the presence of circumstellar material, effective temperature, or luminosity, but we found no evident correlations. More accurate proper motions, combined with radial velocity information, may unveil possible hidden trends with the physical properties of the stars, undetectable with the precision of the proper motions used here.

\end{itemize}

Our work has shown how the use of kinematical information can complement photometric and spectroscopic data to constrain the selection of members of young star-forming clusters, minimizing the contamination of the samples. Currently available astrometric catalogues are deep enough to provide proper motions for a significant number of candidate members of nearby star-forming regions precisely enough to disentangle true members, not only from background sources, but also from young stars in clusters overlapping in the line of sight. In a forthcoming paper \citep{vocha2}, we will report on a search for new candidate members of the Chamaeleon clouds based on proper motion.

Furthermore, the results obtained both in the present work and in our previous Lupus study illustrate the potential of using a VO methodology for analysing heterogeneous datasets in an efficient manner. 

Most of the issues left unsolved in the present study are expected to get a definitive answer with the advent of the {\em Gaia} mission of the European Space Agency, whose launch is foreseen in 2013.
{\em Gaia} will provide astrometric information with unprecedented precision for most of the stars in the samples studied here, remarkably improving our understanding of the kinematics, physical properties and formation process of the objects in the Chamaeleon associations.

 \begin{acknowledgements}

This work was funded by the Spanish MICINN through grant Consolider-CSD2006-00070. It also benefitted from funding from the Spanish government through grants ESP2007-65475-C02-02, AYA2008-02156, AYA2010-21161-C02-02, and AyA2011-24052, and from the Madrid regional government through grant PRICIT-S2009ESP-1496. A.~B. was co-funded under the Marie Curie Actions of the European Commission (FP7-COFUND).
 
This publication made use of VOSA, developed under the Spanish Virtual Observatory project supported by  the Spanish MEC through grants AyA2008-02156 and AyA2011-24052. It greatly benefitted 
from the use of the SIMBAD database and the VIZIER Catalogue Service, both operated at the CDS (Strasbourg, France). We used the VO-compliant tools Aladin, developed at the CDS, and TOPCAT, currently developed within the AstroGrid project. 

This publication makes use of data products from the Two Micron All Sky Survey (2MASS), which is a joint project of the University of Massachusetts and the Infrared Processing and Analysis Center/California Institute of Technology, funded by the US National Aeronautics and Space Administration and National Science Foundation, and from the Wide-field Infrared Survey Explorer (WISE), which is a joint project of the University of California, Los Angeles, and the Jet Propulsion Laboratory/California Institute of Technology, funded by the US National Aeronautics and Space Administration. 

\end{acknowledgements}


\bibliographystyle{aa} 
\bibliography{references}



\footnotesize

\input Chamaeleon_Members.table   

\input Chamaeleontis_Members.table   


\input Background_Members.table 

\input Members-photometry_vosa.table

\input Rosat.table  

\input Previous_Candidates.table  

\input Space-velocity.table

\input Members-properties.table 

\end{document}

%% file: Chamaeleon_Members.table
%

\longtabL{1}{
\begin{landscape}
\begin{longtable}{lll cc c ccccc c}

\caption{Confirmed members of Chamaeleon~I and II with UCAC3 proper motions (excluding wrong measurements and interlopers)}
\label{tab:mem}\\
    \hline \hline 
    \multicolumn{1}{c}{2MASS~J} &
    \multicolumn{1}{c}{Other names} &
    \multicolumn{1}{c}{UCAC3} &
    \multicolumn{1}{c}{$\mu_{\alpha}\cos\delta$} &
    \multicolumn{1}{c}{$\mu_{\delta}$} &
    \multicolumn{1}{c}{$V_r$} &
    \multicolumn{1}{c}{R} &
    \multicolumn{1}{c}{I} &
    \multicolumn{1}{c}{J} &
    \multicolumn{1}{c}{H} &
    \multicolumn{1}{c}{K} &
    \multicolumn{1}{c}{Notes} \\
    \multicolumn{1}{l}{} &
    \multicolumn{1}{c}{} &
    \multicolumn{1}{c}{} &
    \multicolumn{1}{c}{(mas/yr)} &
    \multicolumn{1}{c}{(mas/yr)} &
    \multicolumn{1}{c}{(km/s)} &
    \multicolumn{1}{c}{(mag)} &
    \multicolumn{1}{c}{(mag)} &
    \multicolumn{1}{c}{(mag)} &
    \multicolumn{1}{c}{(mag)} &
    \multicolumn{1}{c}{(mag)} &
    \multicolumn{1}{c}{} \\
      \hline
\endfirsthead

\caption{(continued)}\\
   \hline\hline
    \multicolumn{1}{c}{2MASS~J} &
    \multicolumn{1}{c}{Other names} &
    \multicolumn{1}{c}{UCAC3} &
    \multicolumn{1}{c}{$\mu_{\alpha}\cos\delta$} &
    \multicolumn{1}{c}{$\mu_{\delta}$} &
    \multicolumn{1}{c}{$V_r$} &
    \multicolumn{1}{c}{R} &
    \multicolumn{1}{c}{I} &
    \multicolumn{1}{c}{J} &
    \multicolumn{1}{c}{H} &
    \multicolumn{1}{c}{K} &
    \multicolumn{1}{c}{Notes} \\
    \multicolumn{1}{l}{} &
    \multicolumn{1}{c}{} &
    \multicolumn{1}{c}{} &
    \multicolumn{1}{c}{(mas/yr)} &
    \multicolumn{1}{c}{(mas/yr)} &
    \multicolumn{1}{c}{(km/s)} &
    \multicolumn{1}{c}{(mag)} &
    \multicolumn{1}{c}{(mag)} &
    \multicolumn{1}{c}{(mag)} &
    \multicolumn{1}{c}{(mag)} &
    \multicolumn{1}{c}{(mag)} &
    \multicolumn{1}{c}{} \\
    \hline
\endhead
\hline
\endfoot

  \multicolumn{12}{c}{Chamaeleon~I Cloud}\\

\object{10463795-7736035} & \object{HD~93828}		     		& \object{025-022840} &  --25.2\,$\pm$\,0.9  & --7.1\,$\pm$\,1.0  &		    & 13.298 & 12.637 & 7.844\,$\pm$\,0.02  & 7.655\,$\pm$\,0.06  & 7.477\,$\pm$\,0.02  &  \\
\object{10555973-7724399} & \object{T3A}		     		& \object{026-024097} &  --5.7\,$\pm$\,10.9  &  1.4\,$\pm$\,7.0   & 13.4\,$\pm$\,0.9 & 14.049 & 12.982 & 10.778\,$\pm$\,1.27 & 9.832\,$\pm$\,0.04  & 8.685\,$\pm$\,0.02  &  (1) \\
\object{10563044-7711393} & \object{SY~Cha}, \object{T4} 		& \object{026-024109} &  --19.0\,$\pm$\,4.7  & 7.4\,$\pm$\,2.6    & 12.7\,$\pm$\,0.1 & 13.339 & 12.421 & 9.969\,$\pm$\,0.02  & 9.102\,$\pm$\,0.02  & 8.631\,$\pm$\,0.02  &  \\
\object{10574219-7659356} & \object{Sz~4}, \object{T5} 		     	& \object{027-026396} &  --9.2\,$\pm$\,7.6 & 0.0\,$\pm$\,7.7	   & 16.5\,$\pm$\,1.3 & 14.754 & 13.329 & 10.433\,$\pm$\,0.02 & 9.563\,$\pm$\,0.02  & 9.246\,$\pm$\,0.02  &  (1) \\
\object{10581677-7717170} & \object{T6} 		     		& \object{026-024168} &  --22.5\,$\pm$\,6.9  & 1.7\,$\pm$\,2.7    &		      & 12.182 & 10.893 & 9.253\,$\pm$\,0.03  & 8.406\,$\pm$\,0.04  & 7.758\,$\pm$\,0.03  &  \\
\object{10590108-7722407} & \object{SW~Cha}, \object{T7}		& \object{026-024197} &  --18.1\,$\pm$\,1.8  & 4.6\,$\pm$\,1.8    & 15.7\,$\pm$\,1.2 & 13.496 & 12.891 & 10.135\,$\pm$\,0.02 & 9.232\,$\pm$\,0.02  & 8.616\,$\pm$\,0.02  &  \\
\object{10590699-7701404} & \object{CR~Cha}, \object{T8}		& \object{026-024202} &  --20.0\,$\pm$\,1.3  & 4.7\,$\pm$\,1.9    & 14.9\,$\pm$\,0.8 & 14.231 & 13.295 & 8.462\,$\pm$\,0.03  & 7.815\,$\pm$\,0.04  & 7.31\,$\pm$\,0.02   &  \\
\object{11004022-7619280} & \object{T10}		     		& \object{028-028320} &  --17.1\,$\pm$\,8.2  & --1.1\,$\pm$\,5.5  &		       & 16.027 & 14.561 & 11.859\,$\pm$\,0.03 & 11.238\,$\pm$\,0.02 & 10.871\,$\pm$\,0.02 &  \\
\object{11011875-7627025} & \object{CHXR~9C}		     		& \object{028-028370} &  --14.1\,$\pm$\,4.8  & --7.5\,$\pm$\,13.4 &		   & 13.305 & 11.663 & 10.08\,$\pm$\,0.03  & 9.307\,$\pm$\,0.03  & 8.994\,$\pm$\,0.02  &  \\
\object{11022491-7733357} & \object{CS~Cha}, \object{T11}		& \object{025-023391} &  --20.2\,$\pm$\,1.2  & 5.6\,$\pm$\,2.7    & 14.7\,$\pm$\,0.3 & 11.714 & 11.047 & 9.105\,$\pm$\,0.03  & 8.452\,$\pm$\,0.06  & 8.199\,$\pm$\,0.03  & (4) \\
\object{11022610-7502407} &                		     		& \object{030-032402} &  --17.9\,$\pm$\,4.7  & --4.5\,$\pm$\,4.5  &		   & 16.64  & 14.243 & 11.761\,$\pm$\,0.03 & 11.115\,$\pm$\,0.02 & 10.806\,$\pm$\,0.02 &  \\
\object{11023265-7729129} & \object{CHXR~71}		     		& \object{026-024313} &  --19.3\,$\pm$\,2.7  & 7.0\,$\pm$\,5.9    &		   & 16.357 & 14.764 & 11.268\,$\pm$\,0.03 & 10.459\,$\pm$\,0.02 & 10.129\,$\pm$\,0.03 &  \\
\object{11025504-7721508} & \object{T12}		     		& \object{026-024325} &  --20.7\,$\pm$\,3.2  & 8.5\,$\pm$\,3.3    &		   & 16.084 & 15.282 & 11.565\,$\pm$\,0.03 & 10.858\,$\pm$\,0.03 & 10.451\,$\pm$\,0.02 &  \\
\object{11035682-7721329} & \object{CHXR~12}		     		& \object{026-024352} &  --19.3\,$\pm$\,2.2  & --2.5\,$\pm$\,2.2  &		   & 15.505 & 13.686 & 10.796\,$\pm$\,0.02 & 9.994\,$\pm$\,0.03  & 9.712\,$\pm$\,0.02  &  \\
\object{11040909-7627193} & \object{CT~Cha}, \object{T14}		& \object{028-028608} &  --18.9\,$\pm$\,1.6  & 1.5\,$\pm$\,1.6    & 15.1\,$\pm$\,0.5 & 12.162 & 11.417 & 9.715\,$\pm$\,0.02  & 8.944\,$\pm$\,0.05  & 8.661\,$\pm$\,0.02  &  \\
\object{11045100-7625240} & \object{CHXR~14N}		     		& \object{028-028660} &  --18.1\,$\pm$\,1.9  & 4.3\,$\pm$\,1.9    &		   & 13.463 & 12.7   & 10.543\,$\pm$\,0.03 & 9.831\,$\pm$\,0.03  & 9.598\,$\pm$\,0.02  &  \\
\object{11045285-7625514} & \object{CHXR~14S}		     		& \object{028-028661} &  --18.6\,$\pm$\,2.5  & 1.1\,$\pm$\,2.4    &		   & 14.007 & 13.22  & 10.723\,$\pm$\,0.02 & 9.981\,$\pm$\,0.02  & 9.747\,$\pm$\,0.02  &  \\
\object{11050752-7812063} &                 		     		& \object{024-022792} &  --19.2\,$\pm$\,13.2 & --4.7\,$\pm$\,5.3  &		   & 16.535 & 14.615 & 11.776\,$\pm$\,0.02 & 11.096\,$\pm$\,0.02 & 10.816\,$\pm$\,0.02 &  \\
\object{11051467-7711290} & \object{Hn~4}		     		& \object{026-024389} &  --23.7\,$\pm$\,3.6  & 3.0\,$\pm$\,3.5    &		   & 16.425 & 15.139 & 10.944\,$\pm$\,0.02 & 10.035\,$\pm$\,0.03 & 9.608\,$\pm$\,0.02  &  \\
\object{11052472-7626209} &                 		     		& \object{028-028699} &  --21.1\,$\pm$\,3.0  & 0.4\,$\pm$\,3.1    &		   & 15.539 & 14.009 & 11.474\,$\pm$\,0.02 & 10.741\,$\pm$\,0.02 & 10.522\,$\pm$\,0.02 &  \\
\object{11055261-7618255} & \object{T20}		     		& \object{028-028735} &  --21.9\,$\pm$\,2.0  & 0.7\,$\pm$\,1.9    &		       & 13.367 & 12.807 & 10.312\,$\pm$\,0.02 & 9.593\,$\pm$\,0.02  & 9.339\,$\pm$\,0.02  &  \\
\object{11061540-7721567} & \object{T21}		     		& \object{026-024403} &  --19.2\,$\pm$\,1.0  & 5.0\,$\pm$\,1.6    & 17.0       &	&	 & 7.597\,$\pm$\,0.02  & 6.828\,$\pm$\,0.05  & 6.419\,$\pm$\,0.02  &  \\
\object{11064346-7726343} & \object{T22}		     		& \object{026-024409} &  --21.4\,$\pm$\,3.5  & --0.5\,$\pm$\,3.7  &		       & 16.252 & 14.776 & 10.813\,$\pm$\,0.03 & 9.788\,$\pm$\,0.03  & 9.392\,$\pm$\,0.02  &  \\
\object{11064510-7727023} & \object{CHXR~20}		     		& \object{026-024411} &  --17.9\,$\pm$\,2.4  & 5.6\,$\pm$\,7.1    &		   & 15.535 & 14.781 & 10.184\,$\pm$\,0.02 & 9.197\,$\pm$\,0.02  & 8.88\,$\pm$\,0.02   &  \\
\object{11065906-7718535} & \object{T23}		     		& \object{026-024418} &  --18.3\,$\pm$\,3.4  & 1.7\,$\pm$\,4.6    &		   & 16.199 & 15.368 & 11.204\,$\pm$\,0.03 & 10.423\,$\pm$\,0.02 & 10.003\,$\pm$\,0.02 &  \\
\object{11071148-7746394} & \object{CHXR~21}		     		& \object{025-023408} &  --22.3\,$\pm$\,3.9  & 3.0\,$\pm$\,3.7    &		   & 16.162 & 14.964 & 11.084\,$\pm$\,0.03 & 10.083\,$\pm$\,0.02 & 9.657\,$\pm$\,0.02  &  \\
\object{11071915-7603048} & \object{T25}		     		& \object{028-028831} &  --22.1\,$\pm$\,2.8  & --2.0\,$\pm$\,2.9  &		   & 14.781 & 13.468 & 10.961\,$\pm$\,0.02 & 10.092\,$\pm$\,0.02 & 9.769\,$\pm$\,0.02  &  \\
\object{11072040-7729403} & \object{[LES2004]ChaI~448}, \object{ISO~99} & \object{026-024421} &  --20.8\,$\pm$\,2.8  & 2.8\,$\pm$\,2.5    &                  & 16.241 & 15.254 & 11.134\,$\pm$\,0.02 & 10.547\,$\pm$\,0.02 & 10.259\,$\pm$\,0.02 &  \\
\object{11072074-7738073} & \object{DI~Cha}, \object{T26}		& \object{025-023410} &  --15.8\,$\pm$\,1.6  & --5.1\,$\pm$\,1.7  & 13.5\,$\pm$\,0.6 &      &	     & 7.819\,$\pm$\,0.03  & 6.942\,$\pm$\,0.04  & 6.217\,$\pm$\,0.02  &  \\
\object{11072825-7652118} & \object{VV~Cha}, \object{T27}		& \object{027-026870} &  --20.0\,$\pm$\,1.9  & --1.8\,$\pm$\,1.9  & 15.4\,$\pm$\,1.3 & 14.455 & 13.013 & 10.663\,$\pm$\,0.03 & 9.905\,$\pm$\,0.03  & 9.519\,$\pm$\,0.02  &  \\
\object{11074366-7739411} & \object{T28}		        	& \object{025-023415} &  --19.8\,$\pm$\,2.2  & --0.3\,$\pm$\,3.9  &		   & 15.719 & 14.11  & 10.165\,$\pm$\,0.02 & 8.975\,$\pm$\,0.02  & 8.258\,$\pm$\,0.03  &  \\
\object{11075588-7727257} & \object{CHXR~28}                    	& \object{026-024437} &  --28.8\,$\pm$\,3.3  & 6.1\,$\pm$\,3.4    &		  & 14.907 & 13.734 & 9.215\,$\pm$\,0.03  & 8.052\,$\pm$\,0.03  & 7.691\,$\pm$\,0.03  &  \\
\object{11075792-7738449} & \object{T29}                        	& \object{025-023419} &  --26.7\,$\pm$\,5.7  & 13.7\,$\pm$\,7.7   &		  & 15.038 & 13.725 & 9.502\,$\pm$\,0.03  & 7.919\,$\pm$\,0.04  & 6.83\,$\pm$\,0.02   &  \\
\object{11080148-7742288} & \object{VW~Cha}, \object{T31}               & \object{025-023420} &  --19.7\,$\pm$\,1.3  & --0.8\,$\pm$\,3.4  & 15.1\,$\pm$\,0.1 & 14.915 & 13.48  & 8.703\,$\pm$\,0.03  & 7.637\,$\pm$\,0.04  & 6.962\,$\pm$\,0.03  & (5) \\
\object{11080329-7739174} & \object{HD~97048}, \object{T32}             & \object{025-023421} &  --17.8\,$\pm$\,0.8  & 1.5\,$\pm$\,0.8    &  		&	 &	  & 7.267\,$\pm$\,0.02  & 6.665\,$\pm$\,0.05  & 5.941\,$\pm$\,0.03  &  \\
\object{11081509-7733531} & \object{T33A}                       	& \object{025-023422} &  --2.1\,$\pm$\,5.6	  & --4.4\,$\pm$\,5.5  &		  & 14.726 & 13.679 & 8.638\,$\pm$\,0.03  & 7.729\,$\pm$\,1.27  & 6.876\,$\pm$\,1.27  & (1) \\
\object{11081648-7744371} & \object{T34}                        	& \object{025-023423} &  --17.0\,$\pm$\,3.1  & --1.2\,$\pm$\,4.1  &		  & 15.975 & 14.716 & 11.198\,$\pm$\,0.03 & 10.339\,$\pm$\,0.03 & 10.024\,$\pm$\,0.02 &  \\
\object{11083905-7716042} & \object{T35}                        	& \object{026-024448} &  --19.7\,$\pm$\,3.0  & 1.2\,$\pm$\,3.1    &		  & 15.822 & 17.073 & 11.173\,$\pm$\,0.03 & 9.904\,$\pm$\,0.02  & 9.109\,$\pm$\,0.02  &  \\
\object{11084069-7636078} & \object{CHXR~33}                     	& \object{027-026898} &  --17.3\,$\pm$\,2.6  & --0.8\,$\pm$\,6.0  &		  & 15.044 & 13.57  & 10.558\,$\pm$\,0.02 & 9.658\,$\pm$\,0.02  & 9.276\,$\pm$\,0.02  &  \\
\object{11085367-7521359} & \object{PU~Car}                      	& \object{030-033240} &  --5.5\,$\pm$\,5.2	  & 2.9\,$\pm$\,5.1    &		  &	   &	    & 10.98\,$\pm$\,0.03  & 10.072\,$\pm$\,0.03 & 9.556\,$\pm$\,0.03  &  (1) \\
\object{11085464-7702129} & \object{T38}                         	& \object{026-024450} &  --18.4\,$\pm$\,3.1  & 6.1\,$\pm$\,2.7    &		  & 16.279 & 14.393 & 11.208\,$\pm$\,0.03 & 10.175\,$\pm$\,0.02 & 9.457\,$\pm$\,0.02  &  \\
\object{11091769-7627578} & \object{CHXR~37}, \object{RXJ1109.4-7627}   & \object{028-028916} &  --22.3\,$\pm$\,1.8  & --0.8\,$\pm$\,9.9  & 13.1\,$\pm$\,2.0 & 15.032 & 14.147 & 9.998\,$\pm$\,0.02  & 9.044\,$\pm$\,0.02  & 8.701\,$\pm$\,0.02  & (3) \\
\object{11092379-7623207} & \object{VZ~Cha}, \object{T40}               & \object{028-028926} &  --20.9\,$\pm$\,3.3  & 1.7\,$\pm$\,2.6    & 14.7	  & 13.513 & 11.686 & 10.44\,$\pm$\,0.03  & 9.214\,$\pm$\,0.02  & 8.242\,$\pm$\,0.04  &  \\
\object{11094006-7628391} & \object{CHXR 40}                    	& \object{028-028940} &  --22.4\,$\pm$\,6.8  & 4.0\,$\pm$\,2.6    &		  & 15.058 & 13.993 & 10.072\,$\pm$\,0.02 & 9.23\,$\pm$\,0.02	& 8.961\,$\pm$\,0.02  &  \\
\object{11095003-7636476} & \object{HD~97300}, \object{T41}      	& \object{027-026916} &  --17.7\,$\pm$\,1.0  & --2.7\,$\pm$\,1.1  &		  &	   &	    & 7.641\,$\pm$\,0.03  & 7.347\,$\pm$\,0.06  & 7.149\,$\pm$\,0.02  &  (2) \\
\object{11095407-7629253} & \object{T43}	         		& \object{028-028953} &  --25.5\,$\pm$\,6.4  & 2.5\,$\pm$\,5.8   &		  & 16.736 & 15.258 & 11.295\,$\pm$\,0.03 & 10.003\,$\pm$\,0.02 & 9.254\,$\pm$\,0.02  &  \\
\object{11095873-7737088} & \object{T45}	         		& \object{025-023440} &  --20.9\,$\pm$\,1.7  & --0.6\,$\pm$\,2.3  &		  & 13.104 & 12.147 & 9.835\,$\pm$\,0.03  & 8.768\,$\pm$\,0.05  & 7.97\,$\pm$\,0.02   &  \\
\object{11100469-7635452} & \object{T45a}		 		& \object{027-026921} &  --16.3\,$\pm$\,2.1  & 5.9\,$\pm$\,2.1    &		   &	    &	     & 10.565\,$\pm$\,0.02 & 9.636\,$\pm$\,0.02  & 9.24\,$\pm$\,0.02   &  \\
\object{11100704-7629376} & \object{T46}		 		& \object{028-028961} &  --22.2\,$\pm$\,1.8  & --0.5\,$\pm$\,1.8  & 12.9\,$\pm$\,0.9 &      &	     & 9.911\,$\pm$\,0.02  & 8.964\,$\pm$\,0.04  & 8.451\,$\pm$\,0.03  &  \\
\object{11101141-7635292} & \object{ISO~237}		 		& \object{027-026922} &  --21.1\,$\pm$\,5.6  & 5.1\,$\pm$\,5.7    &  	      &        &	& 10.932\,$\pm$\,0.03 & 9.438\,$\pm$\,0.02  & 8.621\,$\pm$\,0.02  &  \\
\object{11103801-7732399} & \object{CHXR~47}, \object{WKK~F34}		& \object{025-023452} &  --20.3\,$\pm$\,1.8  & 1.7\,$\pm$\,1.8    & 14.0\,$\pm$\,2.0 & 13.826 & 11.99  & 9.741\,$\pm$\,0.03  & 8.687\,$\pm$\,0.05  & 8.277\,$\pm$\,0.03  & (3) \\
\object{11105333-7634319} & \object{T48}		    		& \object{027-026930} &  --21.6\,$\pm$\,2.6  & --0.1\,$\pm$\,2.8  &		   & 15.064 & 13.723 & 11.259\,$\pm$\,0.02 & 10.447\,$\pm$\,0.02 & 10.038\,$\pm$\,0.02 &  \\
\object{11113474-7636211} & \object{CHXR~48}		    		& \object{027-026948} &  --17.9\,$\pm$\,9.9  & --2.4\,$\pm$\,2.2  &		   & 15.741 & 14.367 & 10.864\,$\pm$\,0.03 & 10.081\,$\pm$\,0.02 & 9.802\,$\pm$\,0.02  &  \\
\object{11114632-7620092} & \object{CHX~18N}		    		& \object{028-029063} &  --20.1\,$\pm$\,1.4  & 0.8\,$\pm$\,1.4    & 15.5	   & 11.334 & 10.824 & 9.111\,$\pm$\,0.03  & 8.286\,$\pm$\,0.03  & 7.767\,$\pm$\,0.03  &  (4) \\
\object{11115400-7619311} & \object{CHXR 49NE}  	    		& \object{028-029070} &  --22.4\,$\pm$\,1.8  & --0.2\,$\pm$\,1.8  &	    & 13.893 & 12.16  & 10.203\,$\pm$\,0.02 & 9.531\,$\pm$\,0.03  & 9.233\,$\pm$\,0.02  &  \\
\object{11122441-7637064} & \object{CHX~20}, \object{T51}, \object{RXJ1112.7-7637} & \object{027-026990} &  --29.8\,$\pm$\,4.3  & 4.7\,$\pm$\,1.7	 & 13.9\,$\pm$\,0.4 &	&	 & 9.275\,$\pm$\,0.03  & 8.524\,$\pm$\,0.06  & 7.999\,$\pm$\,0.03  & (3, 4) \\
\object{11122772-7644223} & \object{CV~Cha}, \object{T52}	        & \object{027-026995} &  --21.3\,$\pm$\,1.1  & 1.9\,$\pm$\,1.0	& 15.1\,$\pm$\,0.3 & 16.96  & 12.953 & 8.285\,$\pm$\,0.02  & 7.461\,$\pm$\,0.04  & 6.845\,$\pm$\,0.03  &  \\
\object{11123092-7644241} & \object{CW~Cha}, \object{T53}	        & \object{027-026996} &  --13.6\,$\pm$\,3.2  & 4.2\,$\pm$\,3.4	&		&	 &	  & 10.913\,$\pm$\,0.03 & 9.854\,$\pm$\,0.03  & 9.125\,$\pm$\,0.02  &  (1, 2) \\
\object{11124210-7658400} & \object{CHXR~54}		        	& \object{027-027004} &  --22.4\,$\pm$\,1.2  & 1.9\,$\pm$\,2.5    &		    & 12.693 & 11.816 & 10.409\,$\pm$\,0.02 & 9.721\,$\pm$\,0.02  & 9.499\,$\pm$\,0.02  &  \\
\object{11124268-7722230} & \object{T54}		        	& \object{026-024473} &  --21.6\,$\pm$\,1.0  & 4.8\,$\pm$\,1.0     &		   &	    &	     & 8.645\,$\pm$\,0.03  & 8.042\,$\pm$\,0.05  & 7.88\,$\pm$\,0.03   &  \\
\object{11124299-7637049} & \object{CHXR~55}		        	& \object{027-027005} &  --21.6\,$\pm$\,1.7  & --1.9\,$\pm$\,1.7  &		   &	    &	     & 10.058\,$\pm$\,0.03 & 9.4\,$\pm$\,0.02	 & 9.288\,$\pm$\,0.02  &  \\
\object{11132012-7701044} & \object{CHXR~57}		        	& \object{026-024482} &  --20.9\,$\pm$\,1.4  & 2.7\,$\pm$\,2.0     &		   & 15.836 & 14.439 & 10.884\,$\pm$\,0.02 & 10.245\,$\pm$\,0.03 & 10.013\,$\pm$\,0.03 &  \\
\object{11132112-7625558} &                                     	& \object{028-029155} &  --27.2\,$\pm$\,5.3  & 10.8\,$\pm$\,2.7   &		  & 10.519 & 8.432  & 5.436\,$\pm$\,0.02  & 4.512\,$\pm$\,0.08  & 4.128\,$\pm$\,0.3   &  \\
\object{11132446-7629227} & \object{Hn~18}                      	& \object{028-029161} &  --24.6\,$\pm$\,4.0  & 1.5\,$\pm$\,3.4    &		  & 16.094 & 14.367 & 11.855\,$\pm$\,0.03 & 11.108\,$\pm$\,0.02 & 10.797\,$\pm$\,0.02 &  \\
\object{11132737-7634165} & \object{CHXR~59}		        	& \object{027-027042} &  --21.5\,$\pm$\,1.8  & --0.8\,$\pm$\,1.8  &		   & 13.993 & 12.487 & 10.606\,$\pm$\,0.03 & 9.863\,$\pm$\,0.03  & 9.634\,$\pm$\,0.02  &  \\
\object{11132970-7629012} & \object{CHXR~60}		        	& \object{028-029171} &  --23.9\,$\pm$\,3.5  & --5.3\,$\pm$\,3.4  &		   & 16.029 & 14.136 & 11.569\,$\pm$\,0.03 & 10.856\,$\pm$\,0.02 & 10.577\,$\pm$\,0.02 &  \\
\object{11141565-7627364} & \object{CHXR~62}		        	& \object{028-029208} &  --21.0\,$\pm$\,3.5  & 3.4\,$\pm$\,3.5    &		   & 15.731 & 13.71  & 11.286\,$\pm$\,0.02 & 10.465\,$\pm$\,0.03 & 10.12\,$\pm$\,0.02  &  \\
\object{11145031-7733390} & \object{BYB~53}                        	& \object{025-023572} &  --21.9\,$\pm$\,1.6  & 2.5\,$\pm$\,1.5    &		  & 13.865 & 12.4   & 10.48\,$\pm$\,0.03  & 9.75\,$\pm$\,0.02	& 9.547\,$\pm$\,0.02  &  \\
\object{11155827-7729046} &                             		& \object{026-024615} &  --23.7\,$\pm$\,4.0  & 9.0\,$\pm$\,3.9    &		   & 16.454 & 14.647 & 11.942\,$\pm$\,0.03 & 11.311\,$\pm$\,0.02 & 11.016\,$\pm$\,0.02 &  \\
\object{11173700-7704381} & \object{T56}                	        & \object{026-024719} &  --21.8\,$\pm$\,1.5  & --1.7\,$\pm$\,1.4  &		   & 13.008 & 12.204 & 10.303\,$\pm$\,0.02 & 9.575\,$\pm$\,0.03  & 9.227\,$\pm$\,0.02  &  \\
\object{11175211-7629392} &                             		& \object{028-029527} &  --14.2\,$\pm$\,5.5  & 3.6\,$\pm$\,7.2    &		   & 15.663 & 13.394 & 11.664\,$\pm$\,0.04 & 11.151\,$\pm$\,0.05 & 10.715\,$\pm$\,0.04 &  (1) \\
\object{11242980-7554237} &                             		& \object{029-033027} &  --9.4\,$\pm$\,3.8 & 4.6\,$\pm$\,4.2	&		   & 15.837 & 13.418 & 10.931\,$\pm$\,0.02 & 10.202\,$\pm$\,0.02 & 9.884\,$\pm$\,0.02  &  (1) \\
\object{11253653-7700348} &                             		& \object{026-025416} &  --11.4\,$\pm$\,2.0  & 5.0\,$\pm$\,1.3    &		   & 9.514  & 8.135  & 6.405\,$\pm$\,0.02  & 5.543\,$\pm$\,0.06  & 5.084\,$\pm$\,0.02  &  (1) \\

  \hline
  \multicolumn{12}{c}{Chamaeleon~II Cloud}\\

\object{12563364-7645453} & \object{Sz 46N}                                           & \object{027-035004} &  --18.7\,$\pm$\,5.4  & 23.0\,$\pm$\,19.2  & 15.5$\pm$1.2 & 14.128 & 12.943 & 11.253\,$\pm$\,0.03 & 10.262\,$\pm$\,0.03 & 9.758\,$\pm$\,0.02  &  (1) \\
\object{12565864-7647067} & \object{Sz 47}                                            & \object{027-035025} &  --4.7\,$\pm$\,2.8   & --0.6\,$\pm$\,2.8  &              & 13.711 & 13.342 & 12.84\,$\pm$\,0.03  & 12.496\,$\pm$\,0.02 & 12.247\,$\pm$\,0.02 &  (1) \\
\object{12571172-7640111} & \object{IRAS 12535-7623}, \object{CHIIXR 2}               & \object{027-035032} &  --26.3\,$\pm$\,3.4  & --12.4\,$\pm$\,2.7 & 10.0$\pm$1.5 & 13.383 & 11.791 & 10.037\,$\pm$\,0.02 & 8.896\,$\pm$\,0.02  & 8.395\,$\pm$\,0.02  &  \\
\object{12592638-7747086} & \object{IRAS 12556-7731}                                  & \object{025-031227} &  --7.0\,$\pm$\,1.8   & --4.1\,$\pm$\,1.4  &              & 12.423 & 9.534  & 7.876\,$\pm$\,0.02  & 6.862\,$\pm$\,0.03  & 6.548\,$\pm$\,0.02  &  (1) \\
\object{13005323-7654151} & \object{Sz 49}, \object{Iso-Cha II 55}, \object{CHIIXR 9} & \object{027-035312} &  --21.5\,$\pm$\,6.5  & --9.7\,$\pm$\,21.8 &              & 16.74  &        & 11.823\,$\pm$\,0.02 & 11.071\,$\pm$\,0.02 & 10.633\,$\pm$\,0.02 &  \\
\object{13005532-7710222} & \object{Sz 50}, \object{Iso-Cha II 52}, \object{CHIIXR 8} & \object{026-031723} &  --26.3\,$\pm$\,4.0  & --6.5\,$\pm$\,3.7  & 13.0$\pm$1.2 & 14.51  & 12.123 & 10.314\,$\pm$\,0.02 & 9.325\,$\pm$\,0.03  & 8.849\,$\pm$\,0.02  &  \\
\object{13005534-7708296} & \object{WFIJ13005531-7708295}                             & \object{026-031722} &  --33.3\,$\pm$\,10.1 & --9.8\,$\pm$\,8.6  &              & 15.22  & 13.17  & 11.175\,$\pm$\,0.02 & 10.092\,$\pm$\,0.03 & 9.691\,$\pm$\,0.02  &  \\
\object{13005622-7654021} & \object{RXJ1301.0-7654a}                                  & \object{027-035316} &  --23.0\,$\pm$\,2.6  & --8.3\,$\pm$\,2.5  &              & 11.174 & 10.201 & 9.031\,$\pm$\,0.02  & 8.285\,$\pm$\,0.04  & 8.095\,$\pm$\,0.03  &  \\
\object{13015891-7751218} & \object{Sz 51}                                            & \object{025-031334} &  --20.7\,$\pm$\,1.6  & --11.9\,$\pm$\,1.6 &              & 13.328 & 12.153 & 10.616\,$\pm$\,0.02 & 9.849\,$\pm$\,0.02  & 9.354\,$\pm$\,0.02  &  \\
\object{13030444-7707027} & \object{RXJ1303.1-7706}                  		      & \object{026-031811} &  --31.8\,$\pm$\,2.0  & --7.3\,$\pm$\,1.8  & 12.4$\pm$0.4 & 12.009 & 10.828 & 9.633\,$\pm$\,0.02  & 8.9\,$\pm$\,0.02    & 8.671\,$\pm$\,0.02  &  \\
\object{13042410-7650012} & \object{Hn 23}			     		      & \object{027-035636} &	  5.1\,$\pm$\,23.8 &  18.8\,$\pm$\,21.6 & 15.2$\pm$0.2 & 11.027 & 10.023 & 9.909\,$\pm$\,0.03  & 9.109\,$\pm$\,0.03  & 8.769\,$\pm$\,0.03  &  (1) \\
\object{13045571-7739495} & \object{Hn 24}			     		      & \object{025-031455} &  --25.6\,$\pm$\,4.3  & --7.2\,$\pm$\,3.4  &              & 12.948 & 11.614 & 10.178\,$\pm$\,0.02 & 9.302\,$\pm$\,0.02  & 8.92\,$\pm$\,0.02   &  \\
\object{13052072-7739015} & \object{BF~ Cha}, \object{Sz 54}			      & \object{025-031467} &  --24.2\,$\pm$\,1.7  & --4.3\,$\pm$\,1.8  &              & 14.473 & 12.969 & 9.052\,$\pm$\,0.02  & 8.157\,$\pm$\,0.04  & 7.597\,$\pm$\,0.03  &  \\
\object{13063053-7734001} & \object{Sz 55}			     		      & \object{025-031478} &  --14.1\,$\pm$\,26.8 & --4.9\,$\pm$\,7.0  &              & 16.725 & 14.694 & 12.544\,$\pm$\,0.02 & 11.556\,$\pm$\,0.02 & 10.921\,$\pm$\,0.02 &  \\
\object{13065744-7723415} & \object{Sz 58}, \object{IRAS 13030-7707}, \object{[VCE2001]\,C61}    & \object{026-032018} &  --27.9\,$\pm$\,9.0  & --8.0\,$\pm$\,4.3  & 12.3$\pm$4.3 & 14.348 & 13.68  & 10.841\,$\pm$\,0.02 & 9.579\,$\pm$\,0.03  & 8.755\,$\pm$\,0.02  &  \\
\object{13070922-7730304} & \object{Sz 59}			    		      & \object{025-031492} &  --19.2\,$\pm$\,2.2  & --12.8\,$\pm$\,3.5 & 12.7$\pm$2.7 & 12.535 & 12.069 & 10.517\,$\pm$\,0.03 & 9.26\,$\pm$\,0.03   & 8.381\,$\pm$\,0.02  &  \\
\object{13072241-7737225} & \object{Sz 60W}			    		      & \object{025-031495} &  --52.6\,$\pm$\,20.4 & --7.3\,$\pm$\,4.5  & 13.2$\pm$0.9 & 13.639 & 11.804 & 11.19\,$\pm$\,0.06  & 10.224\,$\pm$\,0.07 & 9.54\,$\pm$\,0.04   & (6) \\
\object{13074851-7741214} & \object{Hn 26}			    		      & \object{025-031505} &  --16.3\,$\pm$\,4.9  & --3.8\,$\pm$\,9.1  &              & 14.92  & 14.354 & 11.689\,$\pm$\,0.02 & 10.62\,$\pm$\,0.03  & 10.019\,$\pm$\,0.02 &  \\
\object{13080628-7755051} & \object{Sz 61}			    		      & \object{025-031510} &  --25.1\,$\pm$\,2.1  & --6.3\,$\pm$\,2.0  & 12.0$\pm$3.0 & 13.721 & 11.932 & 9.888\,$\pm$\,0.02  & 8.761\,$\pm$\,0.02  & 7.948\,$\pm$\,0.02  &  \\
\object{13095036-7757240} & \object{Sz 62}			    		      & \object{025-031562} &  --15.0\,$\pm$\,5.4  & --7.0\,$\pm$\,2.3  &              & 13.736 & 12.248 & 10.522\,$\pm$\,0.03 & 9.653\,$\pm$\,0.03  & 9.119\,$\pm$\,0.02  &  \\
\object{13100415-7710447} & \object{Sz 63}			    		      & \object{026-032239} &  --25.3\,$\pm$\,4.8  & --3.0\,$\pm$\,2.7  &              & 14.369 & 12.645 & 11.419\,$\pm$\,0.03 & 10.707\,$\pm$\,0.03 & 10.393\,$\pm$\,0.02 &  \\
\object{13125238-7739182} &                                         		      & \object{025-031762} &  --15.4\,$\pm$\,4.3  & 9.7\,$\pm$\,4.7	&              & 15.191 & 12.824 & 11.493\,$\pm$\,0.03 & 10.854\,$\pm$\,0.03 & 10.544\,$\pm$\,0.02 &  (1)\\

  \hline
\end{longtable}

 \begin{flushleft}
 {\bf Notes.} \\
      (1) Dubious proper motion. \\
      (2) Discrepant proper motion form Hipparcos.\\
      (3) ROSAT source studied by \citet{covino1997}.\\
      (4) Spectroscopic binary.\\
      (5) Sometimes classified in the literature as a member of the $\epsilon$~Cha cluster (see text for discussion).\\
      (6) Possible interloper from the $\epsilon$~Cha cluster (see text for discussion).
 \end{flushleft}

\end{landscape}
}

%% file: Chamaeleontis_Members.table
\begin{landscape}
\begin{table}
  \caption[]{Kinematic members of the foreground $\epsilon$ and $\eta$ Chamaeleontis clusters (excluding wrong measurements and interlopers)}
  \label{tab:clus}
  \begin{center}
  \begin{tabular}{lll cc c ccccc l}
    \hline
    \hline

    \multicolumn{1}{c}{2MASS~J} &
    \multicolumn{1}{c}{Other names} &
    \multicolumn{1}{c}{UCAC3} &
    \multicolumn{1}{c}{$\mu_{\alpha}\cos\delta$} &
    \multicolumn{1}{c}{$\mu_{\delta}$} &
    \multicolumn{1}{c}{Rv} &
    \multicolumn{1}{c}{R} &
    \multicolumn{1}{c}{I} &
    \multicolumn{1}{c}{J} &
    \multicolumn{1}{c}{H} &
    \multicolumn{1}{c}{K} &
    \multicolumn{1}{c}{Notes} \\
    \multicolumn{1}{l}{} &
    \multicolumn{1}{c}{} &
    \multicolumn{1}{c}{} &
    \multicolumn{1}{c}{(mas/yr)} &
    \multicolumn{1}{c}{(mas/yr)} &
    \multicolumn{1}{c}{(km/s)} &
    \multicolumn{1}{c}{(mag)} &
    \multicolumn{1}{c}{(mag)} &
    \multicolumn{1}{c}{(mag)} &
    \multicolumn{1}{c}{(mag)} &
    \multicolumn{1}{c}{(mag)} &
    \multicolumn{1}{c}{} \\

    \hline

  \multicolumn{12}{c}{$\epsilon$~Chamaeleontis cluster}\\

\object{10574936-6913599} & \object{CP-68 1388}                             & \object{042-083475} & -35.8\,$\pm$\,2.2  & 7.6\,$\pm$\,2.3 & 15.9  	    & 13.941 & 12.741 & 8.484\,$\pm$\,0.03 & 8.009\,$\pm$\,0.05 & 7.792\,$\pm$\,0.02 & (1) \\
\object{11080148-7742288} & \object{VW~Cha}, \object{RXJ1108.1-7742}        & \object{025-023420} & -19.7\,$\pm$\,1.3  & -0.\,$\pm$\,3.4 &		    & 14.915 & 13.48  & 8.703\,$\pm$\,0.03 & 7.637\,$\pm$\,0.04 & 6.962\,$\pm$\,0.03 & (2) \\
\object{11183572-7935548} & \object{$\epsilon$~Cha~13}                      & \object{021-021275} & -46.6\,$\pm$\,4.5  & 4.4\,$\pm$\,2.6 &		    & 14.201 & 12.374 & 10.50\,$\pm$\,0.03 & 9.891\,$\pm$\,0.02 & 9.62\,$\pm$\,0.02  & (3) \\
\object{11493184-7851011} & \object{DZ~Cha}                                 & \object{023-025652} & -43.4\,$\pm$\,2.9  & -8.\,$\pm$\,1.8 & 12.2\,$\pm$\,2.0 & 11.997 & 10.513 & 9.449\,$\pm$\,0.02 & 8.721\,$\pm$\,0.05 & 8.486\,$\pm$\,0.02 & (2) \\
\object{11571348-7921313} & \object{T~Cha}, \object{RXJ1157.2-7921}         & \object{022-024424} & -41.2\,$\pm$\,3.2  & -6.\,$\pm$\,1.4 & 20.0\,$\pm$\,2.0 & 14.981 & 13.477 & 8.955\,$\pm$\,0.03 & 7.86 \,$\pm$\,0.05 & 6.954\,$\pm$\,0.02 & (2) \\
\object{11582681-7754450} & \object{GSC 9415-2676}                          & \object{025-027033} & -38.5\,$\pm$\,2.9  & -7.\,$\pm$\,2.2 & 13.0\,$\pm$\,2.0 &	     &	      & 10.34\,$\pm$\,0.03 & 9.715\,$\pm$\,0.03 & 9.436\,$\pm$\,0.02 &  \\
\object{11582816-7754294} & \object{DW~Cha}, \object{RXJ1158.5-7754}        & \object{025-027034} & -40.4\,$\pm$\,2.0  & -2.\,$\pm$\,1.0 & 13.1\,$\pm$\,2.0 & 14.084 & 12.967 & 8.219\,$\pm$\,0.03 & 7.556\,$\pm$\,0.04 & 7.404\,$\pm$\,0.02 & (2) \\
\object{11594226-7601260} & \object{RXJ1159.7-7601}                         & \object{028-033151} & -40.9\,$\pm$\,2.1  & -6.\,$\pm$\,1.8 & 13.1\,$\pm$\,2.0 & 10.32  & 9.978  & 9.14 \,$\pm$\,0.03 & 8.469\,$\pm$\,0.04 & 8.304\,$\pm$\,0.03 & (2) \\
\object{12000829-7811395} & \object{HD~104237D} 			    & \object{024-027122} & -42.9\,$\pm$\,7.1  & 27.\,$\pm$\,7.1 &		    &	     &	      & 9.615\,$\pm$\,0.06 & 8.744\,$\pm$\,0.11 & 8.124\,$\pm$\,0.21 & (4) \\
\object{12000931-7811424} & \object{HD~104237E} 			    & \object{024-027125} &   7.3\,$\pm$\,5.2  & 5.9\,$\pm$\,5.2 &		    &	     &	      & 9.104\,$\pm$\,1.27 & 8.254\,$\pm$\,1.27 & 7.488\,$\pm$\,0.02 & (4) \\
\object{12013915-7859168} & \object{HD~104467}, \object{RXJ1201.7-7859}     & \object{023-026618} & -41.3\,$\pm$\,1.0  & -5.\,$\pm$\,1.0 & 10.0\,$\pm$\,2.0 &	     &	      & 7.263\,$\pm$\,0.03 & 6.967\,$\pm$\,0.04 & 6.848\,$\pm$\,0.02 & (2) \\
\object{12020369-7853012} & \object{GSC~9420-0948}, \object{RXJ1202.1-7853} & \object{023-026647} & -41.0\,$\pm$\,3.5  & -4.\,$\pm$\,1.7 & 5.0 \,$\pm$\,2.0 & 14.944 & 13.629 & 9.215\,$\pm$\,0.02 & 8.456\,$\pm$\,0.04 & 8.307\,$\pm$\,0.02 & (2, 5) \\
\object{12043615-7731345} & \object{GSC~9416-1029}, \object{RXJ1204.6-7731} & \object{025-027530} & -38.8\,$\pm$\,1.5  & -2.\,$\pm$\,2.4 & 10.4\,$\pm$\,2.0 & 12.879 & 11.306 & 9.765\,$\pm$\,0.02 & 9.125\,$\pm$\,0.02 & 8.881\,$\pm$\,0.02 & (2) \\
\object{12074597-7816064} & \object{2MASS~J12074597-7816064}                & \object{024-027765} & -53.4\,$\pm$\,14.5 & -15\,$\pm$\,7.0 &		    & 15.827 & 14.409 & 11.55\,$\pm$\,0.02 & 10.98\,$\pm$\,0.03 & 10.67\,$\pm$\,0.02 &  \\
\object{12113815-7110360} & \object{HD~105923}, \object{HIP~105923}         & \object{038-075763} & -37.9\,$\pm$\,2.5  & -8.\,$\pm$\,1.6 & 14.2  	    & 8.506  & 5.818  & 7.674\,$\pm$\,0.02 & 7.308\,$\pm$\,0.04 & 7.175\,$\pm$\,0.03 &  \\
\object{12194369-7403572} & \object{RXJ1219.7-7403}, \object{GSC~9239-1495} & \object{032-049103} & -42.4\,$\pm$\,4.2  & -6.\,$\pm$\,1.9 & 13.1\,$\pm$\,2.0 & 12.48  & 11.386 & 9.746\,$\pm$\,0.03 & 9.048\,$\pm$\,0.03 & 8.858\,$\pm$\,0.02 & (2) \\
\object{12202177-7407393} & \object{GSC~9239-1572}, \object{RXJ1220.4-7407} & \object{032-049216} & -40.6\,$\pm$\,2.6  & -5.\,$\pm$\,1.8 & 18.0\,$\pm$\,2.0 & 11.836 & 10.565 & 9.26 \,$\pm$\,0.02 & 8.607\,$\pm$\,0.03 & 8.367\,$\pm$\,0.02 & (2) \\
\object{12392124-7502391} & \object{RXJ1239.4-7502}, \object{CD-74~712}     & \object{030-043659} & -40.7\,$\pm$\,2.8  & -13\,$\pm$\,2.4 & 13.1\,$\pm$\,2.0 & 9.537  & 8.346  & 8.434\,$\pm$\,0.02 & 7.953\,$\pm$\,0.03 & 7.777\,$\pm$\,0.02 & (2) \\
\object{12582559-7028490} & \object{CD-69~1055}                             & \object{040-110662} & -40.9\,$\pm$\,2.6  & -17\,$\pm$\,1.1 & 12.8  	    & 9.558  & 7.938  & 8.184\,$\pm$\,0.03 & 7.702\,$\pm$\,0.06 & 7.545\,$\pm$\,0.03 &  \\
\object{13021351-7637577} & \object{CM~Cha}, \object{IRAS~12584-7621}       & \object{027-035419} & -43.3\,$\pm$\,10.6 & 7.5\,$\pm$\,9.1 &		    & 12.04  & 10.936 & 10.02\,$\pm$\,0.02 & 9.165\,$\pm$\,0.03 & 8.515\,$\pm$\,0.02 & (3) \\
\object{13220753-6938121} & \object{MP~Mus}                                 & \object{041-134775} & -40.4\,$\pm$\,1.0  & -20\,$\pm$\,1.6 & 11.6\,$\pm$\,0.2 &	     &	      & 8.277\,$\pm$\,0.03 & 7.641\,$\pm$\,0.02 & 7.293\,$\pm$\,0.02 &  \\

  \hline
   \multicolumn{12}{c}{$\eta$~Chamaeleontis cluster}\\

\object{08365623-7856454} & \object{EG Cha}, \object{RECX 1}  & \object{023-014604} & $-$27.1\,$\pm$\,1.0  & 25.3\,$\pm$\,1.0  & 18.0\,$\pm$\,2.0 &  9.359 & 8.525  &  8.155\,$\pm$\,0.02 & 7.498\,$\pm$\,0.05 & 7.338\,$\pm$\,0.02 & (2) \\
\object{08413703-7903304} & \object{EH Cha}, \object{RECX 3}  & \object{022-014105} & $-$29.5\,$\pm$\,2.2  & 33.6\,$\pm$\,2.2  &		  & 15.595 & 13.998 & 10.349\,$\pm$\,0.02 & 9.647\,$\pm$\,0.02 & 9.415\,$\pm$\,0.02 &  \\
\object{08414471-7902531} & \object{HD 75505}                 & \object{022-014112} & $-$30.3\,$\pm$\,1.4  & 25.0\,$\pm$\,1.8  &		  &  7.148 & 6.776  &  7.059\,$\pm$\,0.03 & 6.987\,$\pm$\,0.04 & 6.928\,$\pm$\,0.02 &  \\
\object{08422372-7904030} & \object{EI Cha}, \object{RECX 4}  & \object{022-014146} & $-$27.9\,$\pm$\,10.7 & 17.9\,$\pm$\,2.4  &  	          & 14.993 & 13.508 &  9.535\,$\pm$\,0.02 & 8.779\,$\pm$\,0.06 & 8.615\,$\pm$\,0.02 &  \\
\object{08422710-7857479} & \object{EK Cha}, \object{RECX 5}  & \object{023-014912} & $-$30.8\,$\pm$\,2.7  & 35.1\,$\pm$\,4.7  &  	          & 14.229 & 12.114 & 10.777\,$\pm$\,0.02 &10.099\,$\pm$\,0.02 & 9.855\,$\pm$\,0.02 &  \\      
\object{08423879-7854427} & \object{EL Cha}, \object{RECX 6}  & \object{023-014921} & $-$35.7\,$\pm$\,2.3  & 25.0\,$\pm$\,2.4  &  	          & 13.023 & 11.39  & 10.232\,$\pm$\,0.03 & 9.584\,$\pm$\,0.02 & 9.29 \,$\pm$\,0.02 &  \\      
\object{08430723-7904524} & \object{EM Cha}, \object{RECX 7}  & \object{022-014185} & $-$29.7\,$\pm$\,1.9  & 25.3\,$\pm$\,3.1  &  4.3\,$\pm$\,2.0 &	   &	    &  8.42\,$\pm$\,0.02  & 7.758\,$\pm$\,0.03 & 7.635\,$\pm$\,0.03 & (2, 5) \\   
\object{08431857-7905181} & \object{ET Cha}, \object{RECX 15} & \object{022-014197} & $-$14.2\,$\pm$\,20.9 & 36.0\,$\pm$\,28.9 &  	          & 15.144 & 13.742 & 10.505\,$\pm$\,0.03 & 9.834\,$\pm$\,0.02 & 9.431\,$\pm$\,0.02 &  \\
\object{08441637-7859080} & \object{EN Cha}, \object{RECX 9}  & \object{023-015007} & $-$29.5\,$\pm$\,2.4  & 27.5\,$\pm$\,2.7  &  	          & 13.795 & 11.695 & 10.26\,$\pm$\,0.03  & 9.668\,$\pm$\,0.03 & 9.335\,$\pm$\,0.02 &  \\      
\object{08443188-7846311} & \object{EO Cha}, \object{RECX 10} & \object{023-015019} & $-$31.6\,$\pm$\,2.0  & 25.2\,$\pm$\,1.9  & 15.0\,$\pm$\,2.0 & 11.659 & 10.287 & 9.653\,$\pm$\,0.02  & 8.919\,$\pm$\,0.06 & 8.732\,$\pm$\,0.02 & (2) \\   
\object{08470165-7859345} & \object{EP Cha}, \object{RECX 11} & \object{023-015143} & $-$30.7\,$\pm$\,1.2  & 26.0\,$\pm$\,1.2  &  	          & 10.311 & 9.438  &  8.729\,$\pm$\,0.02 & 8.025\,$\pm$\,0.06 & 7.655\,$\pm$\,0.04 &  \\
\object{08475676-7854532} & \object{EQ Cha}, \object{RECX 12} & \object{023-015181} & $-$32.1\,$\pm$\,1.8  & 26.8\,$\pm$\,1.8  & 18.0\,$\pm$\,2.0 & 12.076 & 10.534 &  9.323\,$\pm$\,0.02 & 8.683\,$\pm$\,0.08 & 8.41 \,$\pm$\,0.03 & (2) \\

  \hline
  \end{tabular}
  \end{center}
  \begin{flushleft}
  {\bf Notes.} \\
  (1) Possible interloper (see text for discussion). \\
  (2) ROSAT source studied by \citet{covino1997}.\\
  (3) Classified in the literature as a Chameleon I or II member.\\ 
  (4) Dubious proper motion. \\ 
  (5) Discrepant radial velocity respect the mean group value.  
 \end{flushleft}
\end{table}
\end{landscape}

%% file: Background_Members.table
\begin{table*}
  \caption[]{Background objects.}
  \label{tab:cont}
  \begin{center}
  \begin{tabular}{ll c cc}
    \hline
    \hline

  \multicolumn{1}{c}{2MASS~J} &
   \multicolumn{1}{l}{Other names} &
   \multicolumn{1}{c}{UCAC3} &
 \multicolumn{1}{c}{$\mu_{\alpha}\cos\delta$} &
  \multicolumn{1}{c}{$\mu_{\delta}$} \\
  \multicolumn{1}{l}{} &
   \multicolumn{1}{l}{} &
   \multicolumn{1}{c}{} &
 \multicolumn{1}{c}{(mas/yr)} &
  \multicolumn{1}{c}{(mas/yr)} \\
      \hline


10452780-7715335 &                         & 026-023750 & 1.2\,$\pm$\,2.2   & 7.2\,$\pm$\,4.6	\\
11001473-7757103 &                         & 025-023372 & -10.7\,$\pm$\,4.5 & 15.1\,$\pm$\,2.7  \\
11014949-7412268 &                         & 032-038265 & -4.8\,$\pm$\,3.9  & 1.7\,$\pm$\,3.2	\\
11035902-7743349 & ISO 35                  & 025-023396 & 1.5\,$\pm$\,3.8   & -4.4\,$\pm$\,2.9  \\
11045890-7656550 &                         & 027-026781 & -6.5\,$\pm$\,4.0  & 16.1\,$\pm$\,16.8 \\
11050937-7706578 &                         & 026-024388 & -8.6\,$\pm$\,1.6  & 6.0\,$\pm$\,1.7	\\
11051798-7706565 &                         & 026-024390 & 5.3\,$\pm$\,3.0   & 8.3\,$\pm$\,2.1	\\
11052489-7620373 &                         & 028-028700 & -1.4\,$\pm$\,3.4  & 0.3\,$\pm$\,1.7	\\
11052850-7639489 &                         & 027-026798 & -11.3\,$\pm$\,2.3 & 2.3\,$\pm$\,3.1	\\
11053587-7638034 &                         & 027-026805 & -8.9\,$\pm$\,3.8  & -2.0\,$\pm$\,1.8  \\
11054972-7640462 &                         & 027-026814 & -0.1\,$\pm$\,1.8  & 4.9\,$\pm$\,1.8	\\
11055683-7656438 &                         & 027-026817 & -15.1\,$\pm$\,1.9 & 11.0\,$\pm$\,1.9  \\
11063479-7645404 &                         & 027-026840 & -43.9\,$\pm$\,2.1 & 15.5\,$\pm$\,3.5  \\
11063841-7612032 &                         & 028-028775 & -1.9\,$\pm$\,2.6  & 4.1\,$\pm$\,2.6	\\
11064235-7632450 &                         & 027-026847 & -4.0\,$\pm$\,3.1  & 2.2\,$\pm$\,3.2	\\
11064274-7636124 &                         & 027-026848 & -5.9\,$\pm$\,1.8  & 9.8\,$\pm$\,2.4	\\
11064586-7625317 &                         & 028-028787 & -0.8\,$\pm$\,1.8  & 2.3\,$\pm$\,3.4	\\
11065986-7651181 &                         & 027-026858 & -4.4\,$\pm$\,1.6  & 1.5\,$\pm$\,1.6	\\
11070350-7631443 &                         & 027-026861 & -1.0\,$\pm$\,2.9  & 7.4\,$\pm$\,2.9	\\
11070380-7635440 &                         & 027-026862 & -72.5\,$\pm$\,1.4 & 30.2\,$\pm$\,1.3  \\
11073280-7748594 & ISO 104                 & 025-023413 & 1.5\,$\pm$\,2.4   & 1.9\,$\pm$\,2.3	\\
11074763-7711156 &                         & 026-024434 & -7.1\,$\pm$\,1.6  & 2.1\,$\pm$\,1.6	\\
11075699-7741558 & ISO 118                 & 025-023418 & -21.2\,$\pm$\,6.7 & -1.3\,$\pm$\,5.3  \\
11081069-7637424 &                         & 027-026887 & -0.3\,$\pm$\,2.2  & -3.1\,$\pm$\,2.2  \\
11081916-7656525 & ISO 139                 & 027-026889 & -13.1\,$\pm$\,2.5 & 3.0\,$\pm$\,3.4	\\
11083897-7614457 &                         & 028-028890 & -3.9\,$\pm$\,1.9  & 2.4\,$\pm$\,1.9	\\
11085527-7704502 & ISO 166                 & 026-024451 & -13.4\,$\pm$\,1.6 & 6.3\,$\pm$\,1.6	\\
11085813-7646392 & ISO 170                 & 027-026908 & -34.0\,$\pm$\,2.5 & 12.3\,$\pm$\,2.5  \\
11090332-7700495 & ISO 172                 & 026-024453 & 12.7\,$\pm$\,7.4  & 9.6\,$\pm$\,7.4	\\
11090367-7707456 & ISO 175                 & 026-024454 & 5.3\,$\pm$\,1.9   & 0.6\,$\pm$\,1.9	\\
11090766-7618144 &                         & 028-028907 & -4.3\,$\pm$\,1.4  & 6.2\,$\pm$\,1.4	\\
11091379-7637531 & ISO 183                 & 027-026913 & 4.3\,$\pm$\,2.0   & 10.9\,$\pm$\,2.1  \\
11092068-7713591 & ISO 188                 & 026-024457 & -5.3\,$\pm$\,2.7  & -5.6\,$\pm$\,2.8  \\
11092482-7642073 & ISO 191                 & 027-026915 & -5.5\,$\pm$\,3.5  & 1.2\,$\pm$\,3.5	\\
11095119-7658568 & ISO 213                 & 027-026917 & -82.0\,$\pm$\,1.0 & 17.6\,$\pm$\,1.1  \\
11095465-7625333 &                         & 028-028954 & 1.8\,$\pm$\,3.0   & 1.1\,$\pm$\,3.1	\\
11095822-7659147 & [LM04]ChaI~616, ISO 229 & 027-026919 & -3.2\,$\pm$\,2.6  & 4.8\,$\pm$\,2.6   \\
11102819-7626164 & ISO 242                 & 028-028979 & -22.1\,$\pm$\,1.3 & 5.0\,$\pm$\,1.3	\\
11102990-7736064 & ISO 243                 & 025-023448 & -3.0\,$\pm$\,4.1  & 7.9\,$\pm$\,4.4	\\
11103067-7637186 & ISO 245                 & 027-026923 & -6.3\,$\pm$\,3.3  & 0.5\,$\pm$\,2.8	\\
11105038-7631440 & ISO 255                 & 027-026928 & -6.9\,$\pm$\,5.9  & 8.4\,$\pm$\,2.0	\\
11105215-7709286 & ISO 257                 & 026-024459 & -1.4\,$\pm$\,1.8  & -7.6\,$\pm$\,1.8  \\
11110383-7706261 & ISO 263                 & 026-024460 & 6.2\,$\pm$\,3.2   & 1.7\,$\pm$\,4.4	\\
11110670-7717594 & ISO 264                 & 026-024461 & 6.3\,$\pm$\,2.2   & 10.7\,$\pm$\,2.2  \\
11110714-7722056 & ISO 265                 & 026-024462 & -9.3\,$\pm$\,4.8  & -0.3\,$\pm$\,2.9  \\
11111062-7657205 & ISO 267                 & 027-026938 & -4.9\,$\pm$\,2.1  & 8.5\,$\pm$\,2.1	\\
11111680-7628573 &                         & 028-029023 & 7.3\,$\pm$\,15.6  & 4.8\,$\pm$\,21.1  \\
11131188-7647392 &                         & 027-027030 & -6.7\,$\pm$\,2.5  & -3.3\,$\pm$\,2.4  \\
11173079-7627078 &                         & 028-029481 & -2.6\,$\pm$\,1.2  & 5.8\,$\pm$\,1.0	\\
11195060-7628492 &                         & 028-029694 & -7.5\,$\pm$\,6.9  & 0.1\,$\pm$\,4.3	\\
11213026-7805240 &                         & 024-023578 & -2.8\,$\pm$\,2.5  & 2.6\,$\pm$\,4.5	\\
11225703-7649560 &                         & 027-027766 & -8.7\,$\pm$\,1.1  & 6.4\,$\pm$\,2.6	\\

  \hline
  \end{tabular}
  \end{center}
\end{table*}

%% file: Members-photometry_vosa.table
%

\onltab{4}{
\longtabL{4}{
\begin{longtable}{l l l r r r}

\caption{Compiled photometry for Chamaeleon~I and II members.} \label{tab:phot} \\
\hline 
\hline  
  \multicolumn{1}{l}{Cloud} &
  \multicolumn{1}{l}{2MASS J} &
  \multicolumn{1}{l}{Filter} &
  \multicolumn{1}{c}{$\lambda_{eff}$} &
  \multicolumn{1}{c}{Flux} &
  \multicolumn{1}{c}{Flux error}  \\
  \multicolumn{1}{l}{} &
  \multicolumn{1}{c}{} &
  \multicolumn{1}{c}{} &
  \multicolumn{1}{c}{(\AA)} &
  \multicolumn{1}{c}{(erg/cm$^2$/s/\AA)} &
  \multicolumn{1}{c}{(erg/cm$^2$/s/\AA)}  \\
\hline
\endfirsthead

\caption{(continued)}\\
\hline 
\hline  
  \multicolumn{1}{l}{Cloud} &
  \multicolumn{1}{l}{2MASS J} &
  \multicolumn{1}{l}{Filter} &
  \multicolumn{1}{c}{$\lambda_{eff}$} &
  \multicolumn{1}{c}{Flux} &
  \multicolumn{1}{c}{Flux error}  \\
  \multicolumn{1}{l}{} &
  \multicolumn{1}{c}{} &
  \multicolumn{1}{c}{} &
  \multicolumn{1}{c}{(\AA)} &
  \multicolumn{1}{c}{(erg/cm$^2$/s/\AA)} &
  \multicolumn{1}{c}{(erg/cm$^2$/s/\AA)}  \\
\hline
\endhead
\hline
\endfoot


  ChaI & 10463795-7736035 & GALEX/GALEX.FUV    &    1542.3  &   2.409492947054E-16   &  1.3892712556971E-17 \\  
  ChaI & 10463795-7736035 & GALEX/GALEX.NUV    &    2274.4  &   7.1026990635684E-15  &  3.8503326252756E-17   \\
  ChaI & 10463795-7736035 & TYCHO/TYCHO.B      &    4280.0  &   5.7787397363199E-13  &  1.4370523170994E-14   \\
  ChaI & 10463795-7736035 & TYCHO/TYCHO.V      &    5340.0  &   6.7712367865798E-13  &  1.2473079108729E-14   \\
  ChaI & 10463795-7736035 & DENIS/DENIS.I      &    7862.1  &   2.8492382854225E-13  &  2.0993963527685E-14   \\
  ChaI & 10463795-7736035 & 2MASS/2MASS.J      &   12350.0  &   2.2793213744817E-13  &  3.7787986216975E-15   \\
  ChaI & 10463795-7736035 & 2MASS/2MASS.H      &   16620.0  &   9.8226780523406E-14  &  4.9758614524518E-15   \\
  ChaI & 10463795-7736035 & 2MASS/2MASS.Ks     &   21590.0  &   4.3746979677725E-14  &  9.2672651807888E-16   \\
  ChaI & 10463795-7736035 & WISE/WISE.W1       &   33156.6  &   8.674311371174E-15   &  2.1571207259674E-16   \\
  ChaI & 10463795-7736035 & WISE/WISE.W2       &   45645.0  &   2.5436716474833E-15  &  4.6856163335733E-17   \\
  ChaI & 10463795-7736035 & AKARI/IRC.S9W      &   82283.6  &   3.304074731026E-16   &  3.9850807530467E-17   \\
  ChaI & 10463795-7736035 & WISE/WISE.W3       &  107868.4  &   6.8415691439533E-17  &  1.0082108879075E-18   \\
  ChaI & 10463795-7736035 & WISE/WISE.W4       &  219149.6  &   5.4364535147677E-18  &  4.0557985702824E-19   \\
  ChaI & 10555973-7724399 & GALEX/GALEX.FUV    &    1542.3  &   3.4666792373282E-17  &  6.336030288325E-18	 \\
  ChaI & 10555973-7724399 & GALEX/GALEX.NUV    &    2274.4  &   2.5294576603161E-17  &  3.2778926195947E-18   \\
  \multicolumn{6}{c}{...}\\

  \hline
\end{longtable}

{\bf References:} \\

    GALEX: \citet{bianchi2000}
    
    Johnson: \citet{mermilliod1994}
    
    Tycho-2: \citet{hog2000} 
    
    Stromgren: \citet{hauck1998} 
    
    WFI: \citet{wfichai,spezzi2007} 
    
    DENIS: \citet{denis2005} 
    
    2MASS: \citet{skrutskie2006} 
    
    WISE: \citet{wright2010} 
    
    Spitzer: \citet{alcala2008,luhman2008c} 
    
    AKARI: \citet{ishihara2010}
    
    IRAS: \citet{helou1988}

}

}

%% file: Rosat.table
\setcounter{table}{6}
\begin{table*}[t]
\label{tab:rosat}
\caption{Kinematic properties of ROSAT stars not included in our initial member lists}
\centering
\begin{tabular}{lclc rrc c l}
\hline\hline
  \multicolumn{1}{c}{RXJ} &
  \multicolumn{1}{c}{2MASS~J} &
  \multicolumn{1}{l}{Other names} &
  \multicolumn{1}{c}{UCAC3} &
  \multicolumn{1}{c}{$\mu_{\alpha}\cos\delta$} &
  \multicolumn{1}{c}{$\mu_{\delta}$} &
  \multicolumn{1}{c}{$V_r$} &
  \multicolumn{1}{c}{Evolutionary} &
  \multicolumn{1}{c}{Group$^{\mathrm{b}}$} \\
  \multicolumn{1}{c}{} &
  \multicolumn{1}{c}{} &
  \multicolumn{1}{l}{} &
  \multicolumn{1}{c}{} &
  \multicolumn{1}{c}{(mas/yr)} &
  \multicolumn{1}{c}{(mas/yr)} &
  \multicolumn{1}{c}{(km/s)} &
  \multicolumn{1}{c}{status$^{\mathrm{a}}$} &
  \multicolumn{1}{c}{} \\
\hline

  \object{1129.2-7546}  & \object{11291261-7546263} &   			 & 029-033458 & $-$24.5$\pm$1.5 &  $-$2.1$\pm$1.5 & 11.4$\pm$2.0 & PMS & Cha~I \\
  \object{1158.5-7913}  & \object{11583429-7913175} &                            & 022-024487 & $-$18.3$\pm$3.8 &  $-$3.1$\pm$1.9 & 13.1$\pm$2.0 & PMS & Cha~I? \\

\hline
  \object{1123.2-7924}  & \object{11225562-7924438} & \object{$\epsilon$~Cha~14} & 022-022912 & $-$29.8$\pm$2.1 & $-$19.0$\pm$1.7 & 10.0$\pm$2.0 & PMS & $\epsilon$~Cha? \\
  \object{1150.4-7704}  & \object{11502829-7704380} & \object{$\epsilon$~Cha~19} & 026-027460 & $-$40.3$\pm$1.2 &  $-$8.3$\pm$1.5 & 	      & PMS & $\epsilon$~Cha \\
  \object{1216.8-7753}  & \object{12164593-7753333} &   			 & 025-028496 & $-$37.5$\pm$1.6 &  $-$8.0$\pm$1.6 & 14.0$\pm$2.0 & PMS & $\epsilon$~Cha \\

\hline
  \object{0850.1-7554}  & \object{08500540-7554380} &   			 & 029-020107 & $-$15.9$\pm$1.1 & 33.0$\pm$0.9 & 15.5$\pm$2.0 & PMS & $\eta$~Cha \\
  \object{0902.9-7759}  &   		            &   			 & 025-017479 & $-$34.6$\pm$1.7 & 22.4$\pm$1.7 & 11.0$\pm$2.0 & PMS & $\eta$~Cha \\
  \object{0915.5-7608}  & \object{09152912-7608471} &   			 & 028-021037 & $-$29.3$\pm$2.0 & 18.9$\pm$1.4 & 21.0$\pm$2.0 & PMS & $\eta$~Cha \\
  \object{0928.5-7815}  & \object{09282116-7815352} & \object{HD~82879}          & 024-017885 & $-$25.4$\pm$1.0 & 18.0$\pm$1.5 & 16.8$\pm$2.0 & ?   & $\eta$~Cha \\
  \object{0935.0-7804}  & \object{09345604-7804193} &                   	 & 024-018258 & $-$28.6$\pm$1.3 & 19.2$\pm$1.3 & 14.0$\pm$2.0 & PMS & $\eta$~Cha \\
  \object{0951.9-7901}  & \object{09515069-7901377} & \object{HD~86356}          & 022-017379 & $-$26.4$\pm$0.8 & 38.5$\pm$0.8 & 12.2$\pm$2.0 & PMS & $\eta$~Cha \\
  \object{1001.1-7913}  & \object{10010873-7913074} &				 & 022-017956 & $-$25.0$\pm$1.5 & 38.7$\pm$2.3 & 12.2$\pm$2.0 & PMS & $\eta$~Cha \\
  \object{1140.3-8321}  & \object{11401658-8321003} &				 & 014-014885 & $-$41.7$\pm$1.1 & 27.7$\pm$1.1 & 10.5$\pm$2.0 & ?   & $\eta$~Cha \\

\hline
  \object{1017.9-7431}  & \object{10175369-7431172} &                    		 & 031-030277 &   $-$6.5$\pm$2.1 &     6.6$\pm$1.3 &   104.0$\pm$2.0 & ZAMS & Field \\
  \object{1035.8-7859}  & \object{10354856-7858583} &                    		 & 023-020966 &  $-$65.9$\pm$5.5 &     6.2$\pm$3.0 &	70.8$\pm$2.0 & ZAMS & Field \\
  \object{1044.6-7849}  & \object{10443523-7849244} &                    		 & 023-021597 &  $-$87.6$\pm$2.3 &    25.9$\pm$4.3 &  $-$8.4$\pm$2.0 & ZAMS & Field \\
  \object{0842.4-8345}  & \object{08422284-8345248} &                    		 & 013-008994 &  $-$49.5$\pm$1.3 &    91.2$\pm$1.2 &  $-$8.0$\pm$2.0 & ?    & Field \\
  \object{0849.2-7735}  & \object{08491110-7735585} &                    		 & 025-016705 &  $-$11.5$\pm$1.2 &    16.3$\pm$1.1 &  $-$3.4$\pm$2.0 & ?    & Field \\
  \object{0853.1-8244}  & \object{08530528-8243597} &                    		 & 015-010451 &   0.5$\pm$0.9    & $-$14.4$\pm$1.1 &     7.7$\pm$2.0 & ?    & Field \\
  \object{0936.3-7820}  & \object{09361783-7820417} & \object{HD~84075}  		 & 024-018347 &  $-$73.4$\pm$0.8 &    50.1$\pm$0.8 &	 3.2$\pm$2.0 & ?    & Field \\
  \object{0946.9-8011}  & \object{09465051-8011352} &                    		 & 020-015930 &  $-$16.8$\pm$1.9 &  $-$3.9$\pm$3.9 &	 5.4$\pm$2.0 & ?    & Field \\
  \object{0952.7-7933}a & \object{09531375-7933285} & \object{HD~86588}  		 & 021-016442 &  $-$13.5$\pm$0.9 &     2.0$\pm$0.9 &	 2.0$\pm$2.0 & ?    & Field \\
  \object{1007.7-8504}  & \object{10072939-8504329} & \object{DR~Oct}, \object{HD~89499} & 010-008728 & $-$566.0$\pm$    &   395.0$\pm$    & $-$67.1$\pm$2.0 & ?    & Field \\
  \object{1009.6-8105}  & \object{10093502-8105509} &                   		 & 018-015349 &  $-$15.4$\pm$4.5 &    12.0$\pm$3.8 &	 8.8$\pm$2.0 & ?    & Field \\
  \object{1014.2-7636}  & \object{10140807-7636327} &                   		 & 027-023785 &  $-$47.2$\pm$1.7 &    30.6$\pm$3.6 &	 8.0$\pm$2.0 & ?    & Field \\
  \object{1014.4-8138}  & \object{10142048-8138423} &                   		 & 017-014665 &  $-$36.5$\pm$1.3 &    17.4$\pm$2.0 &	 6.6$\pm$2.0 & ?    & Field \\
  \object{1048.9-7655}  & \object{10485467-7655450} &                   		 & 027-025899 &  $-$95.6$\pm$1.4 &    45.7$\pm$2.1 & $-$19.9$\pm$2.0 & ?    & Field \\
  \object{1207.9-7555}  & \object{12075118-7555159} &                   		 & 029-037031 & $-$156.7$\pm$2.6 &  $-$1.5$\pm$2.0 &  $-$3.4$\pm$2.0 & ?    & Field \\
  \object{1209.8-7344}  & \object{12094282-7344414} &                   		 & 033-053854 &   $-$9.3$\pm$2.3 &  $-$3.2$\pm$1.3 &	 1.0$\pm$2.0 & ?    & Field \\
  \object{1217.4-8035}  & \object{12172694-8035069} & \object{HD~106772}                 & 019-022395 &   $-$0.4$\pm$1.5 & $-$10.4$\pm$1.4 & $-$13.0$\pm$2.0 & ?    & Field \\
  \object{1220.6-7539}  & \object{12203437-7539286} &                                    & 029-038424 & $-$111.2$\pm$1.5 &     4.0$\pm$1.0 &     6.0$\pm$2.0 & ?    & Field \\
  \object{1223.5-7740}  & \object{12232906-7740514} & \object{HD~107722}                 & 025-029020 &  $-$66.8$\pm$1.2 &    11.9$\pm$1.4 &    10.4$\pm$2.0 & ?    & Field \\
  \object{1225.3-7857}  & \object{12251340-7857347} &                                    & 023-028311 &  $-$27.0$\pm$1.0 & $-$16.4$\pm$1.6 &  $-$6.7$\pm$2.0 & ?    & Field \\
  \object{1233.5-7523}  & \object{12332981-7523112} & \object{HD~109138}                 & 030-042831 &  $-$94.4$\pm$1.1 &    16.3$\pm$1.0 &    14.0$\pm$2.0 & ?    & Field \\
  \object{1307.3-7602}  & \object{13072289-7602360} &                                    & 028-039286 &  $-$19.1$\pm$1.5 &     1.7$\pm$1.2 & $-$62.8$\pm$2.0 & ?    & Field \\
  \object{1349.2-7549}E & \object{13491293-7549475} &                                    & 029-049787 &  $-$61.7$\pm$1.0 & $-$32.3$\pm$1.7 &     1.0$\pm$2.0 & ?    & Field \\
\hline\end{tabular}		        										      
\begin{flushleft}
{\bf Notes.} \\
$^{\mathrm{a}}$ \citet{covino1997} \\
$^{\mathrm{b}}$ This work (a question mark if dubious, see text) \\
\end{flushleft}		    
\end{table*}

%% file: Previous_Candidates.table

\begin{landscape}
\begin{table}
\caption[]{UCAC3 proper motions for candidate members of the Chamaeleon dark clouds from the literature}
  \label{tab:cand}
  \begin{center}
  \begin{tabular}{lll cc ccccc ll}
    \hline \hline 
    \multicolumn{1}{c}{2MASS~J} &
    \multicolumn{1}{c}{Other names} &
    \multicolumn{1}{c}{UCAC3} &
    \multicolumn{1}{c}{$\mu_{\alpha}\cos\delta$} &
    \multicolumn{1}{c}{$\mu_{\delta}$} &
    \multicolumn{1}{c}{R$_{phot}$} &
    \multicolumn{1}{c}{I$_{phot}$} &
    \multicolumn{1}{c}{J} &
    \multicolumn{1}{c}{H} &
    \multicolumn{1}{c}{K} &
    \multicolumn{1}{c}{Ref.$^{(a)}$}  &
    \multicolumn{1}{c}{Notes$^{(b)}$} \\
    \multicolumn{1}{l}{} &
    \multicolumn{1}{c}{} &
    \multicolumn{1}{c}{} &
    \multicolumn{1}{c}{(mas/yr)} &
    \multicolumn{1}{c}{(mas/yr)} &
    \multicolumn{1}{c}{mag} &
    \multicolumn{1}{c}{mag} &
    \multicolumn{1}{c}{mag} &
    \multicolumn{1}{c}{mag} &
    \multicolumn{1}{c}{mag} &
    \multicolumn{1}{c}{} &
    \multicolumn{1}{c}{} \\

    \hline

  \multicolumn{12}{c}{Chamaeleon~I Cloud}\\

11094646-7628575  &  \object{C1-14} & \object{028-028944} & -27.4\,$\pm$\,2.0 & 10.2\,$\pm$\,1.0 & 13.977 & 13.111 & 8.432 $\pm$0.04 & 8.035 $\pm$0.06 & 7.854 $\pm$0.02 & 1 & C \\

  \hline
  \multicolumn{12}{c}{Chamaeleon~II Cloud}\\

12482571-7706366 & \object{IRAS 12448-7650}                         & \object{026-031454} &  -5.4\,$\pm$\,1.8  &   4.2\,$\pm$\,1.7  & 13.507 & 9.872  & 6.789 $\pm$0.02 & 5.674 $\pm$0.02 & 5.177 $\pm$0.02 & 2 & R \\
12540028-7624251 & \object{[VCE2001]\,C18}                          & \object{028-038152} &  -1.4\,$\pm$\,4.4  &   0.6\,$\pm$\,5.9  & 14.894 & 11.308 & 8.194 $\pm$0.03 & 7.139 $\pm$0.05 & 6.606 $\pm$0.02 & 3 & R \\
12551447-7622005 & \object{[VCE2001]\,C29}                          & \object{028-038240} &  -6.6\,$\pm$\,10.2 &   7.1\,$\pm$\,4.6  & 15.36  & 12.238 & 10.053$\pm$0.02 & 9.063 $\pm$0.02 & 8.673 $\pm$0.02 & 3 & R \\
13024772-7702399 & \object{IRAS 12589-7646}, \object{Iso-Cha II 89} & \object{026-031795} &   --21.2\,$\pm$\,3.9  & 7.9\,$\pm$\,9.3 & 13.179 & 9.554  & 7.196 $\pm$0.02 & 6.199 $\pm$0.03 & 5.716 $\pm$0.02 & 2 & R \\
13024919-7705448 & \object{[VCE2001]\,X3}                           & \object{026-031797} & -24.6\,$\pm$\,5.3  &   5.9\,$\pm$\,6.0  & 15.889 & 15.483 & 14.315$\pm$0.03 & 13.839$\pm$0.04 & 13.597$\pm$0.04 & 3 & R \\
13030903-7700328 & \object{[VCE2001]\,X4}                           & \object{026-031820} & -11.3\,$\pm$\,5.8  & -13.7\,$\pm$\,4.0  & 13.597 & 12.954 & 12.465$\pm$0.02 & 12.1  $\pm$0.03 & 11.991$\pm$0.03 & 3 & R \\
13074107-7805434 & \object{[VCE2001]\,C64}                          & \object{024-032467} & -12.9\,$\pm$\,5.4  &   0.8\,$\pm$\,14.7 & 15.605 & 13.072 & 10.458$\pm$0.02 & 9.118 $\pm$0.02 & 8.641 $\pm$0.02 & 3 & R \\

  \hline
  \end{tabular}
  \end{center}
  {\bf Notes.} 
   \begin{list}{}{}
	\item[$^{a}$] References: 1) \citet{luhman2004,spezzi2008,vuong2001}
      	\item[$^{b}$] Notes: C$=$confirmed as probable kinematical member; R$=$rejected candidate (proper motion not consistent with membership).
   \end{list}
\end{table}
\end{landscape}

%% file: Space-velocity.table

\begin{table*}[t]
\caption{Space velocities for members and candidate members of the Chamaeleon associations.} 
\label{tab:vel}
\centering

\begin{tabular}{l c c c r r r l}
    \hline 
    \hline     
  \multicolumn{1}{l}{Name} &
  \multicolumn{1}{c}{$U$} &
  \multicolumn{1}{c}{$V$} &
  \multicolumn{1}{c}{$W$} &
  \multicolumn{1}{c}{$U_{LSR}$} &
  \multicolumn{1}{c}{$V_{LSR}$} &
  \multicolumn{1}{c}{$W_{LSR}$} &
  \multicolumn{1}{c}{Notes} \\
  \multicolumn{1}{c}{} &
  \multicolumn{1}{c}{(km/s)} &
  \multicolumn{1}{c}{(km/s)} &
  \multicolumn{1}{c}{(km/s)} &
  \multicolumn{1}{c}{(km/s)} &
  \multicolumn{1}{c}{(km/s)} &
  \multicolumn{1}{c}{(km/s)} &
  \multicolumn{1}{c}{} \\
	\hline
  \multicolumn{8}{c}{Chamaeleon~I cloud}\\
  T3A			 & $-$1.8$\pm$7.3  & $-$13.2$\pm$3.4 &  $-$4.6$\pm$5.8 & $-$10.3 &     0.2 &	 1.9 &  \\
  SY~Cha 		 &    8.5$\pm$3.3  & $-$17.5$\pm$1.4 &  $-$4.8$\pm$2.5 &  $-$0.0 &  $-$4.1 &	 1.7 &  \\
  Sz~4  		 & $-$1.1$\pm$5.3  & $-$16.3$\pm$3.2 &  $-$7.4$\pm$5.6 &  $-$9.6 &  $-$2.9 &  $-$0.9 &  \\
  TW~Cha  		 &    5.9$\pm$1.7  & $-$19.0$\pm$1.3 &  $-$7.0$\pm$1.6 &  $-$2.6 &  $-$5.7 &  $-$0.5 &  \\
  CR~Cha 	         &    7.6$\pm$1.5  & $-$18.8$\pm$1.0 &  $-$7.3$\pm$1.6 &  $-$0.9 &  $-$5.5 &  $-$0.8 &  \\
  CS~Cha 	         &    7.9$\pm$1.5  & $-$19.1$\pm$1.0 &  $-$6.6$\pm$2.1 &  $-$0.6 &  $-$5.7 &  $-$0.1 &  \\
  CT~Cha 	         &    6.1$\pm$1.6  & $-$18.0$\pm$0.8 &  $-$8.6$\pm$1.3 &  $-$2.4 &  $-$4.6 &  $-$2.2 &  \\
  T21			 &   14.3$\pm$3.3  & $-$20.9$\pm$1.4 &  $-$8.8$\pm$2.0 &     5.8 &  $-$7.5 &  $-$2.3 &  \\
  DI~Cha 	         &    3.4$\pm$1.5  & $-$13.8$\pm$1.0 & $-$11.8$\pm$1.3 &  $-$5.1 &  $-$0.4 &  $-$5.3 &  \\
  VV~Cha 	         &    6.0$\pm$1.9  & $-$17.6$\pm$1.4 & $-$11.2$\pm$1.5 &  $-$2.5 &  $-$4.2 &  $-$4.7 &  \\
  VW~Cha 	         &    6.1$\pm$1.6  & $-$17.5$\pm$1.2 & $-$10.5$\pm$2.4 &  $-$2.4 &  $-$4.1 &  $-$4.9 &  \\
  CHXR~37                &    8.7$\pm$2.8  & $-$16.6$\pm$3.4 & $-$10.4$\pm$6.7 &     0.2 &  $-$3.2 &  $-$3.9 &  \\
  VZ~Cha 		 &    6.1$\pm$1.7  & $-$17.5$\pm$1.2 & $-$10.5$\pm$2.4 &  $-$2.4 &  $-$4.1 &  $-$4.0 &  \\
  T46			 &    8.8$\pm$1.9  & $-$16.5$\pm$1.2 & $-$10.0$\pm$1.5 &     0.3 &  $-$3.1 &  $-$3.6 &  \\
  CHXR~47  	         &    7.4$\pm$1.9  & $-$17.5$\pm$1.9 &  $-$8.5$\pm$1.6 &  $-$1.1 &  $-$4.2 &  $-$2.0 &  \\
  CHX~18N		 &    8.5$\pm$3.3  & $-$17.5$\pm$1.4 &  $-$4.8$\pm$2.5 &  $-$0.0 &  $-$4.1 &	 1.7 &  \\
  CHX~20 		 &   14.3$\pm$3.3  & $-$20.9$\pm$1.4 &  $-$8.8$\pm$2.0 &     5.8 &  $-$7.5 &  $-$2.3 &  \\
  CV~Cha 		 &    6.1$\pm$5.0  & $-$18.3$\pm$1.8 &  $-$8.1$\pm$2.9 &  $-$2.4 &  $-$4.9 &  $-$1.6 & (1) \\
  RXJ1129.2-7546	 &   10.8$\pm$2.0  & $-$16.2$\pm$2.0 & $-$10.0$\pm$1.3 &     2.3 &  $-$2.8 &  $-$3.5 &  \\
  RXJ1158.5-7913	 &    5.7$\pm$2.8  & $-$16.1$\pm$2.3 &  $-$8.8$\pm$1.7 &  $-$2.8 &  $-$2.8 &  $-$2.3 & (2) \\
  Mean values ($\sigma$) &    7.0 (3.8)    & $-$17.4 (1.9)   &  $-$8.4 (2.1)   &  $-$1.5 (3.8) & $-$4.1 (1.9) & $-$2.0 (2.1) & \\
\hline
  \multicolumn{8}{c}{Chamaeleon~II cloud}\\
  Sz~46N                 &     2.0$\pm$5.5  &   $-$25.0$\pm$2.9  &    15.4$\pm$15.9 &  $-$6.5 & $-$11.6 &    21.9 & \\         
  IRAS~12535-7623        &    14.7$\pm$3.3  &   $-$18.4$\pm$2.4  & $-$12.0$\pm$2.4  &     6.2 &  $-$5.1 &  $-$5.6 & \\ 
  Sz~50                  &    12.4$\pm$3.5  &   $-$21.9$\pm$2.5  &  $-$7.7$\pm$3.1  &     3.9 &  $-$8.5 &  $-$1.2 & \\ 
  RXJ1303.1-7706         &    16.6$\pm$2.7  &   $-$24.0$\pm$1.8  &  $-$7.8$\pm$1.5  &     8.1 & $-$10.6 &  $-$1.3 & \\ 
  Hn~23                  & $-$14.5$\pm$16.8 &   $-$12.6$\pm$11.4 &    11.5$\pm$18.1 & $-$23.0 &     0.8 &    18.0 & \\   
  Sz~58                  &    14.0$\pm$6.9  &   $-$22.1$\pm$5.7  &  $-$8.2$\pm$3.8  &     5.5 &  $-$8.7 &  $-$1.7 & \\   
  Sz~59                  &    8.5$\pm$2.7   &   $-$17.5$\pm$2.6  & $-$12.7$\pm$3.0  &     0.0 &  $-$4.2 &    $-$6 & \\ 
  Sz~60W                 &   30.5$\pm$14.4  &   $-$34.8$\pm$10.1 &  $-$6.6$\pm$4.8  &    22.0 & $-$21.4 &  $-$0.1 & \\  
  Sz~61                  &   12.0$\pm$2.9   &   $-$20.7$\pm$2.9  &  $-$6.9$\pm$1.8  &	  3.5 &  $-$7.4 &  $-$0.4 & \\  
  Mean values ($\sigma$) &   10.7 (11.4)    &   $-$21.9 (5.8)    &  $-$3.9 (9.5)    &     2.2 (11.4)   & $-$8.5 (5.8)     &     2.6 (9.5) & \\
\hline
  \multicolumn{8}{c}{$\epsilon$~Chamaeleontis cluster}\\
  CM~Cha	         &    9.6$\pm$4.8  & $-$26.3$\pm$3.3 &     1.0$\pm$4.8 &     1.1 & $-$13.0 &	 7.5 & (3) \\
  T~Cha 	         &   18.1$\pm$13.6 & $-$28.9$\pm$7.5 & $-$16.9$\pm$3.6 &     9.6 & $-$15.5 & $-$10.4 & (1, 4) \\
  DW~Cha	         &   12.2$\pm$1.5  & $-$19.9$\pm$1.9 &  $-$9.1$\pm$0.7 &     3.7 &  $-$6.5 &  $-$2.6 &  \\
  RXJ1159.7-7601         &   12.3$\pm$1.6  & $-$19.7$\pm$1.9 & $-$10.3$\pm$1.0 &     3.8 &  $-$6.3 &  $-$3.9 &  \\
  HD~104467	         &   14.0$\pm$1.4  & $-$17.1$\pm$1.8 &  $-$9.5$\pm$0.6 &     5.5 &  $-$3.7 &  $-$3.0 &  \\ 
  RXJ1219.7-7403         &   12.7$\pm$2.3  & $-$21.1$\pm$2.1 &  $-$8.5$\pm$1.1 &     4.2 &  $-$7.7 &  $-$2.0 &  \\
  RXJ1239.4-7502         &   11.9$\pm$1.8  & $-$20.6$\pm$1.9 & $-$10.8$\pm$1.3 &     3.4 &  $-$7.2 &  $-$4.3 &  \\ 
  DZ Cha	         &   13.8$\pm$1.9  & $-$18.3$\pm$1.9 & $-$12.8$\pm$1.1 &     5.3 &  $-$4.9 &  $-$6.3 &  \\
  GSC 9420-0948          &   16.3$\pm$2.0  & $-$13.1$\pm$2.0 &  $-$7.5$\pm$1.1 &     7.8 &     0.3 &  $-$1.0 &  \\
  GSC 9416-1029          &   12.7$\pm$1.4  & $-$17.6$\pm$1.8 &  $-$7.5$\pm$1.2 &     4.2 &  $-$4.2 &  $-$1.0 &  \\
  GSC 9239-1572          &    9.4$\pm$1.8  & $-$24.9$\pm$1.9 &  $-$8.9$\pm$1.0 &     0.9 & $-$11.5 &  $-$2.4 &  \\
  CP-68~1388	         &   11.3$\pm$1.3  & $-$21.2$\pm$0.6 &  $-$6.8$\pm$1.2 &     2.8 &  $-$7.8 &  $-$0.3 &  \\
  HD 105923	         &   10.1$\pm$1.4  & $-$20.6$\pm$0.8 &  $-$9.8$\pm$0.9 &     1.6 &  $-$7.2 &  $-$3.3 &  \\
  GSC 9415-2676          &   11.2$\pm$1.8  & $-$18.5$\pm$1.9 & $-$11.3$\pm$1.2 &     2.7 &  $-$5.1 &  $-$4.8 &  \\
  CD-69 1055	         &   11.7$\pm$1.4  & $-$21.7$\pm$0.9 & $-$10.0$\pm$0.7 &     3.2 &  $-$8.3 &  $-$3.5 &  \\
  MP~Mus	         &   12.0$\pm$0.9  & $-$21.6$\pm$0.6 &  $-$9.3$\pm$0.9 &     3.5 &  $-$8.2 &  $-$2.8 &  \\
  RXJ1123.2-7924         &    7.3$\pm$1.5  &  $-$9.9$\pm$1.9 & $-$17.2$\pm$1.0 &    $-$1.2 &   3.5 & $-$10.7 & (5) \\
  RXJ1216.8-7753         &   10.2$\pm$1.5  & $-$19.8$\pm$1.8 & $-$10.3$\pm$0.8 &     1.7 &  $-$6.4 &  $-$3.8 &  \\
  Mean values ($\sigma$) &   12.1 (2.4)    & $-$20.0 (4.1)   &  $-$9.7 (3.7)   &  3.6 (2.4) & $-$6.6 (4.1) & $-$3.2 (3.7) & \\
\hline
  \multicolumn{8}{c}{$\eta$~Chamaeleontis cluster}\\
   EG~Cha	          &    9.5$\pm$0.9  & $-$20.0$\pm$1.4 & $-$11.1$\pm$1.6 &     1.0 &  $-$6.7 &  $-$4.6 &  \\
   EM~Cha	          &   15.0$\pm$1.2  &  $-$8.4$\pm$1.7 &  $-$6.6$\pm$1.8 &     6.5 &	4.9 &  $-$0.1 &  \\
   EO~Cha	          &   11.8$\pm$1.2  & $-$17.6$\pm$1.5 & $-$11.0$\pm$1.7 &     3.3 &  $-$4.2 &  $-$4.6 &  \\
   EQ~Cha	          &   11.4$\pm$1.1  & $-$20.7$\pm$1.5 & $-$11.7$\pm$1.7 &     2.9 &  $-$7.3 &  $-$5.2 &  \\
   RXJ0850.1-7554         &   10.3$\pm$0.8  & $-$20.2$\pm$1.4 &  $-$2.7$\pm$1.6 &     1.8 &  $-$6.9 &	3.8   &  \\
   RXJ0902.9-7759         &   13.4$\pm$1.1  & $-$14.2$\pm$1.5 &  $-$9.9$\pm$1.6 &     4.9 &  $-$0.8 &  $-$3.4 &  \\
   RXJ0915.5-7608         &    7.7$\pm$1.2  & $-$22.8$\pm$1.5 & $-$11.0$\pm$1.5 &  $-$0.8 &  $-$9.4 &  $-$4.5 &  \\
   HD~82879	          &    7.0$\pm$1.0  & $-$19.2$\pm$1.5 &  $-$8.3$\pm$1.4 &  $-$1.5 &  $-$5.8 &  $-$1.8 &  \\
   RXJ0935.0-7804         &    9.4$\pm$1.0  & $-$17.4$\pm$1.6 &  $-$7.5$\pm$1.4 &     0.9 &  $-$4.0 &  $-$1.1 &  \\
   HD~86356	          &   13.4$\pm$0.9  & $-$20.7$\pm$1.6 &     0.9$\pm$1.4 &     4.9 &  $-$7.3 &   7.4   &  \\
   RXJ1001.1-7913         &   12.5$\pm$1.1  & $-$20.9$\pm$1.8 &     2.0$\pm$1.6 &     4.0 &  $-$7.6 &	8.6   &  \\
   RXJ1140.3-8321         &   12.7$\pm$1.2  & $-$21.7$\pm$1.8 &     2.5$\pm$1.2 &     4.2 &  $-$8.3 &	8.9   &  \\
   Mean values ($\sigma$) &   11.2 (2.3)    & $-$18.7 (3.8)   &  $-$6.2 (5.2)   & 2.7 (2.3) & $-$5.3 (3.8) & 0.3 (5.2) & \\
\hline
\end{tabular}			    
\begin{flushleft}
{\bf Notes.} \\
(1) The distance inferred from the Hipparcos parallax (Table~5) 
was used in the calculation. \\
(2) Assuming the distance to $\epsilon$~Cha for this star, the following space velocities are obtained $(U, V, W)=(2.0\pm2.0, -14.5\pm2.0, -7.2\pm1.2)$~km/s \\
(3) If the distance to Cha~II is assumed for this object, the resulting space velocities are $(U, V, W)=(20.6\pm8.0, -34.5\pm5.5, 3.9\pm7.7)$~km/s \\
(4) With a parallax $\pi=15.06\pm3.31$, we obtain for this star $(U, V, W)=(1.5\pm7.0, -21.5\pm8.1, -10.2\pm3.7)$~km/s \\
(5) Assuming the distance to Cha~II for this star, we obtain $(U, V, W)=(14.5\pm3.2, -10.8\pm2.4, -25.7\pm1.6)$~km/s 
\end{flushleft}	

\end{table*}
	    

%% file: Members-properties.table
\longtabL{10}{
\begin{longtable}{l c r r}

\caption{~~~~~~~~~~~~~~~~~~~~~~~~~~~~~~~~Table~\ref{tab:pars}: Physical parameters for kinematic members of the Chamaeleon clouds$^{\mathrm{a}}$}
\label{tab:pars}\\
    \hline \hline 
    \multicolumn{1}{c}{2MASS~J} &
    \multicolumn{1}{l}{Class} &
    \multicolumn{1}{c}{T$_{eff}$} &
    \multicolumn{1}{c}{L$_{bol}$} \\ 
    \multicolumn{1}{c}{} &
    \multicolumn{1}{c}{} &
    \multicolumn{1}{c}{K} &
    \multicolumn{1}{c}{Lsun} \\
      \hline
\endfirsthead

\caption{~~~~~~~~~~~~~~~~~~~~~~~~~~~~~~~~Table~\ref{tab:pars}: continued.}\\
   \hline\hline
    \multicolumn{1}{c}{2MASS~J} &
    \multicolumn{1}{l}{Class} &
    \multicolumn{1}{c}{T$_{eff}$} &
    \multicolumn{1}{c}{L$_{bol}$} \\
    \multicolumn{1}{c}{} &
    \multicolumn{1}{c}{} &
    \multicolumn{1}{c}{K} &
    \multicolumn{1}{c}{Lsun} \\
    \hline
\endhead
\hline
\endfoot

  \multicolumn{4}{c}{Chamaeleon~I Cloud}\\

11022491-7733357 & II   & 4205 & 1.2   \\
11023265-7729129 & II   & 3415 & 0.17  \\
11025504-7721508 & II   & 3198 & 0.081 \\
11035682-7721329 & III  & 3342 & 0.21  \\
11040909-7627193 & II   & 4350 & 0.95  \\
11045100-7625240 & III  & 3955 & 0.35  \\
11045285-7625514 & III  & 3596 & 0.26  \\
11051467-7711290 & III  & 3379 & 0.31  \\
11055261-7618255 & III  & 3632 & 0.35  \\
11061540-7721567 & III  & 5770 & 16.0  \\
11064346-7726343 & III  & 3415 & 0.42  \\
11064510-7727023 & II   & 4205 & 1.1   \\
11065906-7718535 & II   & 3234 & 0.11  \\
11071148-7746394 & III  & 3415 & 0.34  \\
11071915-7603048 & III  & 3488 & 0.24  \\
11072074-7738073 & II   & 5860 & 12.0  \\
11074366-7739411 & II   & 3850 & 1.4   \\
11075588-7727257 & III  & 4205 & 4.0   \\
11075792-7738449 & Flat & 4205 & 2.4   \\
11080148-7742288 & II   & 3955 & 3.0   \\
11080329-7739174 & II   &10010 & 70.0  \\
11081648-7744371 & III  & 3306 & 0.15  \\
11083905-7716042 & II   & 3955 & 0.53  \\
11084069-7636078 & III  & 3488 & 0.74  \\
11085464-7702129 & II   & 3778 & 0.34  \\
11091769-7627578 & III  & 4060 & 1.1   \\
11092379-7623207 & II   & 4205 & 0.53  \\
11094006-7628391 & III  & 3669 & 0.68  \\
11095407-7629253 & II   & 3560 & 0.48  \\
11095873-7737088 & II   & 3669 & 0.84  \\
11100469-7635452 & II   & 3705 & 0.43  \\
11100704-7629376 & II   & 3850 & 1.4   \\
11101141-7635292 & II   & 4278 & 1.2   \\
11103801-7732399 & II   & 4730 & 2.6   \\
11105333-7634319 & II   & 3306 & 0.13  \\
11113474-7636211 & III  & 3488 & 0.21  \\
11114632-7620092 & II   & 4205 & 1.3   \\
11115400-7619311 & III  & 3488 & 0.36  \\
11122441-7637064 & II   & 4660 & 1.0   \\
11122772-7644223 & II   & 5410 & 5.0   \\
11124268-7722230 & II   & 5520 & 4.2   \\
11124299-7637049 & III  & 4470 & 0.61  \\
11132446-7629227 & II   & 3342 & 0.073 \\
11132737-7634165 & III  & 3451 & 0.24  \\
11132970-7629012 & III  & 3234 & 0.089 \\
11141565-7627364 & III  & 3306 & 0.15  \\
11145031-7733390 & III  & 3451 & 0.24  \\

  \hline
  \multicolumn{4}{c}{Chamaeleon~II Cloud}\\

12571172-7640111  & II  & 3850 & 1.38  \\
13005323-7654151  & II  & 3777 & 0.20  \\
13005532-7710222  & II  & 3415 & 1.15  \\
13005534-7708296  & III & 3687 & 0.355  \\
13015891-7751218  & II  & 3955 & 0.44  \\
13030444-7707027  & III & 3850 & 1.26  \\
13045571-7739495  & II  & 3850 & 1.047  \\
13052072-7739015  & II  & 4350 & 1.95  \\
13063053-7734001  & II  & 3560 & 0.125  \\
13070922-7730304  & II  & 4060 & 0.89  \\
13072241-7737225  & III & 3705 & 0.29  \\
13080628-7755051  & II  & 4350 & 1.175  \\
13095036-7757240  & II  & 3487 & 0.33  \\
13100415-7710447  & II  & 3415 & 0.23  \\


\end{longtable}
\begin{flushleft}
 {\bf Notes.} \\
	$^{\mathrm{a}}$ References: \citet{luhman2008b}; \citet{alcala2008}; \citet{spezzi2008}
 \end{flushleft}
}